\documentclass[pra,twocolumn]{revtex4-2}

\usepackage{graphicx}
\usepackage{bm}
\usepackage{physics}
\usepackage{amssymb}
\usepackage{xcolor}
\usepackage[hidelinks, colorlinks=true, linkcolor=blue, citecolor=blue, urlcolor=blue]{hyperref}
\usepackage{mhchem}
\usepackage{tipa}

\newcommand{\e}{\mathrm{e}}
\newcommand{\ii}{\mathrm{i}}
\newcommand{\<}{\left\langle}
\renewcommand{\>}{\right\rangle}
\newcommand{\nn}{\nonumber}
\renewcommand{\and}{\quad \mathrm{and} \quad} 

\newcommand{\1}{a}	
\newcommand{\2}{b}
\newcommand{\3}{b^\dagger}
\newcommand{\4}{a^\dagger}
\newcommand{\tb}[4]{\left(t_{#1}, t_{#2}, t_{#3}, t_{#4}\right) }
\newcommand{\tc}[2]{\left(t_{#1}, t_{#2}\right) }

\newcommand{\DD}[1]{\left[ \mathcal{D} {#1}^*  \right]\left[ \mathcal{D} {#1}  \right]}

\begin{document}

\title{Out-of-Equilibrium Dynamics in the Two-Component Bose-Hubbard Model}

\author{Florian R. B\"{a}r}
\author{Malcolm P. Kennett}
\affiliation{Department of Physics, Simon Fraser University, 8888 University Drive, Burnaby, British Columbia, V5A 1S6, Canada}

\date{\today}
\begin{abstract}
We study the out-of-equilibrium dynamics of the Bose-Hubbard model for two-component bosons using a strong-coupling approach within the closed-time-path formalism and develop an effective theory for the action of this problem. We obtain equations of motion for the superfluid order parameters of both boson species for both the superfluid and Mott-insulating phases and study these in the low-frequency, long-wavelength limit during a quantum quench for various initial conditions. We find that an additional degree of freedom for bosons leads to a richer phase diagram and out-of-equilibrium dynamics than the single-component situation.
\end{abstract}

\maketitle

\section{Introduction}

Over the last twenty five years, there has been intense interest in studying cold atoms in optical lattices, both experimentally and theoretically \cite{Bloch2005, Jaksch2005, Morsch2006, Lewenstein2007, Bloch2008, Kennett2013}. The ability to tune system parameters in real time enables the study of out-of-equilibrium dynamics for varied initial conditions, and has spurred interest in the use of cold atoms in optical lattices as quantum simulators \cite{Bloch2012, Choi2016, Takasu2020}. 

Early work on cold atoms in optical lattices focused on realizing the single-species Bose-Hubbard model (BHM) \cite{Fisher1989} in a square optical lattice \cite{Jaksch1998, Greiner2002, Morsch2006}. The BHM is the simplest model that describes interacting bosons on a lattice. The model has a quantum phase transition from a superfluid (SF) phase to a Mott-insulating (MI) phase as the ratio of intersite hopping $J$ and on-site interaction $U$ is varied \cite{Fisher1989}. Studies of the BHM on a square lattice have now been extended to a variety of lattice geometries \cite{Nakafuji2017}, tilted lattices \cite{Simon2011, Sachdev2002}, disorder \cite{Choi2016, Rubio2019}, multiple boson species \cite{Jaksch1998, Altman2003, Kuklov2003, Kuklov2004, Isacsson2005, Mathey2007, Catani2008, Thalhammer2008, Iskin2010, Iskin2010_2, Barman2015, Powell2009, Buonsante2008, Damski2003, Ziegler2003}, and combinations of these \cite{Kuno2014, Rubio2019, Bai2020}.

In this paper, we focus on multiple-component Bose-Hubbard systems, specifically when there are two species, which can be viewed as bosons with an internal degree of freedom. Allowing interactions between bosons of different species gives a larger phase space to be explored and the possibility of new quantum phases \cite{Altman2003, Kuklov2004, Isacsson2005}.

A two-species or two-component realization of the BHM can be achieved in several ways. Techniques include (i) two different types of bosons, e.g. \ce{^{41}K}-\ce{^{87}Rb} mixtures \cite{Catani2008, Thalhammer2008}, (ii) internal states of a single species of bosons to create pseudo-spin-1/2 bosons, e.g. spin states $\ket{F = 2,m_F =-2}$ and $\ket{F = 1, m_F =1}$ of \ce{^{87}Rb} atoms \cite{Gadway2010, Rubio2019},  (iii) bilayers in the limit of no inter-layer tunnelling \cite{Kantian2018}, and (iv) polaritons in optical cavities \cite{Zhang2015}. These spinor bosons are a novel physical system and in the case of cold atoms their inter-species interaction strength can be tuned via Feshbach resonances \cite{Catani2008, Thalhammer2008}.

Studies of the two-component BHM have shown that due to the additional inter-species interaction, a much richer phase space can be obtained than in the single-component case. In addition to the SF and MI phases, other possibilities are a pair SF \cite{Kuklov2004, Hu2009, Chen2010, Menotti2010}, a counterflow SF \cite{Altman2003, Kuklov2003, Hu2009}, charge-density waves \cite{Mishra2008, Hu2009}, a supersolid \cite{Isacsson2005, Hubener2009}, a molecular SF \cite{Lin2020}, and non-integer MI phases \cite{Lin2020}, as well as mixed phases in which each species is in a different quantum phase to the other.  These phases are generally most favoured when the interaction strength for each species of boson is the same \cite{Menotti2010}.  In this work we primarily focus on the situation in which the interaction strengths for each species of boson are unequal.

The BHM realized by cold atoms in optical lattices is an attractive platform to study the out-of-equilibrium (OOE) dynamics of strongly interacting quantum many-body systems. There has been much theoretical and experimental work on this for the single species BHM \cite{Bloch2005, Jaksch2005, Morsch2006, Lewenstein2007, Bloch2008, Kennett2013}. However, despite considerable effort on equilibrium properties, the OOE dynamics and time-dependent phenomena of two-component Bose-Hubbard systems have been explored much less thoroughly \cite{Colussi2022}.

Some of the time-dependent phenomena and OOE dynamics that have been studied include the adiabatic melting of a two-component MI via slow ramping of the lattice potential \cite{Rodriguez2008}, the formation of quantum droplets through control of the trapping potential \cite{Machida2022}, exploration of phase decoherence during a gradual loading of state-dependent optical lattices  \cite{Shim2016}, exploring the effect of quantum fluctuations on SF drag and density/spin fluctuations \cite{Colussi2022}, and studying many-body-localization (MBL) \cite{Choi2016, Rubio2019}. Our interest here is on quench dynamics, in which Hamiltonian parameters are changed faster than the system can respond adiabatically.

Previous work by one of us has focused on the equilibrium properties and OOE dynamics of a single species of bosons in an optical lattice \cite{Kennett2011, Fitzpatrick2018PRA, Fitzpatrick2018NPB, Kennett2020, Mokhtari2021, Mokhtari2023}, including the order parameter dynamics after a quantum quench \cite{Kennett2011} and the spreading of single-particle correlations \cite{Fitzpatrick2018PRA, Mokhtari2021}. This work has combined a strong-coupling approach \cite{Sengupta2005} with the closed-time-path formalism to allow the study of OOE dynamics in the BHM.

Our work in this paper is inspired by the experimental results of Rubio-Abadal \textit{et al.} \cite{Rubio2019}, who studied MBL and the OOE dynamics after a quantum quench in a mixture of two bosons in a two dimensional optical lattice with disorder. They reported that as the population of the second species of bosons was increased, the MBL phase ceased to exist, and argued that the second species acted as a quantum bath that helped thermalize the system as a whole. To gain insight into these observations a theory for OOE dynamics for two-component bosons in disordered optical lattices is required.

The results in this paper are a first step in extending previous work using a strong-coupling approach to the OOE dynamics in the BHM \cite{Kennett2011} to include a second species of bosons. We use a one-particle irreducible strong-coupling approach to the BHM using a closed-time-path method to treat OOE dynamics and derive an expression for the action of the effective theory.  This is our main result. The resulting theory can be applied in both the SF and MI phases. We obtain the saddle-point equations of motion, which we simplify to obtain the mean-field phase boundary at zero and finite temperatures and the mean-field equations for the SF order parameter dynamics during a quantum quench from SF to the MI phases.

This paper is organized as follows: in Sec.~\ref{sec:EffectiveAction}, we derive an effective theory for the two-component BHM using the Schwinger-Keldysh technique, in Sec.~\ref{sec:MFPB}, we discuss the mean-field phase boundary, in Sec.~\ref{sec:EOMs}, we study the saddle-point equations of motion for the SF order parameter dynamics, and in Sec.~\ref{sec:Disc}, we discuss our results and conclude.

\section{Effective Theory}
\label{sec:EffectiveAction}

In this section we discuss the application of the Schwinger-Keldysh or closed-time-path technique to the two-component BHM and derive an effective theory from a strong-coupling approach to the model. The Hamiltonian for the two-component BHM takes the form
\begin{equation}
\hat{H}_\mathrm{BHM} = \hat{H}_J + \hat{H}_0,
\end{equation}
with
\begin{align*}
\hat{H}_J = {} & - \sum_{\< ij\>} J_{a,ij} \left( \hat{a}^\dagger_i \hat{a}^{\phantom{\dagger}}_j  + \hat{a}^\dagger_j \hat{a}^{\phantom{\dagger}}_i \right) -  \sum_{\< ij\>} J_{b,ij} \left( \hat{b}^\dagger_i \hat{b}^{\phantom{\dagger}}_j +  \hat{b}^\dagger_j \hat{b}^{\phantom{\dagger}}_i \right), 
\end{align*}
and
\begin{align*}
\hat{H}_0 = {} & \hat{H}_U - \mu\hspace*{-0.2cm} \sum_{i, x=a,b}\hspace*{-0.2cm} \hat{n}_{x i}.
\end{align*}
It contains single-site terms with
\begin{align*}
\hat{H}_U = {} & \frac{1}{2}\hspace*{-0.1cm}\sum_{i,x=a,b}\hspace*{-0.2cm}U_x \hat{n}_{x i}\left( \hat{n}_{x i} -1 \right) + V \sum_i \hat{n}_{a i} \hat{n}_{b i},
\end{align*}
where $\hat{a}^\dagger_i ( \hat{b}^\dagger_i )$ and $\hat{a}_i (\hat{b}_i)$ are the creation and annihilation operators on site $i$ for bosons of species $a (b)$, respectively, $\hat{n}_{xi} = \hat{x}^\dagger_i \hat{x}^{\phantom{\dagger}}_i$ is the number operator for species $x$, $U_x$ is the intra-species interaction strength between species $x$ bosons, $V$ is the inter-species interaction strength, and $\mu$ is the chemical potential. The hopping parameter for species $a$($b$) is given by $J_{a,ij}$ ($J_{b,ij}$), and $\<ij\>$ indicates that the sum is to be taken over neighbouring sites $i$ and $j$. We choose $J_{a,ij} = J_a$ for nearest neighbour sites and 0 otherwise, and similarly for $J_{b,ij}$. We also allow for the possibility that $J_{a,ij}$ and $J_{b,ij}$ are time-dependent. In experiments, there is a trapping potential, which can be included in a local density approximation via a position dependent chemical potential \cite{DeMarco2005, Gerbier2007}. We do not consider the effects of a trap here.

We restrict ourselves to repulsive interactions that satisfy $0<V<U_a,U_b$ to prevent phase separation as discussed in Sec.~\ref{sec:MFPB}. This is a stronger restriction than the usual phase separation criterion, $V^2<U_a U_b$ \cite{Ho1996, Ao1998, Trippenbach2000, Chen2003, Suthar2017}. We also will generally assume $U_a \neq U_b$. Additionally, the Hamiltonian implies number conservation for each species separately. 

\subsection{Schwinger-Keldysh technique}
\label{sec:Schwinger}

\begin{figure}[b]
\centering
\includegraphics[width=0.47\textwidth]{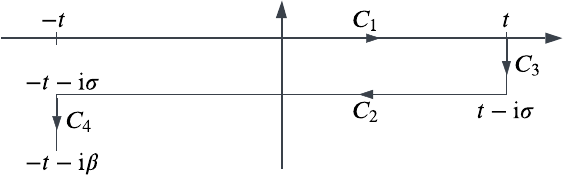}
\caption{Contour for the Schwinger-Keldysh technique for a system with inverse temperature $\beta$. The value of $\sigma$ is arbitrary in the interval $\left[ 0, \beta \right]$ \cite{Landsman1987}.} \label{fig:Schwinger_Keldysh_contour}
\end{figure}

The Schwinger-Keldysh  \cite{Schwinger1961, Keldysh1964, Rammer1986, Niemi1984, Landsman1987, Chou1985} technique is a convenient formalism to describe OOE dynamics of quantum many-body systems. In this formalism, time is promoted to a complex variable lying along a contour in the complex plane and the notion of time-ordering is replaced by contour-ordering in order to calculate Green's functions \cite{Niemi1984}. We evolve the system in time from $-\infty$ to $+\infty$ and back to $-\infty$ along a contour like the one shown in Fig.~\ref{fig:Schwinger_Keldysh_contour}.  This makes it possible to describe both zero and finite temperature systems, as well as both equilibrium properties and OOE dynamics all within one formalism. The number of fields in the theory is doubled, with the additional fields propagating backwards in time. 

For a thermal initial state, as we will assume here, the generating functional factorizes \cite{Landsman1987}:
\begin{equation}
\mathcal{Z} = \mathcal{Z}_{C_1 \cup C_2}\mathcal{Z}_{C_3 \cup C_4},
\end{equation}
with $C_1$, $C_2$, $C_3$, and $C_4$ contour segments as shown in Fig. \ref{fig:Schwinger_Keldysh_contour}. The value of $0 \leq \sigma \leq \beta = (k_\mathrm{B}T)^{-1}$, where $T$ is temperature, is arbitrary \cite{Landsman1987} and we choose $\sigma = 0$ for simplicity.

\subsection{Effective theory for the two-component Bose-Hubbard model}

We can write the generating functional as
\begin{equation}
\mathcal{Z} = \int \DD{a} \DD{b} \e^{\ii S_\mathrm{BHM}\left[ a^*, a, b^*, b \right]},
\end{equation}
where $a$ and $b$ are bosonic fields, 
and we omit source fields and set $\hbar = 1$. The action for the BHM takes the form
\begin{widetext}
\begin{align}
S_\mathrm{BHM} = {}&  \int_{-\infty}^\infty \dd{t} \sum_i \left[ a^*_{i \alpha}(t) \left(\ii \partial_t  \right) \tau^3_{\alpha \beta}a^{\phantom{*}}_{i \beta}(t) \right]
+  \int_{-\infty}^\infty \dd{t} \sum_i \left[ b^*_{i \alpha}(t) \left(\ii \partial_t  \right) \tau^3_{\alpha \beta} b^{\phantom{*}}_{i \beta}(t) \right] 
+ S_J + S_U,
\end{align}
where
\begin{align}
S_J = {}& - \int_{-\infty}^\infty \dd{t} \sum_{\< ij \>} J_{a,ij} \left[ a^*_{i \alpha}(t) \tau^3_{\alpha \beta}a^{\phantom{*}}_{j \beta}(t) + a^*_{j \alpha}(t) \tau^3_{\alpha \beta}a^{\phantom{*}}_{i \beta}(t) \right] 
- \int_{-\infty}^\infty \dd{t} \sum_{\< ij \>} J_{b,ij} \left[ b^*_{i \alpha}(t) \tau^3_{\alpha \beta}b^{\phantom{*}}_{j \beta}(t) + b^*_{j \alpha}(t) \tau^3_{\alpha \beta}b^{\phantom{*}}_{i \beta}(t) \right],
\end{align}
\end{widetext}
and $S_U$ is the action associated with $\hat{H}_0$. $a_{i \alpha}$ and $b_{i \alpha}$ are the fields on site $i$ on contour $\alpha$, with $\alpha = 1$ or $2$. The Pauli matrices are denoted by $\tau^i_{\alpha \beta}$ and act in Keldysh space rather than spin space. We use the Einstein summation convention for Keldysh indices.

We perform a Keldysh rotation for the $a$ fields such that
\begin{align}
\begin{pmatrix} a_1(t)  \\ a_2(t) \end{pmatrix} \longrightarrow \begin{pmatrix}
a_q (t) \\ a_c (t) \end{pmatrix} = \hat{L} \begin{pmatrix} a_1 (t)  \\ a_2 (t) \end{pmatrix},
\end{align}
with
\begin{equation}
\hat{L} = \frac{1}{\sqrt{2}} \begin{pmatrix}
1 & -1\\ 1 & 1
\end{pmatrix}.
\end{equation}
and similarly for the $b$ fields, where the subscripts $q$ and $c$ indicate the quantum and classical components of the field \cite{Cugliandolo1999}, respectively. This has the effect of transforming $\tau^3_{\alpha \beta}$ in the $(1,2)$ basis into $\tau^1_{\alpha \beta}$ in the $(q,c)$ basis. Thus, after the Keldysh rotation

\begin{widetext}
\begin{align}
S_J = {} & - \int_{-\infty}^\infty \dd{t} \sum_{\< ij \>} J_{a,ij} \left[ a^*_{i \alpha}(t) \tau^1_{\alpha \beta}a^{\phantom{*}}_{j \beta}(t) + a^*_{j \alpha}(t) \tau^1_{\alpha \beta}a^{\phantom{*}}_{i \beta}(t) \right]
- \int_{-\infty}^\infty \dd{t} \sum_{\< ij \>} J_{b,ij} \left[ b^*_{i \alpha}(t) \tau^1_{\alpha \beta}b^{\phantom{*}}_{j \beta}(t) + b^*_{j \alpha}(t) \tau^1_{\alpha \beta}b^{\phantom{*}}_{i \beta}(t) \right].
\end{align}
\end{widetext}

We are interested in studying quantum quenches in which the hopping varies as a function of time and the system crosses from the SF to the MI phase. This requires a formalism that is valid in both phases. The generalization of the strong-coupling method used in imaginary time by Sengupta and Dupuis \cite{Sengupta2005} to real time \cite{Kennett2011} can achieve this. This results in a normalized spectral function and enables the calculation of the excitation spectrum and momentum distribution in the SF phase, while also giving a good description in the MI phase \cite{Kennett2011}. Here, we generalize from a single boson  species to two species of bosons.

The approach requires two Hubbard-Stratonovich (HS) transformations. The first of these decouples the hopping terms. We introduce HS fields $\psi$ and $\chi$ for species $a$ and $b$, respectively, and use the identity \cite{Kennett2011}
\begin{align}
\e^{-\ii k \left( \xi^* \eta + \xi \eta^* \right)} = \int & \overline{\mathcal{D}} (\varphi^{\phantom{*}}_1, \varphi_1^*)  \mathcal{D}(\varphi^{\phantom{*}}_2, \varphi_2^*) \e^{\frac{\ii}{k} \left( \varphi_2^* \varphi^{\phantom{*}}_1 + \varphi_1^* \varphi^{\phantom{*}}_2 \right)} \nn \\
& \times \e^{\ii\left( \varphi_1^* \xi + \varphi^{\phantom{*}}_1 \xi^* + \varphi_2^* \eta + \varphi^{\phantom{*}}_2 \eta^* \right) },
\end{align}
to write 
\begin{align}
\mathcal{Z} 
= & \int \DD{\psi} \DD{\chi}  \e^{\ii W\left[ \psi^*, \psi, \chi^*, \chi \right]} \nn \\
& \times \e^{-\frac{\ii}{2} \int_{-\infty}^\infty \dd{t} \sum\limits_{\< ij \>} \left[ \psi^*_{i \alpha}(t) J^{-1}_{a, ij} \tau^1_{\alpha \beta} \psi^{\phantom{*}}_{j \beta}(t) + \chi^*_{i \alpha}(t) J^{-1}_{b, ij} \tau^1_{\alpha \beta} \chi^{\phantom{*}}_{j \beta}(t) \right]},  \label{eq:Generating_Functional_HS1}
\end{align}
with
\begin{equation}
\e^{\ii W\left[ \psi^*, \psi, \chi^*, \chi \right]} = \< \e^{\ii \int_{-\infty}^\infty \dd{t} \sum_i \left[ R_i(t) + Q_i(t) \right]} \>_0,
\end{equation}
where
\begin{align}
R_i(t) = {} & \psi^*_{i\alpha}(t) \tau^1_{\alpha \beta} a^{\phantom{*}}_{i \beta}(t) + \psi^{\phantom{*}}_{i \alpha}(t) \tau^1_{\alpha \beta} a^*_{i \beta}(t), \\
Q_i(t) = {} & \chi^*_{i\alpha}(t) \tau^1_{\alpha \beta} b^{\phantom{*}}_{i \beta}(t) + \chi^{\phantom{*}}_{i \alpha}(t) \tau^1_{\alpha \beta} b^*_{i \beta}(t),
\end{align}
and the average $\< \dots \>_0 $ is taken with respect to
\begin{align}
S_0 = \int_{-\infty}^\infty \dd{t} \sum_i & \left[ a^*_{i \alpha}(t) \left( \ii \partial_t \right) \tau^1_{\alpha \beta} a^{\phantom{*}}_{i \beta}(t) \right. \nn \\
& \left. {} + b^*_{i \alpha}(t) \left( \ii \partial_t \right) \tau^1_{\alpha \beta} b^{\phantom{*}}_{i \beta}(t) \right] + S_U \nonumber.
\end{align}

We can now use the generator of connected Green's functions $W\left[ \psi^*, \psi, \chi^*, \chi \right]$ to calculate the $2(n_a+n_b)$-point connected Green's function where $2n_a$ and $2n_b$ are the number of $a$ and $b$ fields in the correlator, respectively:
\begin{widetext}
\begin{align}
& G^{(n_a+n_b),c}_{ \alpha_1 ...\alpha_{n_a}  \alpha''_1...\alpha''_{n_b} \alpha'''_{n_b}...\alpha'''_1 \alpha'_{n_a} ... \alpha'_1} (t^{\phantom{'}}_1, ..., t^{\phantom{'}}_{n_a}, t''_1, ..., t''_{n_b}, t'''_{n_b}, ..., t'''_1, t'_{n_a}, ..., t'_1) \nn \\
= & \, \ii \e^{-\ii W[0]} \tau^1_{\alpha_1 \beta_1} ...\tau^1_{\alpha_{n_a} \beta_{n_a}}  \tau^1_{\alpha''_1 \beta''_1} ...\tau^1_{\alpha''_{n_b}  \beta''_{n_b}} \tau^1_{\alpha'''_{n_b} \beta'''_{n_b}} ...\tau^1_{\alpha'''_1 \beta'''_1} \tau^1_{\alpha'_{n_a} \beta'_{n_a}} ... \tau^1_{\alpha'_1  \beta'_1} \nn \\
& \times \left. \frac{\delta^{2(n_a+n_b)} \e^{\ii W\left[ \psi^*, \psi, \chi^*, \chi \right]} }{\delta \psi^*_{i\beta_1}(t^{\phantom{'}}_1) ... \delta \psi^*_{i \beta_{n_a}}(t^{\phantom{'}}_{n_a})  \delta \chi^*_{i \beta''_1}(t''_1) ... \delta \chi^*_{i \beta''_{n_b}}(t''_{n_b}) \delta \chi^{\phantom{*}}_{i \beta'''_{n_b}}(t'''_{n_b}) ... \delta \chi^{\phantom{*}}_{i \beta'''_1}(t'''_1) \delta \psi^{\phantom{*}}_{i\beta'_{n_a}}(t'_{n_a}) ... \delta \psi^{\phantom{*}}_{i \beta'_1}(t'_1)  } \right|_{ \begin{matrix} \psi^* = \psi = 0 \\ \chi^* = \chi = 0 
\end{matrix}} \nn \\
= & \, \ii (-1)^{n_a+n_b} \< a^{\phantom{*}}_{i \alpha_1}(t^{\phantom{'}}_1)... a^{\phantom{*}}_{i \alpha_{n_a}}(t^{\phantom{'}}_{n_a})  b^{\phantom{*}}_{i \alpha''_1}(t''_1) ... b^{\phantom{*}}_{i \alpha''_{n_b}}(t''_{n_b}) b^*_{i \alpha'''_{n_b}}(t'''_{n_b}) ... b^*_{i \alpha'''_1} (t'''_1) a^*_{i \alpha'_{n_a}} (t'_{n_a}) ... a^*_{i \alpha'_1}(t'_1) \>^c_0, \label{eq:G_ab_definition} 
\end{align}
where the superscript $c$ indicates a connected Green's function. We can now invert Eq. \eqref{eq:G_ab_definition} to find an expression for $W$ in terms of the connected Green's functions
\begin{align}
\ii W\left[ \psi^*, \psi, \chi^*, \chi \right] = & -\ii \sum_i \sum_{n_a+n_b > 0} \frac{1}{[(n_a+n_b)!]^2} \int_{-\infty}^\infty \left[ \prod_{l=1}^{n_a} \dd{t^{\phantom{'}}_l} \dd{t'_l} \tau^1_{\alpha_l \beta_l} \tau^1_{\alpha'_l \beta'_l} \psi^*_{i\alpha_l}(t^{\phantom{'}}_l) \psi^{\phantom{*}}_{i \alpha'_l}(t'_l)  \right] \left[ \prod_{l=1}^{n_b} \dd{t''_l} \dd{t'''_l} \tau^1_{\alpha''_l \beta''_l} \tau^1_{\alpha'''_l \beta'''_l} \right. \nn \\
& \times \left. \vphantom{\prod_{l=1}^{n_b}} \chi^*_{i \alpha''_l} (t''_l) \chi^{\phantom{*}}_{i \alpha'''_l}(t'''_l)  \right] G^{(n_a+n_b),c}_{ \beta_1\ldots \beta_{n_a} \beta''_1 \ldots \beta''_{n_b} \beta'''_{n_b} \ldots \beta'''_1 \beta'_{n_a} \ldots \beta'_1}  (t^{\phantom{'}}_1,  \ldots , t^{\phantom{'}}_{n_a}, t''_1,  \ldots , t''_{n_b}, t'''_{n_b},  \ldots , t'''_1, t'_{n_a},  \ldots , t'_1)  \nn \\
= {} & \ii \sum_i \sum_{n_a+n_b>0} S^{n_a+n_b}_{\mathrm{int}} \left[ \psi^*, \psi, \chi^*, \chi \right]. \label{eq:iW}
\end{align}

After the first HS transformation the action obtained from Eq.~\eqref{eq:Generating_Functional_HS1} truncated to quartic order is
\begin{align}
S^I_\mathrm{eff} \left[ \psi^*, \psi, \chi^*, \chi \right] 
= {} & - \int_{-\infty}^\infty \dd{t} \sum_{\< ij \>} \psi^*_{i \alpha}(t) \left( 2J_{a,ij} \right)^{-1} \tau^1_{\alpha \beta} \psi^{\phantom{*}}_{j \beta}(t)
 - \int_{-\infty}^\infty \dd{t} \sum_{\< ij \>} \chi^*_{i \alpha}(t) \left( 2J_{b,ij} \right)^{-1} \tau^1_{\alpha \beta} \chi^{\phantom{*}}_{j \beta}(t) \nn \\
& - \int_{-\infty}^\infty \dd{t} \dd{t'} \sum_i \psi^*_{i \alpha} (t) \tau^1_{\alpha \beta} G^{1a,c}_{ \beta \beta'}(t, t') \tau^1_{\alpha' \beta'} \psi^{\phantom{*}}_{i \alpha'}(t') - \int_{-\infty}^\infty \dd{t''} \dd{t'''} \sum_i \chi^*_{i \alpha} (t'') \tau^1_{\alpha \beta} G^{1b,c}_{ \beta \beta'}(t'', t''') \tau^1_{\alpha' \beta'} \chi^{\phantom{*}}_{i \alpha'}(t''') \nn \\
& -  \frac{1}{4} \int_{-\infty}^\infty \dd{t^{\phantom{*}}_1} \dd{t'_1} \dd{t^{\phantom{1}}_2} \dd{t'_2} \sum_i \psi^*_{i \alpha^{\phantom{1}}_1}(t^{\phantom{1}}_1) \psi^*_{i \alpha^{\phantom{*}}_2}(t^{\phantom{*}}_2) \tau^1_{\alpha^{\phantom{1}}_1 \beta^{\phantom{1}}_1} \tau^1_{\alpha^{\phantom{1}}_2 \beta^{\phantom{1}}_2}  G^{2a,c}_{ \beta^{\phantom{1}}_1  \beta^{\phantom{1}}_2 \beta'_2 \beta'_1}(t^{\phantom{1}}_1, t^{\phantom{1}}_2, t'_2, t'_1) \tau^1_{\alpha'_2 \beta'_2} \tau^1_{\alpha'_1 \beta'_1} \psi^{\phantom{*}}_{i \alpha'_2}(t'_2) \psi^{\phantom{*}}_{i \alpha'_1}(t'_1) \nn \\
& -  \frac{1}{4} \int_{-\infty}^\infty \dd{t''_1} \dd{t'''_1} \dd{t''_2} \dd{t'''_2} \sum_i \chi^*_{i \alpha^{\phantom{1}}_1}(t''_1) \chi^*_{i \alpha^{\phantom{1}}_2}(t''_2) \tau^1_{\alpha^{\phantom{1}}_1 \beta^{\phantom{1}}_1} \tau^1_{\alpha^{\phantom{1}}_2 \beta^{\phantom{1}}_2}  G^{2b,c}_{ \beta^{\phantom{1}}_1 \beta^{\phantom{1}}_2 \beta'_2  \beta'_1 }(t''_1, t''_2, t'''_2, t'''_1) \tau^1_{\alpha'_2 \beta'_2} \tau^1_{\alpha'_1 \beta'_1} \chi^{\phantom{*}}_{i \alpha'_2}(t'''_2) \chi^{\phantom{*}}_{i \alpha'_1}(t'''_1) \nn \\
& - \frac{1}{4} \int_{-\infty}^\infty \dd{t} \dd{t'} \dd{t''} \dd{t'''} \sum_i \psi^*_{i \alpha}(t) \chi^*_{i \alpha''}(t'') \tau^1_{\alpha \beta} \tau^1_{\alpha'' \beta''}  G^{1a+1b,c}_{\beta \beta'' \beta''' \beta'}(t, t'', t''', t') \tau^1_{\alpha''' \beta'''} \tau^1_{\alpha' \beta'}\chi^{\phantom{*}}_{i \alpha'''}(t''')  \psi^{\phantom{*}}_{i \alpha'}(t'). \label{eq:S^1_eff}
\end{align}
Only the term on the last line in Eq.~\eqref{eq:S^1_eff}, which describes the interaction between the two different boson species, is not present in the single-species case \cite{Kennett2011, Fitzpatrick2018PRA, Fitzpatrick2018NPB}, however other terms that are present in the single-species model now have two copies (one per species).
As discussed by Sengupta and Dupuis \cite{Sengupta2005} in the single component case, the action after a single HS transformation gives the correct mean field phase boundary.  However, they also showed that this form of the action displays an unphysical excitation spectrum in the SF phase, which they rectified by performing a second HS transformation. We follow Refs.~\cite{Kennett2011, Sengupta2005} and perform a second HS transformation, introducing two additional HS fields $y$ and $z$, such that
\begin{equation}
\e^{ -\frac{\ii}{2} \int_{-\infty}^\infty \dd{t} \sum\limits_{\<ij\>} \psi^*_{i\alpha}(t) J^{-1}_{a,ij} \tau^1_{\alpha\beta} \psi^{\phantom{*}}_{j\beta}(t)} = \int \DD{y} \e^{\ii \int_{-\infty}^\infty \dd{t} \sum\limits_{\<ij\>}  y^*_{i\alpha}(t) 2J_{a,ij} \tau^1_{\alpha\beta} y^{\phantom{*}}_{j \beta}(t) }  \e^{\ii \int_{-\infty}^\infty \dd{t} \sum\limits_i \left[ y^*_{i\alpha}(t) \tau^1_{\alpha \beta} \psi^{\phantom{*}}_{i\beta}(t) + y^{\phantom{*}}_{i\alpha}(t) \tau^1_{\alpha \beta} \psi^*_{i\beta}(t)  \right]  }, \nn \\
\end{equation}
and
\begin{equation}
\e^{ -\frac{\ii}{2} \int_{-\infty}^\infty \dd{t} \sum\limits_{\<ij\>} \chi^*_{i\alpha}(t) J^{-1}_{b,ij} \tau^1_{\alpha \beta} \chi^{\phantom{*}}_{j\beta}(t)}  = \int \DD{z} \e^{\ii \int_{-\infty}^\infty \dd{t} \sum\limits_{\<ij\>}  z^*_{i\alpha}(t) 2J_{b,ij} \tau^1_{\alpha \beta} z^{\phantom{*}}_{j \beta}(t)}  \e^{\ii \int_{-\infty}^\infty \dd{t} \sum\limits_i \left[ z^*_{i\alpha}(t) \tau^1_{\alpha \beta} \chi^{\phantom{*}}_{i\beta}(t) + z^{\phantom{*}}_{i\alpha}(t) \tau^1_{\alpha \beta} \chi^*_{i\beta}(t)  \right]  }. \nn \\
\end{equation}
As discussed in Refs.~\cite{Kennett2011, Sengupta2005, Fitzpatrick2018NPB}, it can be shown that the $y$ and $z$ fields have the same correlations as the original $a$ and $b$ fields, respectively. We then have 
\begin{align}
\mathcal{Z} 
= &\int \DD{y} \DD{z} \e^{\ii \int_{-\infty}^\infty \dd{t} \sum\limits_{\<ij\>}  y^*_{i\alpha}(t) 2J_{a,ij} \tau^1_{\alpha\beta} y^{\phantom{*}}_{j \beta}(t) }  \e^{\ii \int_{-\infty}^\infty \dd{t} \sum\limits_{\<ij\>}  z^*_{i\alpha}(t) 2J_{b,ij} \tau^1_{\alpha \beta} z^{\phantom{*}}_{j \beta}(t)} \e^{\ii \tilde{W}[y^*, y, z^*, z]}, 
\end{align}
where
\begin{align}
\e^{\ii \tilde{W} [y^*, y, z^*, z ]} = & \int \DD{\psi} \DD{\chi}  \e^{\ii \int_{-\infty}^\infty \dd{t} \sum\limits_i \left[ \tilde{R}_i(t) + \tilde{Q}_i(t) \right]  } \e^{\ii W[\psi^*,\psi,\chi^*, \chi]} \nn \\
= & \<  \e^{\ii \int_{-\infty}^\infty \dd{t} \sum\limits_i \left[ \tilde{R}_i(t) + \tilde{Q}_i(t) \right] + \ii \sum\limits_{n_a+n_b = 2} S^{n_a+n_b}_\mathrm{int}\left[ \psi^*, \psi, \chi^*, \chi \right] } \>_{S_{G_a}+S_{G_b}}, \label{eg:eiWtildetruncated}
\end{align}
with
\begin{equation}
\tilde{R}_i(t) = y^*_{i\alpha} (t) \tau^1_{\alpha\beta} \psi^{\phantom{*}}_{i\beta}(t) + y^{\phantom{*}}_{i\alpha} (t) \tau^1_{\alpha\beta} \psi^*_{i\beta}(t), 
\end{equation}   
\begin{equation}
\tilde{Q}_i(t) = z^*_{i\alpha} (t) \tau^1_{\alpha \beta} \chi^{\phantom{*}}_{i\beta}(t) + z^{\phantom{*}}_{i\alpha} (t) \tau^1_{\alpha \beta} \chi^*_{i\beta}(t),
\end{equation}
\begin{equation}
S_{G_a} = - \sum_i \int_{-\infty}^\infty \dd{t} \dd{t'}   \psi^*_{i\alpha}(t) \tau_{\alpha \beta}^1 G^{1a,c}_{\beta \beta'}(t,t') \tau^1_{\alpha' \beta'} \psi^{\phantom{*}}_{i \alpha'}(t'),
\end{equation}
and
\begin{equation}
S_{G_b} = - \sum_i \int_{-\infty}^\infty \dd{t} \dd{t'}  \chi^*_{i\alpha}(t) \tau_{\alpha \beta}^1 G^{1b,c}_{\beta \beta'}(t,t') \tau^1_{\alpha' \beta'} \chi^{\phantom{*}}_{i \alpha'}(t').
\end{equation}
We perform a cumulant expansion for $\tilde{W}\left[ y^*, y, z^*, z \right]$, keeping only terms that are not anomalous, i.e. tree level diagrams \cite{Fitzpatrick2018NPB, Sengupta2005}, to obtain
\begin{equation}
\mathcal{Z} = \int \DD{y} \DD{z} \e^{\ii S^{II}_\mathrm{eff} \left[y^*, y, z^*, z \right]}.
\end{equation}
The action of the effective theory truncated to quartic order is
\begin{align}
S^{II}_\mathrm{eff}\left[ y^*, y, z^*, z \right] = & \int_{-\infty}^\infty \dd{t} \sum\limits_{\<ij\>}  y^*_{i\alpha}(t) 2J_{a,ij} \tau^1_{\alpha\beta} y^{\phantom{*}}_{j \beta}(t) + \int_{-\infty}^\infty \dd{t} \sum\limits_{\<ij\>}  z^*_{i\alpha}(t) 2J_{b,ij} \tau^1_{\alpha \beta} z^{\phantom{*}}_{j \beta}(t) \nn \\
& +  \int_{-\infty}^\infty \dd{t_1} \dd{t_2} \sum_i y^*_{i\alpha_1}(t_1)  \left[ G^a_{\alpha_2 \alpha_1}(t_2, t_1) \right]^{-1} y^{\phantom{*}}_{i\alpha_2}(t_2) + \int_{-\infty}^\infty \dd{t_1}\dd{t_2}  \sum_i z^*_{i\alpha_1}(t_1)\left[ G^b_{\alpha_2 \alpha_1}(t_2, t_1) \right]^{-1} z^{\phantom{*}}_{i\alpha_2}(t_2)  \nn \\
& + \frac{1}{4}  \int_{-\infty}^\infty \dd{t_1} \dd{t_2} \dd{t_3} \dd{t_4} \sum_i u^a_{\alpha_1 \alpha_2 \alpha_3 \alpha_4}(t_1, t_2, t_3, t_4)  y^*_{i\alpha_1}(t_1) y^*_{i\alpha_2}(t_2) y^{\phantom{*}}_{i\alpha_3}(t_3) y^{\phantom{*}}_{i\alpha_4}(t_4)   \nn\\
& + \frac{1}{4}  \int_{-\infty}^\infty \dd{t_1} \dd{t_2} \dd{t_3} \dd{t_4} \sum_i u^b_{\alpha_1 \alpha_2 \alpha_3 \alpha_4}(t_1, t_2, t_3, t_4)  z^*_{i\alpha_1}(t_1) z^*_{i\alpha_2}(t_2) z^{\phantom{*}}_{i\alpha_3}(t_3) z^{\phantom{*}}_{i\alpha_4}(t_4) \nn \\
& + \int_{-\infty}^\infty \dd{t_1} \dd{t_2} \dd{t_3} \dd{t_4}  \sum_i u^{ab}_{\alpha_1 \alpha_2 \alpha_3 \alpha_4}(t_1, t_2, t_3, t_4)  y^*_{i\alpha_1}(t_1) y^{\phantom{*}}_{i\alpha_2}(t_2) z^*_{i\alpha_3}(t_3) z^{\phantom{*}}_{i\alpha_4}(t_4), \label{eq:S2_eff}
\end{align}
where
\begin{align}
\left[ G^a_{\alpha_2 \alpha_1}(t_2, t_1)  \right]^{-1} = {} & \ii \tau^1_{\alpha_1 \beta_1} \tau^1_{\alpha_2 \beta_2} \< \psi^{\phantom{*}}_{i\beta_1}(t_1) \psi^*_{i\beta_2}(t_2) \>, \nn \\
\left[ G^b_{\alpha_2 \alpha_1}(t_2, t_1)  \right]^{-1} = {} & \ii \tau^1_{\alpha_1 \beta_1} \tau^1_{\alpha_2 \beta_2} \< \chi^{\phantom{*}}_{i\beta_1}(t_1) \chi^*_{i\beta_2}(t_2) \>. \label{eq:Greens_functions_for_PD}
\end{align}
\end{widetext}
The truncation to quartic order in the fields in the derivation of both Eq.~(\ref{eq:S^1_eff}) and the effective theory $S^{II}_{\rm eff}$ are uncontrolled approximations.  However, in the single component case, the effective theory has been found to give 
excellent agreement with exact diagonalization and exact analytical results \cite{Fitzpatrick2018PRA} giving us confidence in its use here.  The theory is exact in the limit of $J = 0$ and $U=0$ 
and in the single component case, is most accurate for large $U/J$, but still provides good accuracy for intermediate values of $U/J$ \cite{Mokhtari2021}.

Note that the quartic couplings $u^a$, $u^b$, and $u^{ab}$ generated in the cumulant expansion are non-local in time, explicit expressions for them can be found in Appendix~\ref{app:couplings}.

From the definition in Eq.~\eqref{eq:G_ab_definition} we can see that (under re-labelling of $a \leftrightarrow b$)
\begin{equation}
G^{1a+1b,c}_{\alpha_5 \alpha_6 \alpha'_6 \alpha'_5}(t_5, t_6, t'_6, t'_5) = G^{1b+1a,c}_{\alpha_6 \alpha_5 \alpha'_5 \alpha'_6}(t_6, t_5, t'_5, t'_6), \nn
\end{equation}
which is a slightly more restrictive symmetry compared to the single species equivalents (here for species $a$) \cite{Kennett2013, Kennett2011}
\begin{align*}
G^{2a,c}_{\alpha_5 \alpha_6 \alpha'_6 \alpha'_5} (t_5, t_6, t'_6, t'_5)  = {} & G^{2a,c}_{\alpha_6 \alpha_5 \alpha'_6 \alpha'_5}(t_6, t_5, t'_6, t'_5)  \\
= {} & G^{2a,c}_{\alpha_5 \alpha_6 \alpha'_5 \alpha'_6} (t_5, t_6, t'_5, t'_6). 
\end{align*}
This leaves us with nine independent components that need to be evaluated: $G^{1a+1b,c}_{cccc}$, $G^{1a+1b,c}_{qccc}$, $G^{1a+1b,c}_{cccq}$, $G^{1a+1b,c}_{qqcc}$, $G^{1a+1b,c}_{qcqc}$, $G^{1a+1b,c}_{qccq}$, $G^{1a+1b,c}_{ccqq}$, $G^{1a+1b,c}_{cqqq}$, $G^{1a+1b,c}_{qqqc}$, and the remaining four-point function $G^{1a+1b,c}_{qqqq} = 0$ by causality \cite{Chou1985}.
Explicit expressions for each of the non-trivial four-point functions are given in Appendix~\ref{app:four-point}. For our calculations of the simplified equations of motion, we find that we require only $G^{1a+1b,c}_{qccc}$, but the expressions in Appendix~\ref{app:four-point} allow for a more general study, e.g. when calculating correlation functions \cite{Fitzpatrick2018NPB}.

The mean-field phase boundaries for species $a$ and $b$ can be determined from Eq.~\eqref{eq:S2_eff} from the vanishing of the coefficients of $y^*_q y^{\phantom{*}}_c$ and $z^*_q z^{\phantom{*}}_c$, respectively, by using the definitions in Eq.~\eqref{eq:Greens_functions_for_PD} and noting that the matrix Green's function takes  the form
\begin{equation}
\hat{G}^x(t_1, t_2) = \begin{pmatrix}
0 & \mathcal{G}^{x,A}_0(t_1, t_2) \\
\mathcal{G}^{x,R}_0(t_1, t_2) & \mathcal{G}^{x,K}_0(t_1, t_2)
\end{pmatrix}, \nn
\end{equation} 
where $\mathcal{G}^{x,A}_0(t_1, t_2)$, $\mathcal{G}^{x,R}_0(t_1, t_2)$, and $\mathcal{G}^{x,K}_0(t_1, t_2)$ are the advanced, retarded, and Keldysh Green's functions for species $x$, determined using the single site Hamiltonian $\hat{H}_0$, respectively. We then find the inverse matrix Green's function to be
\begin{equation}
\left[   \hat{G}^x (t_1, t_2)   \right]^{-1} = \begin{pmatrix}
\left[ \left( \mathcal{G}_0^x \right)^{-1} \right]^{K} (t_1, t_2)  &  \left[ \left( \mathcal{G}_0^x \right)^{-1} \right]^{R} (t_1, t_2)  \\
\left[ \left( \mathcal{G}_0^x \right)^{-1} \right]^{A} (t_1, t_2)  & 0
\end{pmatrix}, \nn
\end{equation}
where
\begin{align}
\left[ \left( \mathcal{G}_0^x \right)^{-1} \right]^{R} (t_1, t_2) = {} & \left[ \mathcal{G}_0^{x,R} (t_1, t_2)  \right]^{-1}, \\
\left[ \left( \mathcal{G}_0^x \right)^{-1} \right]^{A} (t_1, t_2) = {} & \left[ \mathcal{G}_0^{x,A} (t_1, t_2)  \right]^{-1}, \\
\left[ \left( \mathcal{G}_0^x \right)^{-1} \right]^{K} (t_1, t_2) = {} & - \int \dd{t'} \dd{t''} \left[  \mathcal{G}^{x,R}_0  (t_1,  t') \right]^{-1} \nn \\
& \times \mathcal{G}_0^{x,K}(t', t'') \left[ \mathcal{G}_0^{x,A}  (t'', t_2) \right]^{-1},
\end{align}
and
\begin{equation}
\mathcal{G}^{x,R}_0(t_1, t_2) = \mathcal{G}^{x,A}_0(t_2, t_1),
\end{equation}
which allow us to determine the equations for the mean-field phase boundary in Sec.~\ref{sec:MFPB}.

\section{Mean-Field Phase Boundary}
\label{sec:MFPB}

We determine the mean-field phase boundary between the SF and MI phases by determining when the coefficients of the quadratic terms in the action, Eq. \eqref{eq:S^1_eff}, vanish \cite{Kennett2011}. In this section, we highlight the most important results, and derive an accurate expression for the zero and non-zero temperature phase boundary. Full details of this derivation are given in Appendix~\ref{app:MFPB}.

Taking the low-frequency, long-wavelength limit, we can locate the phase boundary. The coefficients of the quadratic terms vanish (here for species $a$) when
\begin{equation}
\frac{1}{2d J_{a}} + \mathcal{G}^{a,R}_0(\omega = 0) = 0, \label{eq:MFPB_maintext}
\end{equation}
where $d$ is the number of dimensions of the lattice for a cubic lattice and the retarded propagator at non-zero temperature is
\begin{align}
\mathcal{G}^{a,R}_0 (\omega) = \frac{1}{Z} \sum_{p,q = 0}^\infty \e^{-\beta E_{p,q}} & \left[ \frac{p+1}{\mu - U_a p - Vq + \omega} \right. \nn \\
& \left. {} - \frac{p}{\mu - U_a (p-1) - V q + \omega} \right], \label{eq:G_retarted_fourier_maintext}
\end{align}
with the partition function
\begin{equation}
Z = \Trace\left(\e^{-\beta \hat{H}_0}\right) = \sum_{p,q=0}^\infty \e^{-\beta E_{p,q}},
\end{equation} 
and
\begin{equation}
E_{p,q} = -\mu (p+q) + \frac{U_a}{2}p(p-1) + \frac{U_b}{2}q(q-1) + Vpq.
\end{equation}
Equation~\eqref{eq:G_retarted_fourier_maintext} is derived in Appendix~\ref{app:MFPB}. Similar expressions can be obtained for species $b$.

At non-zero temperature, the phase boundary can be obtained straightforwardly by solving Eq.~\eqref{eq:MFPB_maintext} numerically. In order to find the phase boundary at zero temperature, one might expect to follow the single species approach, where only one term in the sum in Eq.~\eqref{eq:G_retarted_fourier_maintext} contributes, and the exponential factor is cancelled by the partition function. Here, this would result in the equation
\begin{equation}
0= \frac{1}{2dJ_{a}} + \frac{p+1}{\mu - U_a p - Vq} - \frac{p}{\mu - U_a (p-1) - Vq}, \label{eq:MFPB_not_simplified_incorrect}
\end{equation}
which may be rearranged to
\begin{equation}
\bar{\mu} = \frac{1}{2} \left[ 2\left( p + \bar{V} q \right) - 1 - \bar{J} \pm \sqrt{1 - 2 \bar{J} (2p + 1) +\bar{J}^2}  \right], \label{eq:MFPB_incorrect}
\end{equation}
with $\bar{\mu} = \mu/U_a$, $\bar{V} = V/U_a$, and $\bar{J} = 2dJ_{a}/U_a$. Equation~\eqref{eq:MFPB_incorrect} is equal to the single species boundary \cite{Oosten2001} up to a shift of $\bar{V}q$ along the $\bar{\mu}-\mathrm{axis}$. Some previous authors \cite{Iskin2010, Chen2003, Bai2020} have presented Eq.~\eqref{eq:MFPB_incorrect} as the MI-SF phase boundary, but as we show below, this result is not consistent with the zero temperature limit of Eq.~\eqref{eq:MFPB_maintext} for some Mott lobes.

Unlike the single species case, the Green's functions do not simplify to the same extent in the zero temperature limit. In the single species derivation, there is only one term in the sum for each possible occupation number. In the two-species case, however, for any occupation number $n=p+q$, there are $n+1$ possible combinations of $p$ and $q$ that may contribute to the sum. Depending on the values of the interaction strengths, one or more terms may be non-zero, and thus may be important in determining the actual phase boundary.

We determine the phase boundary of each Mott-lobe individually, by solving Eq.~\eqref{eq:MFPB_maintext} for each occupation number separately. Below, we derive analytic expressions for the Mott-lobes of species $a$ with $n=1$ and $n=2$ at zero temperature, and state a general expression for the $n$-th Mott-lobe. The phase boundary for species $b$ follows in the same way. Detailed expressions for the $n=0$, $n=3$, and $n=4$ Mott-lobes are given in Appendix~\ref{app:MFPB}. 

Note that the calculations presented in this paper are a one-particle irreducible theory that can capture SF and MI phases. Other work has shown that in the symmetric case, $U_a = U_b$, the odd-numbered Mott-lobes are replaced by a counter-flow SF (pair SF) phase for attractive (repulsive) interspecies interactions \cite{Altman2003, Kuklov2003, Kuklov2004, Hu2009, Chen2010, Menotti2010, Mishra2008}. To capture these phases, should they exist in the $U_a \neq U_b$ case, the theory needs to be extended to a two-particle irreducible formalism, which we intend to do in future work.

\subsection{First Mott-lobe}

For the first Mott-lobe, we solve Eq.~\eqref{eq:MFPB_maintext} and as $\beta \rightarrow \infty$ we only need to keep the two terms with $(p,q) = (1,0)$ and $(p,q) = (0,1)$:
\begin{align}
2dJ_a = & - \frac{ \sum_{p+q=1} \e^{-\beta E_{p,q}}}{\sum_{p+q=1} \e^{-\beta E_{p,q}} \left[ \frac{p+1}{\mu - U_a p - Vq}  - \frac{p}{\mu - U_a (p-1) - Vq}\right] },
\end{align}
and as $\beta \rightarrow \infty$
\begin{align}
2dJ_a \rightarrow & - \frac{\e^{-\beta E_{1,0}} + \e^{-\beta E_{0,1}}}{\e^{-\beta E_{1,0}} \left( \frac{2}{\mu - U_a} -\frac{1}{\mu} \right) + \e^{-\beta E_{0,1}}\left(\frac{1}{\mu - V}\right)  } \nn \\
= & - \frac{2}{\frac{2}{\mu - U_a} - \frac{1}{\mu} + \frac{1}{\mu - V}}, \label{eq:First_Mott_lobe_maintext}
\end{align}
where we note that $E_{1,0} = E_{0,1} = -\mu$ to cancel the exponentials. This shows that for the first Mott-lobe \textit{both} $(p,q)=(1,0)$ and $(p,q)=(0,1)$ always contribute to the sum, regardless of the values of the interaction strengths. 

We can immediately see that Eq.~\eqref{eq:First_Mott_lobe_maintext} differs from Eq.~\eqref{eq:MFPB_not_simplified_incorrect}. To compare them, we may rewrite Eq.~\eqref{eq:First_Mott_lobe_maintext} as
\begin{align}
0 = {} & \left( \frac{1}{2dJ_{a}} + \frac{2}{\mu - U_a} - \frac{1}{\mu} \right) + \left(\frac{1}{2dJ_{a}} + \frac{1}{\mu - V}\right),
\end{align}
which shows that the phase boundary is given by a sum of two copies of Eq.~\eqref{eq:MFPB_not_simplified_incorrect}, one for $(p,q)=(1,0)$ and one for $(p,q) = (0,1)$.

\begin{figure*}
\centering
\includegraphics[width=0.49\textwidth]{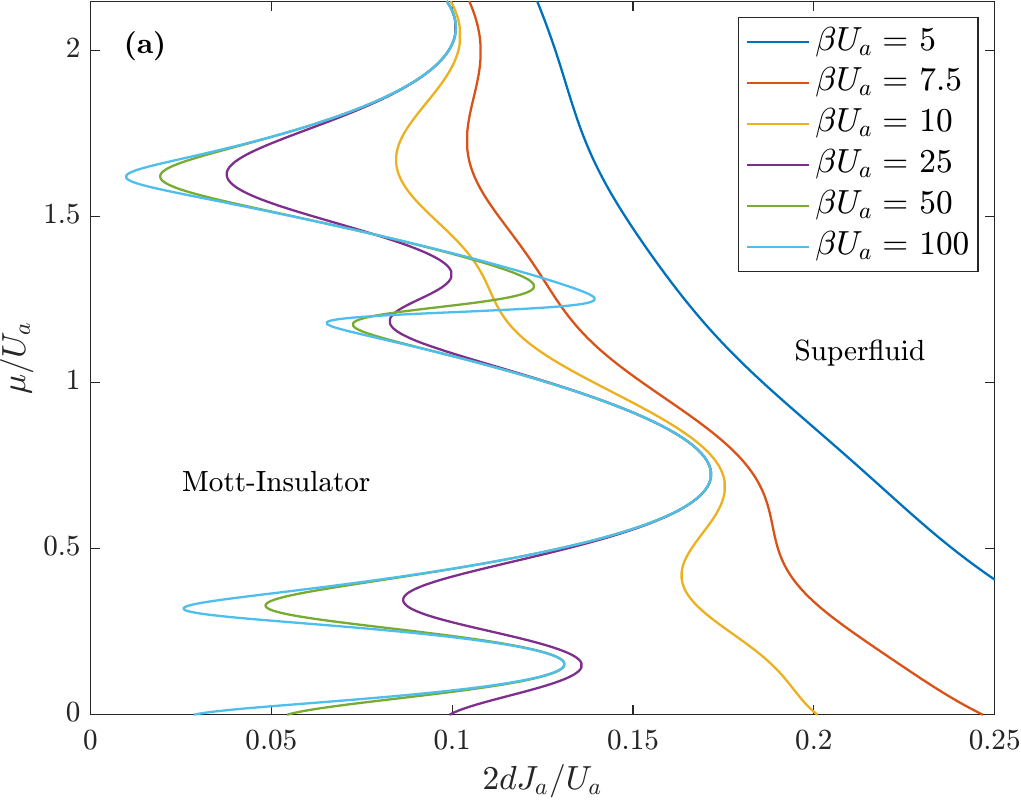} \hfill
\includegraphics[width=0.49\textwidth]{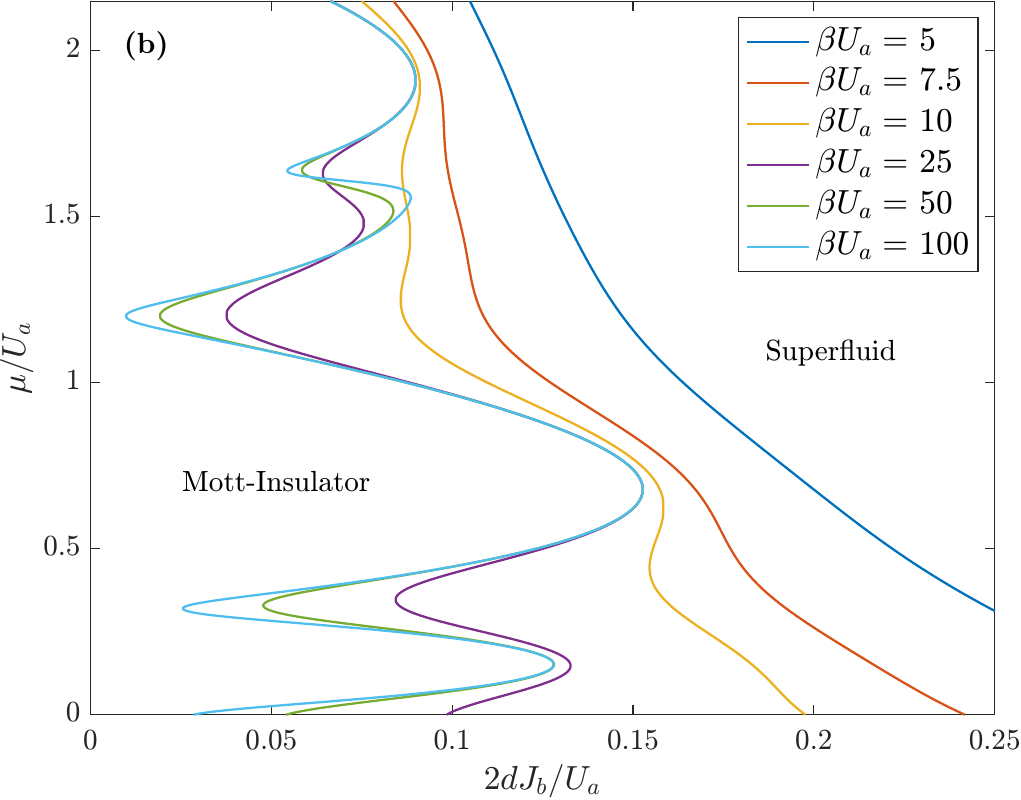}
\caption{Superfluid to Mott-insulator phase boundaries at various temperatures for (a) species $a$ and (b) species $b$ with $U_b = 0.89\, U_a$ and $V = 0.31 U_a$. } \label{fig:MFPB_nonzeroT} 
\end{figure*}

\begin{widetext}
\subsection{Second Mott-lobe} \label{sec:Second_Lobe}
For the second Mott-lobe we include the three terms with $p+q=2$ which can contribute in the $\beta \rightarrow \infty$ limit:
\begin{align}
\lim_{\beta \rightarrow \infty} 2dJ_a \simeq & - \frac{\e^{-\beta E_{2,0}} + \e^{-\beta E_{1,1}} + \e^{-\beta E_{0,2}}}{\e^{-\beta E_{2,0}} \left( \frac{3}{\mu - 2U_a} - \frac{2}{\mu - U_a} \right) + \e^{-\beta E_{1,1}} \left( \frac{2}{\mu - U_a - V} - \frac{1}{\mu - V} \right) + \e^{-\beta E_{0,2}} \left( \frac{1}{\mu - 2V} \right)  } \nn \\
= & - \frac{1}{ \left( \frac{3}{\mu - 2U_a} - \frac{2}{\mu - U_a} \right) + \e^{- \beta (V - U_a)} \left( \frac{2}{\mu - U_a - V} - \frac{1}{\mu - V} \right) + \e^{- \beta (U_b - U_a)} \left( \frac{1}{\mu - 2V} \right)  } \nn \\ 
& - \frac{1}{  \e^{- \beta (U_a - V)} \left( \frac{3}{\mu - 2U_a} - \frac{2}{\mu - U_a} \right) + \left( \frac{2}{\mu - U_a - V} - \frac{1}{\mu - V} \right) + \e^{- \beta (U_b - V)} \left( \frac{1}{\mu - 2V} \right)  } \nn \\ 
& - \frac{1}{ \e^{- \beta (U_a - U_b)} \left(   \frac{3}{\mu - 2U_a} - \frac{2}{\mu - U_a} \right) + \e^{- \beta (V - U_b)} \left( \frac{2}{\mu - U_a - V} - \frac{1}{\mu - V} \right) + \left( \frac{1}{\mu - 2V} \right)  }. \label{eq:Second_Mott_lobe_maintext_beta}
\end{align}
\end{widetext}
Equation~\eqref{eq:Second_Mott_lobe_maintext_beta} clearly shows that it matters what values $U_a$, $U_b$, and $V$ take. The first of the three fractions contains the two exponentials, $\e^{-\beta (V - U_a)}$ and $\e^{- \beta (U_b - U_a)}$. If $U_a>U_b$ and/or $U_a>V$, the exponentials will diverge in the zero temperature limit (i.e. $\beta \rightarrow \infty$), which means that the whole fraction vanishes. Thus this term only contributes if $U_a \leq U_b, V$. Using this argument, we can see that the second and third fractions survive only if $V \leq U_a, U_b$ and $U_b \leq U_a, V$, respectively.

If at least two of the three interaction strengths are the same, we will end up with multiple identical terms on the RHS of Eq. \eqref{eq:Second_Mott_lobe_maintext} which may be non-zero. We can summarize the various contributions as follows:
\begin{widetext}
\begin{equation}
2dJ_a = \left\lbrace \begin{array}{ll}
- \left(1 + \delta_{U_a,V} + \delta_{U_a, U_b}\right) \left[ \left( \frac{3}{\mu - 2U_a} - \frac{2}{\mu - U_a} \right) + \delta_{U_a,V} \left( \frac{2}{\mu - U_a - V} - \frac{1}{\mu - V} \right) + \delta_{U_a,U_b} \left( \frac{1}{\mu - 2V} \right) \right]^{-1}, & \mathrm{if}~ U_a \leq U_b, V \\
- \left( 1 + \delta_{U_a,V} + \delta_{U_b, V}\right) \left[ \delta_{U_a,V} \left( \frac{3}{\mu - 2U_a} - \frac{2}{\mu - U_a} \right) + \left( \frac{2}{\mu - U_a - V} - \frac{1}{\mu - V} \right) + \delta_{U_b, V} \left( \frac{1}{\mu - 2V} \right) \right]^{-1}, & \mathrm{if}~ V \leq U_a, U_b \\
- \left( 1 + \delta_{U_a,V} + \delta_{U_a, U_b} \right) \left[ \delta_{U_a,U_b} \left( \frac{3}{\mu - 2U_a} - \frac{2}{\mu - U_a} \right) + \delta_{U_a,V} \left( \frac{2}{\mu - U_a - V} - \frac{1}{\mu - V} \right) +  \left( \frac{1}{\mu - 2V} \right) \right]^{-1}, & \mathrm{if}~ U_b\leq U_a, V 
\end{array}\right. . \label{eq:Second_Mott_lobe_maintext}
\end{equation}
\end{widetext}
Equation~\eqref{eq:Second_Mott_lobe_maintext} shows that unless $V\leq U_a, U_b$ there will be phase separation, since either the $p=2$, $q=0$ state or the $p=0$, $q=2$ state will result if $V$ is larger than $U_a$ or $U_b$, respectively.

We restrict ourselves to $V<U_a,U_b$, so Eq.~\eqref{eq:Second_Mott_lobe_maintext} further reduces to 
\begin{equation}
2dJ_a =  \left[\frac{1}{\mu - V} -\frac{2}{\mu - U_a - V} \right]^{-1},
\end{equation}
which agrees with Eq.~\eqref{eq:MFPB_incorrect} for $p=q=1$ and with Eq.~(24) from Ref.~\cite{Colussi2022}. Equation~\eqref{eq:Second_Mott_lobe_maintext} also demonstrates that we must employ the more stringent restriction $V \leq U_a, U_b$ rather than $V^2 \leq U_a U_b$ \cite{Ho1996, Trippenbach2000}.

\subsection{$n$-th Mott-lobe}

For larger $n$, the expressions for the phase boundaries grow quickly in size and the restrictions on $U_a$, $U_b$, and $V$ become less obvious, as can be seen in Eqs.~\eqref{eq:MottLobe3} and \eqref{eq:Fourth_Mott_Lobe}. Thus, a general expression for any $n$ is useful for analytic and numerical evaluation. The zero-temperature phase boundary equation for the $n$-th Mott-lobe can be written as
\begin{equation}
2dJ_a = - \sum_{m=0}^n \frac{\prod_{k=0}^n \theta(E_{n-k,k} - E_{n-m,m}) }{\sum_{k=0}^n \delta_{E_{n-k,k},E_{n-m,m}} M(n-k,k)},
\end{equation}
where $M(p,q) = \left[ \frac{p+1}{\mu - U_a p - Vq}  - \frac{p}{\mu - U_a (p-1) - Vq}\right]$. The step functions, $\theta(E_{n-k,k} - E_{n-m,m})$, enforce the relevant restrictions on $U_a$, $U_b$, and $V$. Specifically, for a given $m$, only the terms for which $E_{n-m,m}$ is smaller than or equal to all other $E_{p,q}$ may be non-zero. 

\begin{figure}[b]
\centering
\includegraphics[width=0.47\textwidth]{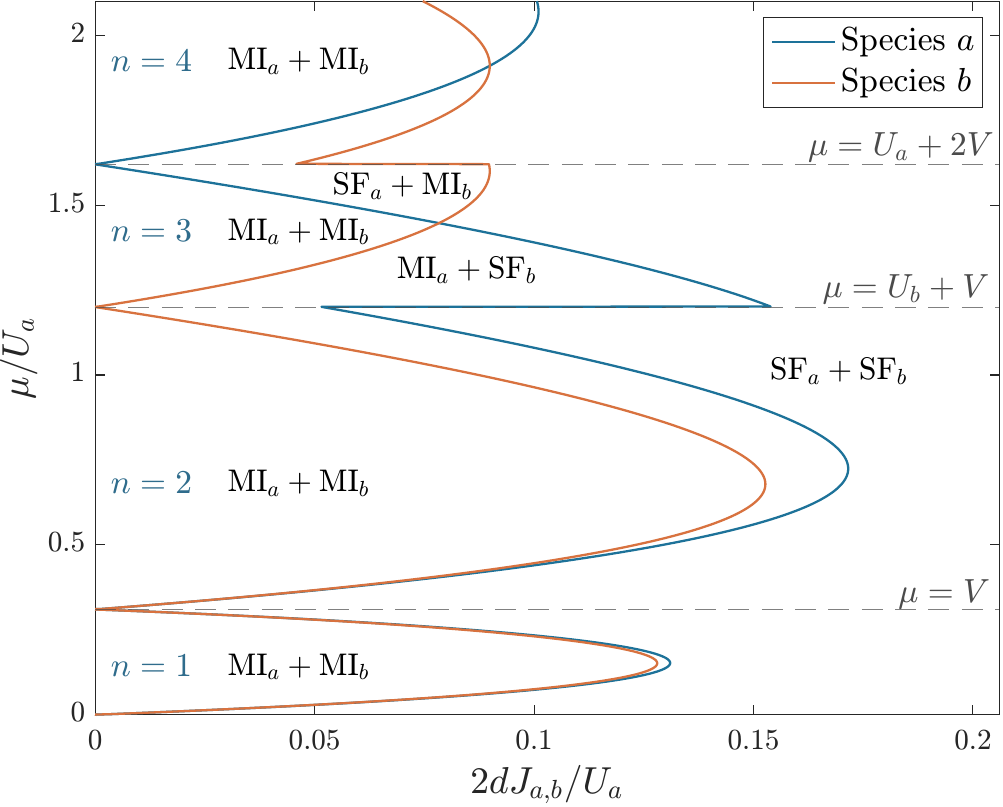}
\caption{Zero temperature phase boundaries between superfluid and Mott-insulating phases for species $a$ and $b$ with $U_b = 0.89\, U_a$ and $V =  0.31 U_a$. The average filling per site, $n$, in the MI phase is indicated.} \label{fig:MFPB_zeroT} 
\end{figure}

The temperature dependence of the mean field phase boundary for both species is shown in Fig.~\ref{fig:MFPB_nonzeroT}. In the third Mott-lobe, we can see that the phase boundary does not approach its zero temperature shape monotonically as the temperature is decreased. Instead, the tips of the lobes reach their minimum $J_x$ at $\beta U_a \sim 25$, before increasing again. This demonstrates that the terms in the Green's function, Eq.~\eqref{eq:G_retarted_fourier_maintext}, contribute to the sum to varying extents at different temperatures as the weight of the exponential factors changes with temperature.

In the zero temperature limit for $U_a \neq U_b$, the phase diagram appears as in Fig.~\ref{fig:MFPB_zeroT}, and we find several features that are not present in the single species equivalent. Rather than having only two separated phases, both species in the SF phase, $\mathrm{SF}_a+\mathrm{SF}_b$, or both in the MI phase, $\mathrm{MI}_a+\mathrm{MI}_b$, there are two new mixed phases. In one, species $a$ is SF while species $b$ is MI, $\mathrm{SF}_a+\mathrm{MI}_b$, and vice versa in the second phase \cite{Altman2003, Kuklov2004, Isacsson2005}.

Figure~\ref{fig:MFPB_zeroT} also shows two abrupt jumps in the phase boundaries; species $a$ has a jump between the $n=2$ and $n=3$ Mott lobes at $\mu = U_b+V$ and species $b$ has a jump between the $n=3$ and $n=4$ Mott lobes at $\mu = U_a + 2V$. Additionally, these jumps make a $\mathrm{MI}(n)\rightarrow \mathrm{MI}(n+1)$ transition at non-zero $J_x$ possible, which is a new feature compared to the single species case.
The jumps we observe may be an artifact of our mean-field like determination of the phase boundaries which may be rounded by a more thorough examination of the phase boundaries, e.g. as in Ref.~\cite{Anufriiev2016}.

\section{Equations of Motion}
\label{sec:EOMs}

We can determine the equations of motion (EOMs) of the order parameters from the saddle-point conditions on the action:
\begin{equation}
\fdv{S^{II}_\mathrm{eff}}{y^*_{iq}} =0,\quad \fdv{S^{II}_\mathrm{eff}}{y^*_{ic}} = 0, \quad\fdv{S^{II}_\mathrm{eff}}{z^*_{iq}} = 0, \quad\fdv{S^{II}_\mathrm{eff}}{z^*_{iq}} = 0. \nn
\end{equation}
Using the identities
\begin{align}
\left[ G^x_{cc}(t_1, t_2) \right]^{-1} = 0; ~ \left[ G^x_{qq}(t_1, t_2) \right]^{-1} = \left[ \left(\mathcal{G}^x_0\right)^{-1} \right]^K(t_1, t_2); \nn \\
\left[ G^x_{qc}(t_1, t_2) \right]^{-1} = \left[ \mathcal{G}_0^{x,R} (t_1, t_2) \right]^{-1} = \left[ \mathcal{G}_0^{x,A} (t_2, t_1) \right]^{-1}, \nn
\end{align}
and
\begin{equation}
\left[ G^x_{cq}(t_1, t_2) \right]^{-1} = \left[ \mathcal{G}_0^{x,A} (t_1, t_2) \right]^{-1} = \left[ \mathcal{G}_0^{x,R} (t_2, t_1) \right]^{-1}, \nn
\end{equation}
we obtain the EOMs for species $a$:
\begin{widetext}
\begin{subequations}
\begin{align}
0 = & \sum_{\langle ij \rangle} \left(2 J_{a,ij}\right) y_{jc}(t) + \int_{-\infty}^\infty \dd{t_2}  \left[ \left(\mathcal{G}^a_{0}\right)^{-1} \right]^K(t,t_2) y_{iq}(t_2) + \int_{-\infty}^\infty \dd{t_2}   \left[ \left(\mathcal{G}^a_{0}\right)^{-1} \right]^R(t, t_2) y_{ic}(t_2) \nn \\
& + \frac{1}{2} \int_{-\infty}^\infty \dd{t_2} \dd{t_3} \dd{t_4}  u^a_{q \alpha_2 \alpha_3 \alpha_4}(t, t_2, t_3, t_4) y^*_{i\alpha_2} (t_2) y^{\phantom{*}}_{i\alpha_3}(t_3) y^{\phantom{*}}_{i\alpha_4}(t_4)  \nn \\
& + \int_{-\infty}^\infty \dd{t_2} \dd{t_3} \dd{t_4}  u^{ab}_{q \alpha_2 \alpha_3 \alpha_4}(t, t_2, t_3, t_4) y^{\phantom{*}}_{i\alpha_2} (t_2) z^*_{i\alpha_3}(t_3) z^{\phantom{*}}_{i\alpha_4}(t_4), \label{eq:dS/dy_q} \\
0 = {} &  \sum_{\< ij \>} 2 J_{a,ij} y_{jq}(t) + \int_{-\infty}^\infty \dd{t_2}  \left[ \left(\mathcal{G}^a_0\right)^{-1} \right]^{A}(t, t_2) y_{iq}(t_2) \nn \\
&+ \frac{1}{2} \int_{-\infty}^\infty \dd{t_2} \dd{t_3} \dd{t_4}  u^a_{c \alpha_2 \alpha_3 \alpha_4} (t, t_2, t_3, t_4)  y^*_{i\alpha_2}(t_2) y^{\phantom{*}}_{i\alpha_3}(t_3) y^{\phantom{*}}_{i\alpha_4}(t_4) \nn \\
& + \int_{-\infty}^\infty \dd{t_2} \dd{t_3} \dd{t_4}  u^{ab}_{c \alpha_2 \alpha_3 \alpha_4} (t, t_2, t_3, t_4)  y^{\phantom{*}}_{i \alpha_2}(t_2) z^*_{i \alpha_3}(t_3) z^{\phantom{*}}_{i \alpha_4}(t_4). \label{eq:dS/dy_c}
\end{align}
\end{subequations}
\end{widetext} 
The EOMs for species $b$ have a similar form and can be obtained by replacing $a, y, z \rightarrow b, z, y$  in Eqs.~\eqref{eq:dS/dy_q} and \eqref{eq:dS/dy_c}.

\subsection{Simplified equations of motion}

Solving the EOMs, Eqs. \eqref{eq:dS/dy_q} and \eqref{eq:dS/dy_c}, is rather challenging due to the presence of multiple time integrals, hence we focus on the low-frequency, long-wavelength limit to obtain simpler equations for the mean-field dynamics of the order parameters.

We assume that in the limit $t\rightarrow -\infty$ the system is in the SF phase and the hopping $J_x(t)$ does not change with time. These initial conditions require $y_1 = y_2$ and $z_1=z_2$, which implies that $y_q = z_q = 0$, and $y_c = \sqrt{2} y$ and $z_c = \sqrt{2}z$, where $y = y_1$ and $z = z_1$ are the SF order parameters. If $y_q$ and $z_q$ remain small under evolution in time, we can focus on the EOMs for $y_c$ and $z_c$ only, i.e. Eq. \eqref{eq:dS/dy_q}. This can be seen from the Keldysh Green's function obtained in Appendix~\ref{app:MFPB}. In the $\omega \rightarrow 0$ limit, terms involving $\mathcal{G}^{a(b),K}_0(\omega)$ will contribute only when $\mu \sim U_{a(b)} p + Vq$ for integers $p$ and $q$. These represent points where the coefficients $\lambda_x$ and $\kappa_x^2$ (introduced below) diverge, and we restrict ourselves to values of $\mu$ away from divergences, for which $\mathcal{G}^{a,K}_0$ and $\mathcal{G}^{b,K}_0$ can be ignored as $\omega \rightarrow 0$, so we only need to keep terms that contain $\mathcal{G}^{a,R}_0$ or $\mathcal{G}^{b,R}_0$.

In order for $y_q$ to become appreciable, the terms $u^a_{cccc}y^*_q y^{\phantom{*}}_q y^{\phantom{*}}_q$ and $u^{ab}_{cccc} y^{\phantom{*}}_q z^*_q z^{\phantom{*}}_q$ in Eq.~\eqref{eq:dS/dy_c} must be appreciable. These terms contain the two-particle connected Green's functions $G^{2a,c}_{cccc}$ and $G^{1a+1b,c}_{cccc}$, respectively. Similarly to $\mathcal{G}_0^{a(b),K}$, these only contribute to the low frequency dynamics when $\mu \sim U_{a(b)} p + V q$. We can therefore ignore $y_q$ and $z_q$ and focus on the dynamical equations for $y_c$ and $z_c$ only. Considering which terms contribute to the dynamics in the $\omega \rightarrow 0$ limit, we find that for values of the chemical potential away from $\mu \sim U_{a(b)} p + Vq$, the only connected Green's functions that we need to evaluate are $G^{2a,c}_{qccc}$, $G^{2b,c}_{qccc}$, and $G^{1a+1b,c}_{qccc}$.

\begin{figure*}
\includegraphics[width=0.49\textwidth]{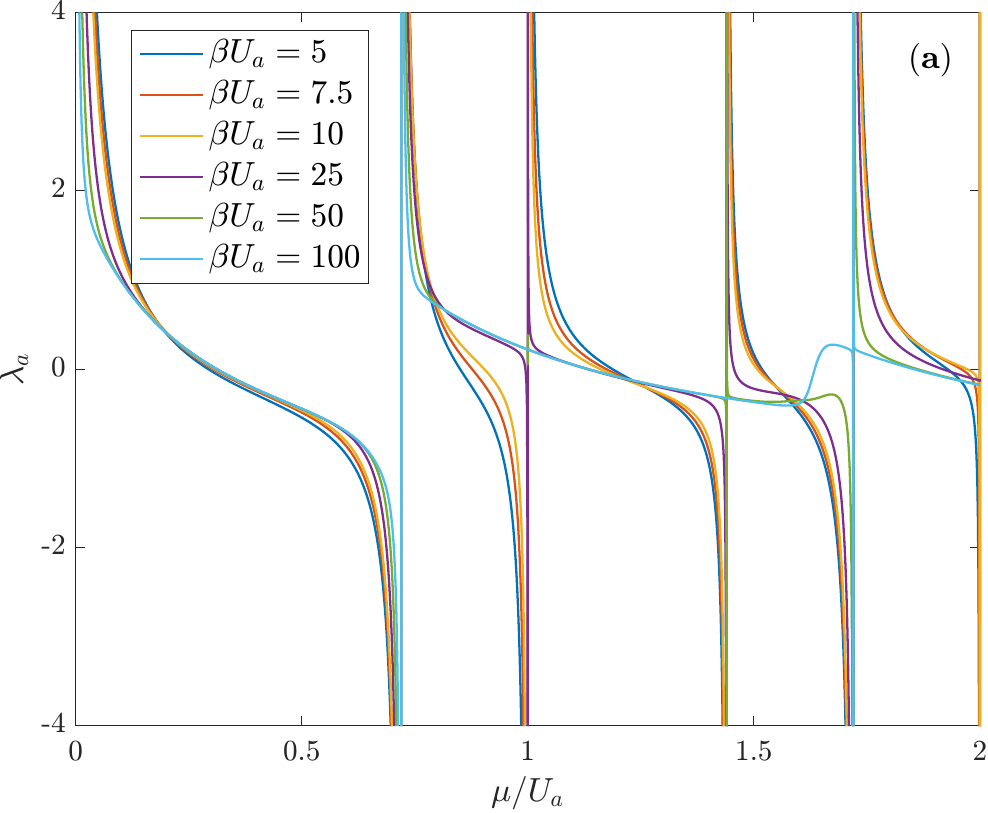} \hfill \includegraphics[width=0.49\textwidth]{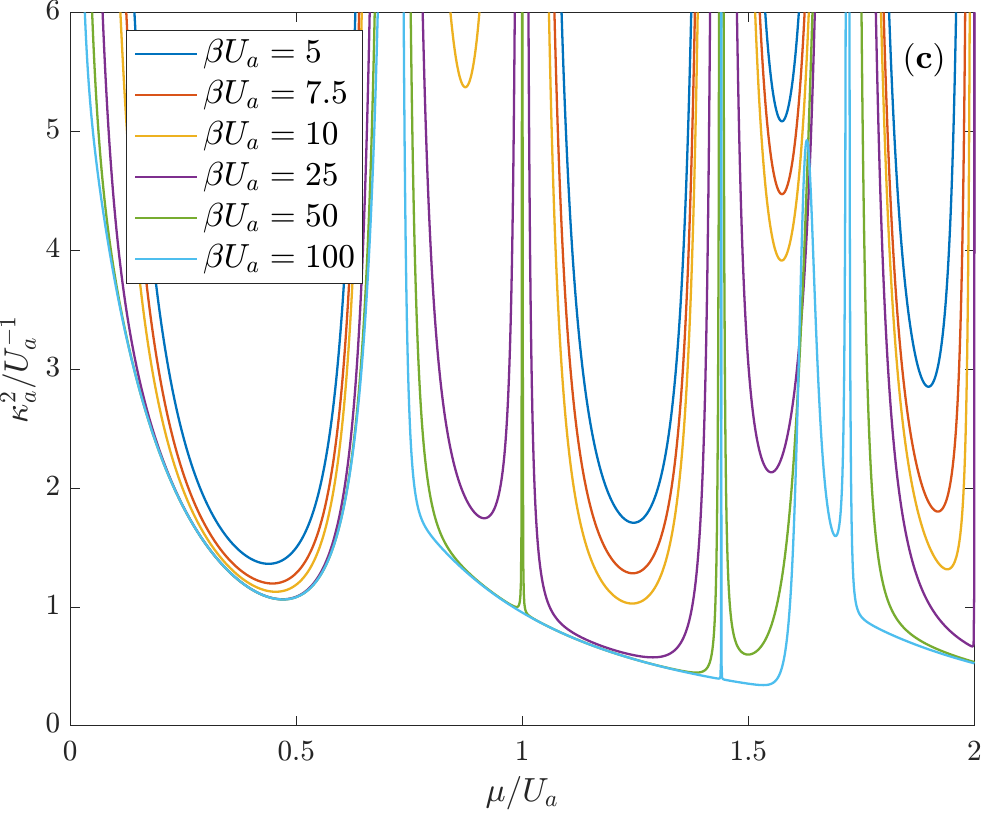} \\
\includegraphics[width=0.49\textwidth]{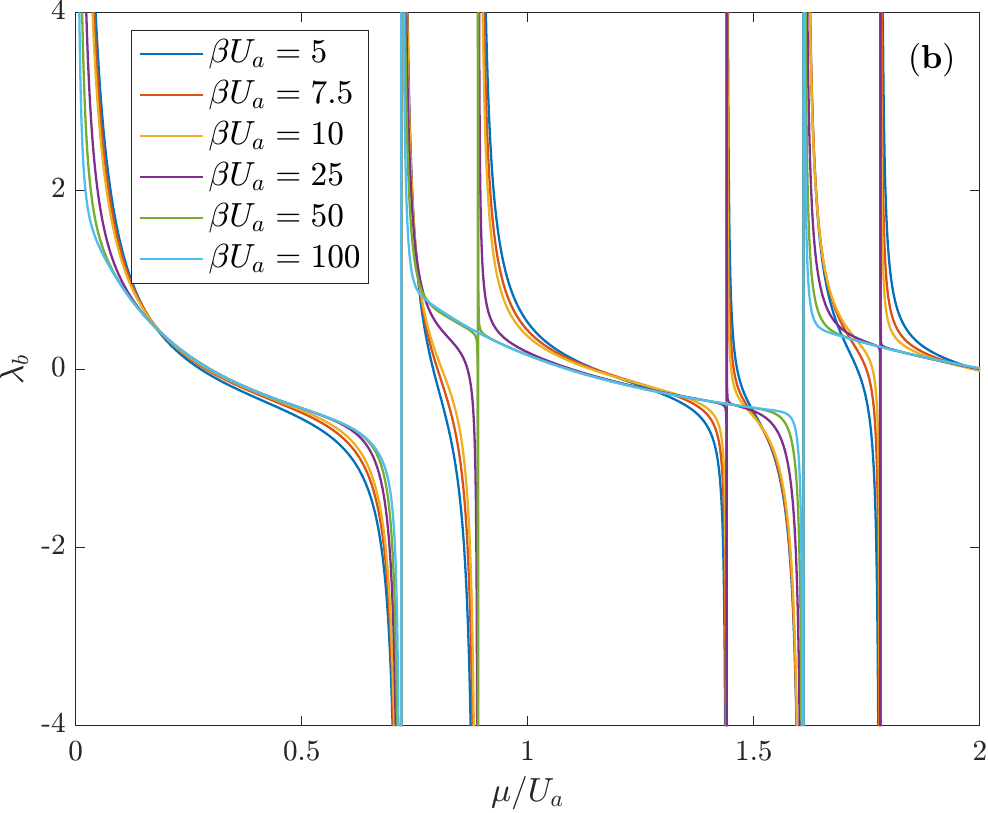} \hfill \includegraphics[width=0.49\textwidth]{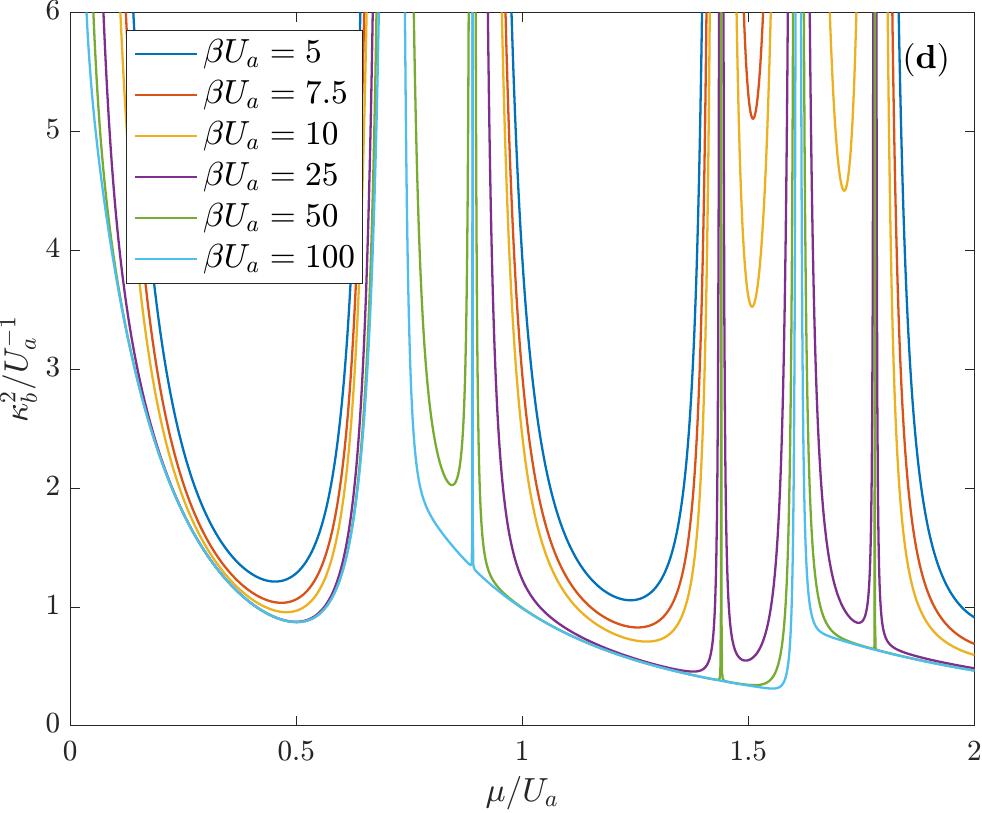} \\
\caption{Plots of (a) $\lambda_a$, (b) $\lambda_b$, (c) $\kappa_a^2$, and (d) $\kappa_b^2$ as functions of $\mu/U_a$ and inverse temperature $\beta U_a$ with $U_b = 0.89 U_a$ and $V = 0.72 U_a$.} \label{fig:lambda_and_kappa}
\end{figure*}

We can expand the inverse retarded Green's function as a power series in frequency, and so
\begin{align}
\int \dd{t_2} & \left[ \left( \mathcal{G}^a_0 \right)^{-1} \right]^R \left( t, t_2 \right) y_{ic} \left( t_2 \right) \nn \\ 
& = \int \frac{\dd{\omega}}{2\pi} \e^{-\ii \omega t} \left[ \left(\mathcal{G}^{a,R}_{0}\right)^{-1} \right](\omega) y(\omega)  \nn \\ 
& \simeq \nu_a y(t) -\ii \lambda_a \pdv{y(t)}{t} - \kappa_a^2 \pdv[2]{y(t)}{t},
\label{eq:FTGRapprox}
\end{align}
where
\begin{align*}
\nu_a = \left. \left[ \left(\mathcal{G}^a_{0}\right)^{-1} \right]^R(\omega) \right\vert_{\omega = 0}&;  \,\, \lambda_a =  - \left. \pdv{\omega}   \left[ \left(\mathcal{G}^a_{0}\right)^{-1} \right]^R(\omega) \right\vert_{\omega = 0}; \\
\kappa_a^2 = \frac{1}{2} & \left. \pdv[2]{\omega}  \left[ \left(\mathcal{G}^a_{0}\right)^{-1} \right]^R(\omega) \right\vert_{\omega = 0}. 
\end{align*}
Explicit expressions for $\nu_x$, $\lambda_x$, and $\kappa_x^2$ can be computed from Eq. \eqref{eq:G_retarded_AppendixA} and are given in Appendix~\ref{app:EOMs}. The temperature and chemical potential dependences of $\lambda_a$, $\lambda_b$, $\kappa_a^2$, and $\kappa_b^2$ are shown in Fig.~\ref{fig:lambda_and_kappa}. 
The coefficients for species $x$ have divergences when
the chemical potential $\mu \sim U_x p + Vq$ for integers $p$ and $q$.  To avoid the effects of such divergences on the equations of motion, we focus on values of the chemical potential away from these divergences.

Next, we take a long-wavelength expansion of the hopping terms on a cubic lattice
\begin{align}
2 J_{a,ij}(t) y_{j,c}(t) \rightarrow {} & 2 J_{a,ij}(t) \sum_{j=1}^d \cos(k_j \tilde{a})y(\mathbf{k}, t) \nn \\
\approx {} & 2 J_{a,ij}(t) \left( d - \frac{1}{2} k^2 \tilde{a}^2 \right) y(\mathbf{k}, t),
\end{align}
for $k \tilde{a} \ll 1$, where $d$ is the number of dimensions, $k_j$ is the momentum in the $j$-direction, and $\tilde{a}$ is the lattice spacing. Since we work in the low-frequency, long-wavelength limit, we ignore terms of order $k\tilde{a}$ or higher. We take the low-frequency limit of the interaction terms by expanding the two-particle connected Green's functions and the retarded and advanced Green's functions about $\omega =0$ (details in Appendix~\ref{app:EOMs}).

\begin{figure}
\includegraphics[width=0.48\textwidth]{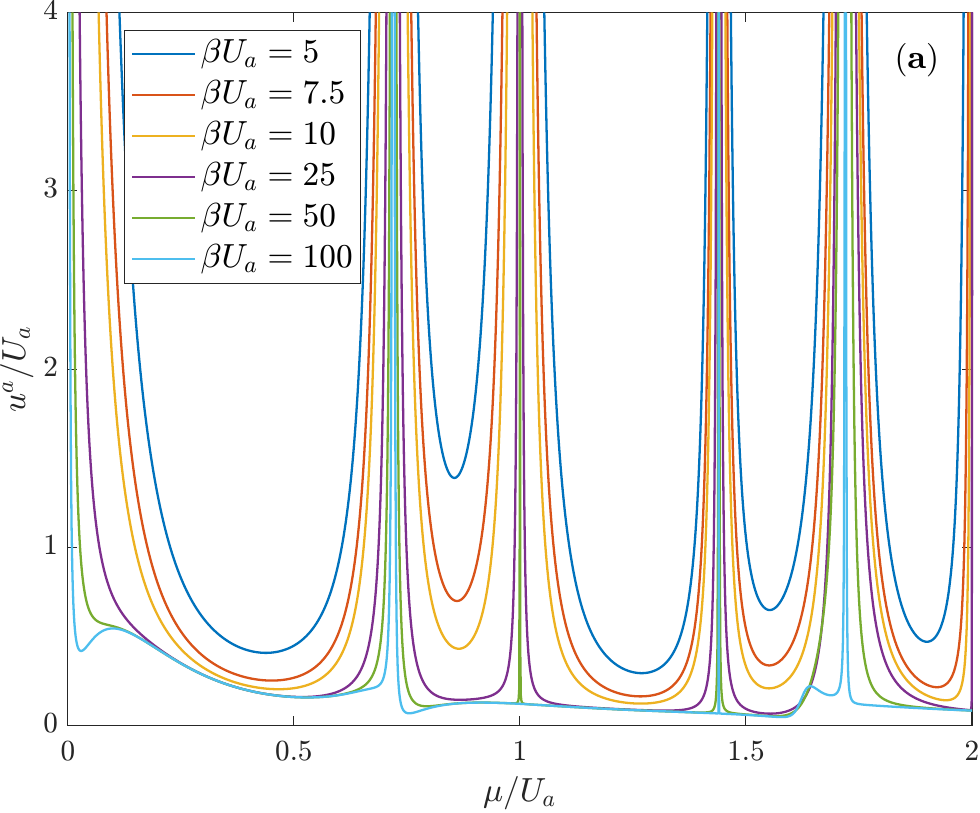} \\
\includegraphics[width=0.48\textwidth]{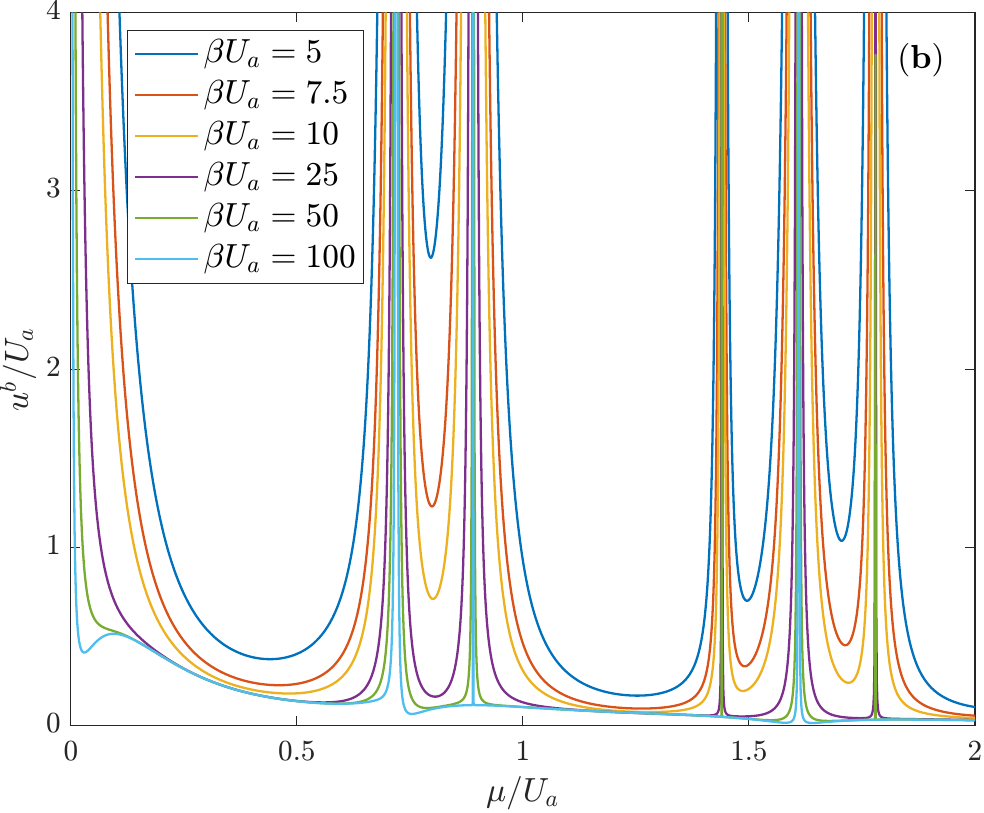} \\
\hspace{-0.4cm} \includegraphics[width=0.49\textwidth]{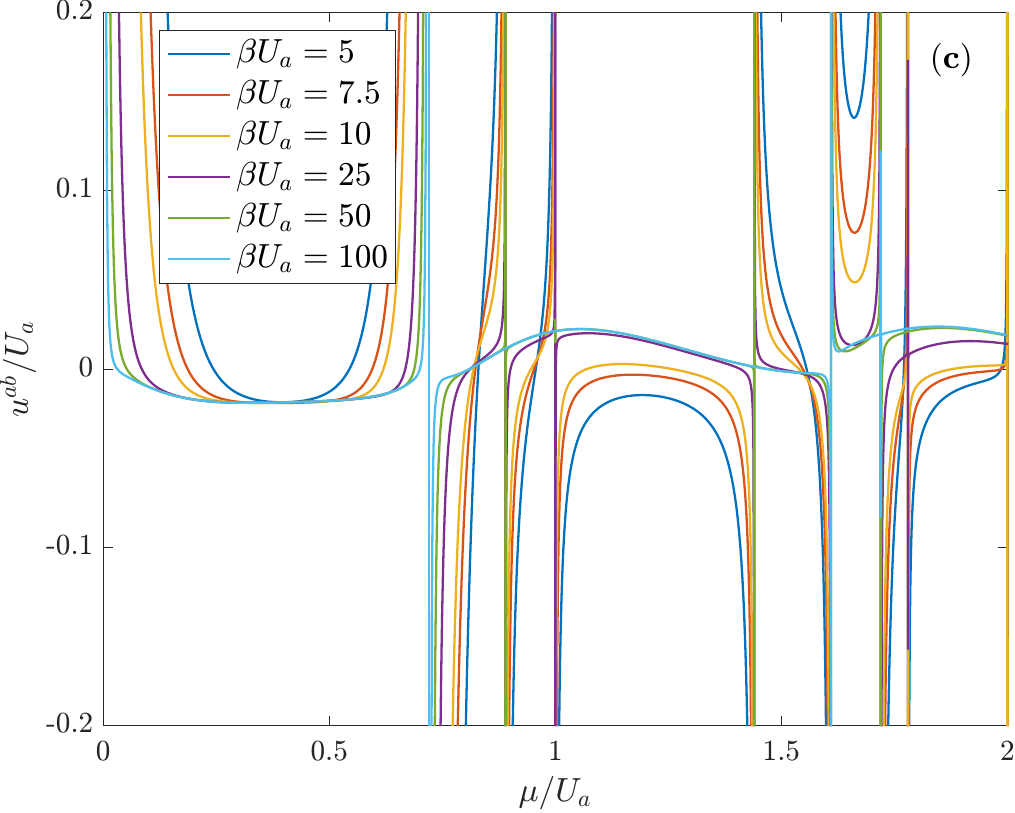}
\caption{Plots of (a) $u^a$, (b) $u^b$, and (c) $u^{ab}$ as a function of $\mu/U_a$ and the inverse temperature $\beta U_a$ with $U_b = 0.89\, U_a$ and $V = 0.72\, U_a$.} \label{fig:u_figures}
\end{figure}

Recalling that our initial state is deep in the SF phase, i.e. $y_q = z_q = 0$, we have $\abs{y_c(t)}^2 = 2 \abs{y(t)}^2$ and $\abs{z_c(t)}^2 = 2 \abs{z(t)}^2$. In the low-frequency limit we approximate the intra-species interaction terms by $- u^a \abs{y}^2y$ and $- u^b \abs{z}^2 z$, and the inter-species interaction terms by $-2 u^{ab}\abs{z}^2 y$ and $-2 u^{ab} \abs{y}^2 z$ for species $a$ and $b$, respectively. The expressions for $u^a$, $u^b$, and $u^{ab}$ are given in Eqs. \eqref{eq:u^a_low_frequency} -- \eqref{eq:u^ab_low_frequency} and are shown graphically in Fig.~\ref{fig:u_figures}. We can see that $u^a$ and $u^b$ are most sensitive to thermal effects close to $\mu = U_a p + Vq$ and $\mu = U_b p + Vq$, respectively, for integers $p$ and $q$. 
The inter-species term $u^{ab}$ is sensitive to thermal effects close to both $\mu = U_a p + Vq$ and $\mu = U_b p + Vq$. We can also see that the single species interaction terms are strictly positive (repulsive), whereas the inter-species interaction term takes both positive (repulsive) and negative (attractive) values as a function of $\mu/U_a$, changing sign even within a single Mott-lobe. Thus, we have as our approximate EOMs:
\begin{subequations}
\begin{align}
0 & = \left[ 2dJ_a(t) + \nu_a \right] y - \ii \lambda_a \dot{y} - \kappa_a^2 \ddot{y} - u^a \abs{y}^2 y - 2u^{ab}y \abs{z}^2 , \\
0 & = \left[ 2dJ_b(t) + \nu_b \right] z - \ii \lambda_b \dot{z} - \kappa_b^2 \ddot{z} - u^b \abs{z}^2 z - 2u^{ab}\abs{y}^2 z.
\end{align}
\end{subequations}

We take $J_x(t) = J_{x,0} + j_x(t)$, where $J_{x,0}$ is chosen such that $2dJ_{x,0} + \nu_x = 0$, i.e. $J_{x,0}$ is chosen to lie on the MF phase boundary for the SF for a given $\mu$. Thus, the EOMs become
\begin{subequations}
\begin{align}
0 & = \kappa_a^2 \ddot{y} + \ii \lambda_a \dot{y} + \delta_a(t) y + u^a \abs{y}^2y + 2u^{ab} y\abs{z}^2, \label{eq:simplified_EOMs_a} \\
0 & = \kappa_b^2 \ddot{z} + \ii \lambda_b \dot{z} + \delta_b(t) z + u^b \abs{z}^2z + 2u^{ab} \abs{y}^2 z, \label{eq:simplified_EOMs_b}
\end{align}
\end{subequations}
where $\delta_x (t) = -2dj_x(t)$. Even after all these simplifications, the equations for the order parameter dynamics are two coupled non-linear  second-order differential equations, for which we have not been able to find analytic solutions. Below we discuss numerical solutions of these equations for fixed $\mu$ and time-varying $J_x$.

\begin{figure}
\centering
\includegraphics[width=0.4\textwidth]{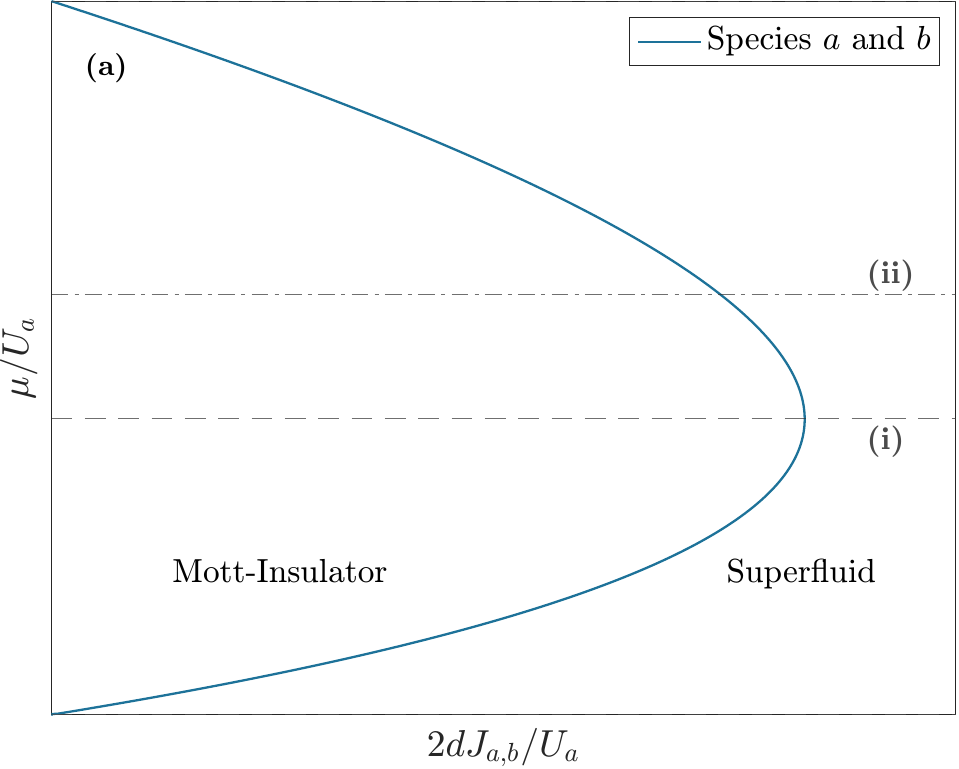}
\includegraphics[width=0.4\textwidth]{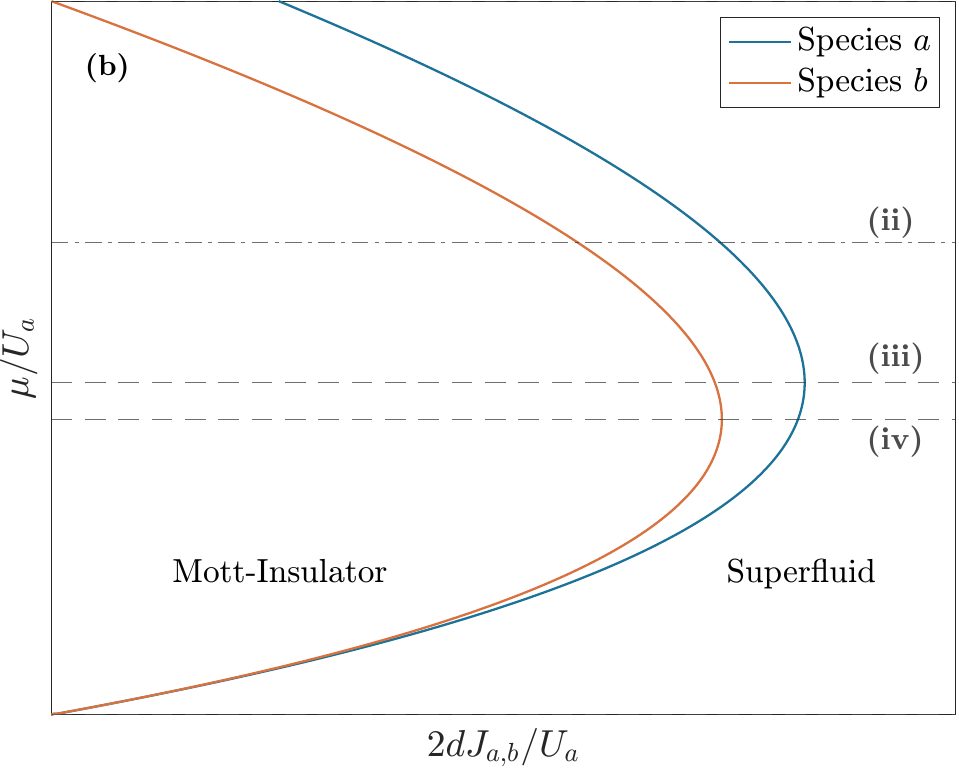}
\caption{Four possibilities for quantum quenches at constant $\mu$ for (a) $U_a=U_b$ and (b) $U_b=0.89U_a$: (i) particle-hole symmetric case, $\lambda_a = \lambda_b = 0$, (ii) generic case, $\lambda_a \lambda_b \neq 0$, (iii) and (iv) two hybrid cases, $\lambda_a \lambda_b = 0$ and $\lambda_a \neq \lambda_b$.}\label{fig:quench_options}
\end{figure}

If we  fix $\mu$, there are four possibilities for the dynamics we should consider: (i) the particle-hole symmetric case in which $\lambda_a = \lambda_b = 0$, (ii) the generic case in which $\lambda_a \lambda_b \neq 0$, and (iii) and (iv), two hybrid cases in which $\lambda_a \lambda_b = 0$ and $\lambda_a \neq \lambda_b$. For $\lambda_x = 0$, the transition occurs at the tip of the Mott-lobe of species $x$, as is illustrated in Fig.~\ref{fig:quench_options}.

We consider traversal of the quantum critical region as $\delta_x(t)$ varies with time. We demand that
\begin{equation}
\lim_{t\rightarrow -\infty} \delta_x(t) = -\delta_{x,0} \quad \mathrm{and} \quad \lim_{t \rightarrow \infty} \delta_x(t) = \delta_{x,1}. \nn
\end{equation}
For our numerical solutions we use the form \cite{Kennett2011}
\begin{equation}
\delta_x(t) = \left( \frac{\delta_{x,0} + \delta_{x,1}}{2} \right) \tanh(\frac{t}{\tau_{x,Q}}) + \frac{\delta_{x,1} - \delta_{x,0}}{2}, \label{eq:delta}
\end{equation}
where $\tau_{x,Q}$ is the characteristic time for $\delta_x(t)$ to cross from $-\delta_{x,0}$ to $\delta_{x,1}$ for species $x$. We choose the form in Eq. \eqref{eq:delta} as a smooth function that satisfies the limits as $t \rightarrow \pm \infty$ that is also linear in $t$ close to the transition at $t=0$. Note that we allow for the case of $\tau_{a,Q} \neq \tau_{b,Q}$, i.e. the two species experience the quench over different time-scales, which introduces another degree of freedom in studying the dynamics.

\subsection{Particle-hole symmetric case}
The particle-hole symmetric case only occurs for $U_a = U_b$, which leads to $\nu_a = \nu_b = \nu$, $\kappa_a^2 = \kappa_b^2 = \kappa^2$, and $u^a = u^b = u$. 
While our focus in this paper is on $U_a \neq U_b$, we consider the case $U_a = U_b$ here to contrast to the generic and hybrid cases.  In the 
particle-hole symmetric case the phase boundaries of both species coincide and a quench is through the tip of the Mott-lobes of both species. Additionally, in the particle-hole symmetric case, $\lambda_a = \lambda_b = 0$, so Eqs. \eqref{eq:simplified_EOMs_a} and \eqref{eq:simplified_EOMs_b} simplify to 
\begin{subequations}
\begin{alignat}{5}
0 = {} & \kappa^2 \ddot{y} && + \delta_a(t) y && + u \abs{y}^2y && + 2u^{ab} y \abs{z}^2, \label{eq:PH_coupled_a} \\
0 = {} & \kappa^2 \ddot{z} && + \delta_b(t) z && + u \abs{z}^2 z && + 2 u^{ab} z \abs{y}^2. \label{eq:PH_coupled_b}
\end{alignat}
\end{subequations}
We seek solutions of the form $y(t) = \rho(t) \e^{\ii \theta(t)}$ and $z(t) = \sigma(t) \e^{\ii \phi(t)}$, which allows us to split Eqs.~\eqref{eq:PH_coupled_a} and \eqref{eq:PH_coupled_b} into real and imaginary parts. For species $a$ we find
\begin{subequations}
\begin{align}
0 & = \ddot{\rho} - \rho \left( \dot{\theta} \right)^2 +  \overline{\delta}_a \rho + u \rho^3 + 2u^{ab} \rho \sigma^2, \label{eq:Real_a} \\
0 & = 2 \dot{\rho}\,\dot{\theta} + \rho\,\ddot{\theta}, \label{eq:Imaginary_a}
\end{align}
\end{subequations}
and for species $b$
\begin{subequations}
\begin{align}
0 & = \ddot{\sigma} - \sigma \left( \dot{\phi} \right)^2 + \overline{\delta}_b \sigma + u \sigma^3 + 2u^{ab} \sigma \rho^2, \label{eq:Real_b} \\
0 & = 2\dot{\sigma} \dot{\phi} +  \sigma \ddot{\phi}, \label{eq:Imaginary_b}
\end{align}
\end{subequations}
where we rescaled
\begin{equation}
t = \kappa \overline{t}; \quad \overline{\delta}_x(\overline{t}) = \delta(\kappa \overline{t}) = \delta (t), \label{eq:rescaledTime}
\end{equation}
and wrote the equations in terms of the rescaled time $\overline{t}$, following Ref.~\cite{Kennett2011} for the one-component case. The imaginary parts, Eqs. \eqref{eq:Imaginary_a} and \eqref{eq:Imaginary_b}, are no longer coupled, allowing us to easily integrate to give
\begin{align*}
\ln(\dot{\theta}) & = -2 \ln(\rho) + c_1, \\
\ln(\dot{\phi}) & = -2 \ln(\sigma) + c_2,
\end{align*}
or equivalently, for constants $c_a$ and $c_b$
\begin{equation}
\dot{\theta} \rho^2 = c_a \quad \mathrm{and} \quad \dot{\phi} \sigma^2 = c_b. \label{eq:c_ab}
\end{equation}
Substituting Eq.~\eqref{eq:c_ab} into the equations for the real parts, Eqs. \eqref{eq:Real_a} and \eqref{eq:Real_b}, respectively, gives
\begin{subequations}
\begin{align}
0 & = \ddot{\rho} - \frac{c_a}{\rho^3} + \overline{\delta}_a \rho + u \rho^3 + 2u^{ab} \rho \sigma^2, \label{eq:PH_real_rho} \\
0 & = \ddot{\sigma} - \frac{c_b}{\sigma^3} + \overline{\delta}_b \sigma + u \sigma^3 + 2u^{ab} \sigma \rho^2. \label{eq:PH_real_sigma}
\end{align}
\end{subequations}

If we initialize the system deep in the SF phase, we have for $t \rightarrow -\infty$, $\dot{\theta}\rightarrow 0$, $\dot{\rho} \rightarrow 0 $, $\dot{\phi}\rightarrow 0$, and $\dot{\sigma} \rightarrow 0 $ and thus we find for $\rho_0 = \rho(t\rightarrow -\infty)$ and $\sigma_0 = \sigma(t\rightarrow -\infty)$
\begin{equation}
\rho_0 = \sqrt{\frac{\abs{\overline{\delta}_{a,0}} - 2 u^{ab} \sigma_0^2}{u}}, \quad
\sigma_0 = \sqrt{\frac{\abs{\overline{\delta}_{b,0}} - 2 u^{ab} \rho_0^2}{u}}, \nn
\end{equation}
and $c_a = c_b = 0$ (we choose $\theta = \phi = 0$ initially without loss of generality). We can decouple these expressions to get
\begin{equation}
\rho_0 = \sqrt{\frac{\abs{\overline{\delta}_{a,0}} u - 2 u^{ab} \abs{\overline{\delta}_{b,0}}}{u^2 - (2 u^{ab})^2 }}, \label{eq:rho_initial}
\end{equation}
and
\begin{equation}
\sigma_0 = \sqrt{\frac{\abs{\overline{\delta}_{b,0}} u - 2 u^{ab} \abs{\overline{\delta}_{a,0}}}{u^2 - (2 u^{ab})^2 }}. \label{eq:sigma_initial}
\end{equation}
Note that in the limit of no inter-species interactions, i.e. $u^{ab} \rightarrow 0$, we recover the single species solutions $\rho_0 = \sqrt{\frac{\abs{\overline{\delta}_{a,0}}}{u}}$ and $\sigma_0 = \sqrt{\frac{\abs{\overline{\delta}_{b,0}}}{u}}$ \cite{Kennett2011}.

\begin{figure}
\includegraphics[width = 0.47\textwidth]{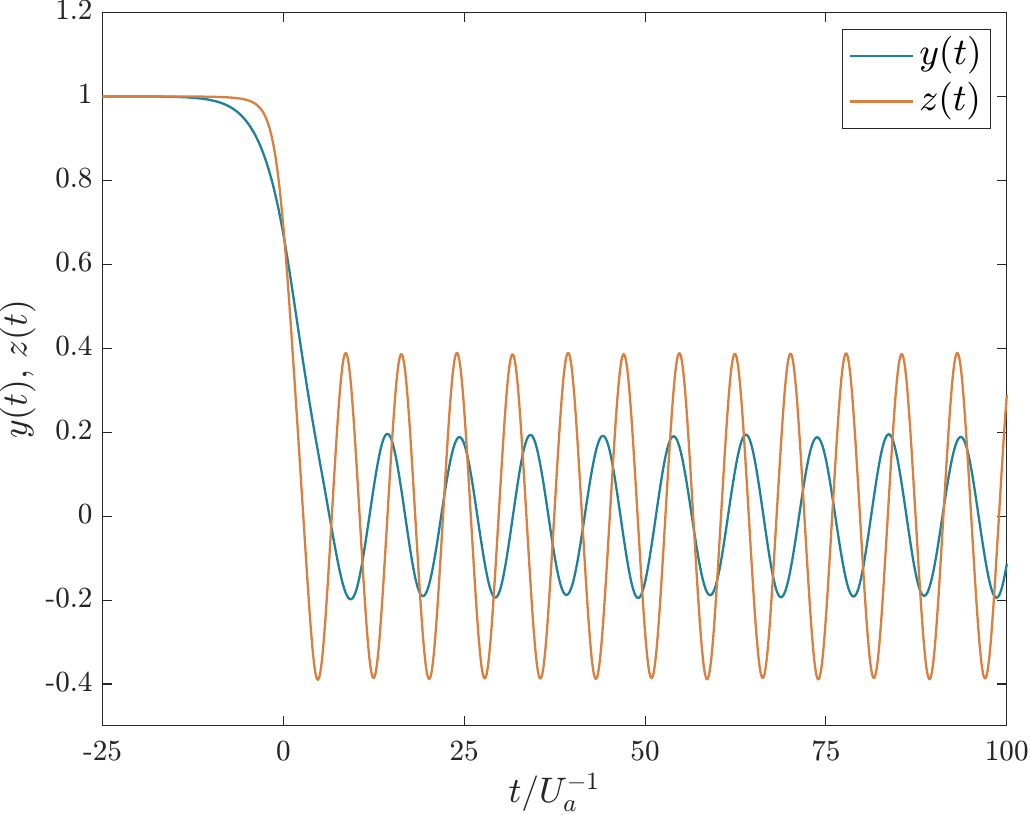}
\caption{Dynamics of $y(t)$ and $z(t)$ with initial amplitude normalized to unity in the particle-hole symmetric case. The parameters are $\beta U_a = 100$, $U_a = U_b$, $V= 0.72 U_a$, $\mu = 0.312363 U_a$, $\tau_{a,Q} = 5/U_a$, $\tau_{b,Q}=2/U_a$, and we take $\delta_{a,0} = \delta_{b,0} = 1.85 U_a$, and $\delta_{a,1} = \delta_{b,1} = J_0(\mu) = 0.2336 U_a$.}
\label{fig:Particle_Hole}
\end{figure}

Equations \eqref{eq:PH_real_rho} and \eqref{eq:PH_real_sigma} then simplify to
\begin{subequations}
\begin{align}
0 & = \ddot{\rho} + \delta_a \rho + u \rho^3 + 2 u^{ab} \sigma^2 \rho, \label{eq:PH_simplified_rho} \\
0 & = \ddot{\sigma} + \delta_b \sigma + u \sigma^3 + 2 u^{ab} \rho^2 \sigma. \label{eq:PH_simplified_sigma}
\end{align}
\end{subequations}
We obtain numerical solutions of Eqs. \eqref{eq:PH_simplified_rho} and \eqref{eq:PH_simplified_sigma} with $\delta_x$ taking the form given in Eq. \eqref{eq:delta}, with $\tau_{a,Q}=5/U_a$ and $\tau_{b,Q}=2/U_a$, at the particle-hole symmetric point around the first Mott-lobes, which is shown in Fig. \ref{fig:Particle_Hole}. We can see that for $t \gg \tau_{x,Q}$, the order parameters oscillate periodically about zero. When we average over a period $T$ at times $t\gg \tau_{a,Q}$,
\begin{equation}
\< y\>_T = \int_t^{t+T} \dd{\tilde{t}} y(\tilde{t}) \simeq  0,
\end{equation}
(and similarly $\<z\>_T \simeq 0$) as is expected in the Mott-insulating state. Different values for $\tau_{x,Q}$ change the dynamics in a similar way to the single species case \cite{Kennett2011}; for larger $\tau_{x,Q}$, the amplitude and frequency of the post-quench oscillation of species $x$ decrease.

\subsection{Generic case}

In the generic case, we have $\lambda_a \lambda_b \neq 0$, and $U_a \neq U_b$. We start with Eqs.~\eqref{eq:simplified_EOMs_a} and \eqref{eq:simplified_EOMs_b} and try a solution of the form $y(t) = \rho(t) \e^{\ii \theta(t)}$ and $z(t) = \sigma(t) \e^{\ii \phi(t)}$, which splits the equations into real and imaginary parts. For species $a$ we find
\begin{subequations}
\begin{align}
0 & = \kappa_a^2 \left[ \ddot{\rho} - \left( \dot{\theta}\right)^2 \rho \right] - \lambda_a \rho\,\dot{\theta} + \delta_a \rho + u^a \rho^3 + 2u^{ab} \rho\,\sigma^2, \label{eq:generic_case_real_species_a} \\
0 & = \kappa_a^2 \left[ 2\dot{\rho}\,\dot{\theta} + \rho\,\ddot{\theta} \right] + \lambda_a \dot{\rho}, \label{eq:generic_case_imaginary_species_a}
\end{align}
\end{subequations}
and for species $b$
\begin{subequations}
\begin{align}
0 & = \kappa_b^2 \left[ \ddot{\sigma} - \left( \dot{\phi}\right)^2 \sigma \right] - \lambda_b \sigma \dot{\phi} + \delta_b \sigma + u^b \sigma^3 + 2u^{ab} \rho^2 \sigma, \label{eq:generic_case_real_species_b} \\
0 & = \kappa_b^2 \left[ 2\dot{\sigma} \dot{\phi} + \sigma \ddot{\phi} \right] + \lambda_b \dot{\sigma}.\label{eq:generic_case_imaginary_species_b}
\end{align}
\end{subequations}
Just as in the particle-hole symmetric case, the imaginary parts can be decoupled, allowing us to find the solutions
\begin{equation}
\dot{\theta}  =  \frac{c_a - \frac{\lambda_a}{2}\rho^2}{\kappa_a^2 \rho^2} \quad \mathrm{and} \quad \dot{\phi}  = \frac{c_b - \frac{\lambda_b}{2}\sigma^2}{\kappa_b^2 \rho^2}.
\end{equation}

In the $t\rightarrow -\infty$ limit, we again have $\dot{\rho} = \dot{\sigma} = \dot{\theta} = \dot{\phi} = 0$, and $\rho = \rho_0$ and $\sigma = \sigma_0$, so we can solve for the integration constants
\begin{eqnarray}
c_a = \frac{\lambda_a}{2} \rho_0^2 \quad \mathrm{and} \quad c_b = \frac{\lambda_b}{2} \sigma_0^2   ,
\end{eqnarray}
with $\rho_0$ and $\sigma_0$ given by 
\begin{subequations}
\begin{align}
\rho_0 & = \sqrt{\frac{\abs{\overline{\delta}_{a,0}} u^b - 2 u^{ab} \abs{\overline{\delta}_{b,0}}}{u^a u^b - (2 u^{ab})^2 }}, \label{eq:rho_initial} \\
\sigma_0 & = \sqrt{\frac{\abs{\overline{\delta}_{b,0}} u^a - 2 u^{ab} \abs{\overline{\delta}_{a,0}}}{u^a u^b - (2 u^{ab})^2 }}. \label{eq:sigma_initial}
\end{align}
\end{subequations}
Note that in the $u^{ab}\rightarrow 0$ limit, the integration constants reduce to the single species solution $c_x = \frac{\lambda_x \delta_{x,0}}{2u^x}$ \cite{Kennett2011}.

\begin{figure}
\includegraphics[width = 0.47\textwidth]{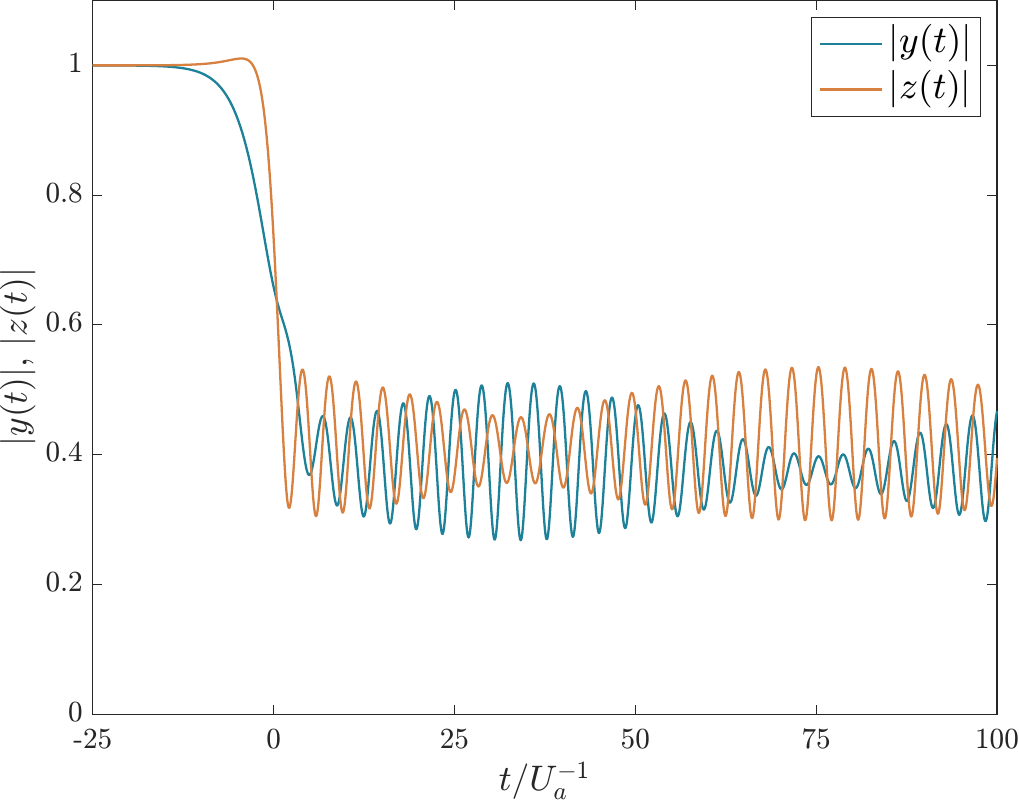}
\caption{Dynamics of $\abs{y(t)}$ and $\abs{z(t)}$ normalized to unity in the generic case. The parameters are $\beta U_a= 100$, $U_b = 0.89\, U_a$, $V = 0.72\, U_a$, $\mu = 0.396\, U_a$, $\tau_{a,Q} = 5/U_a$, $\tau_{b,Q}=2/U_a$, and we take $\delta_{a,0} = \delta_{b,0} = 1.85\, U_a$, $\delta_{a,1} = J_{a,0}(\mu) = 0.2241\, U_a$ and $\delta_{b,1} = J_{b,0}(\mu) = 0.2070\, U_a $.} 
\label{fig:Generic}
\end{figure}

With these results, we can rewrite Eqs. \eqref{eq:generic_case_real_species_a} and \eqref{eq:generic_case_real_species_b} as
\begin{subequations}
\begin{align}
0 & = \kappa_a^2 \ddot{\rho} - \frac{\lambda_a^2 \rho_0^4}{4\kappa_a^2 \rho^3} + \frac{\lambda_a^2}{4\kappa_a^2}\rho + \delta_a \rho + u^a \rho^3 + 2u^{ab} \rho\,\sigma^2, \label{eq:numeric_form_generic_case_a} \\
0 & = \kappa_b^2 \ddot{\sigma} - \frac{\lambda_b^2 \sigma_0^4}{4\kappa_b^2 \sigma^3} + \frac{\lambda_b^2}{4\kappa_b^2} \sigma + \delta_b \sigma + u^b \sigma^3 + 2u^{ab} \rho^2 \sigma. \label{eq:numeric_form_generic_case_b}
\end{align}
\end{subequations}
When $u^{ab} = 0$, Eqs.~\eqref{eq:numeric_form_generic_case_a} and \eqref{eq:numeric_form_generic_case_b} correct Eq. (28) in Ref. \cite{Kennett2011} by 
a factor of $\rho_0^2$ ($\sigma_0^2$) in the second term and the inclusion of the third term on the RHS. We solve Eqs. \eqref{eq:numeric_form_generic_case_a} and \eqref{eq:numeric_form_generic_case_b} numerically for $\mu = 0.396\,U_a$ (well away from both divergences and the particle-hole symmetric case of either species) and display $\abs{y(t)}$ and $\abs{z(t)}$ in Fig. \ref{fig:Generic}. 

Unlike the particle-hole symmetric case, the post-quench oscillations are about a non-zero magnitude of the order parameter. Note that Figs. \ref{fig:Particle_Hole} and \ref{fig:Generic} show different quantities; while in the particle-hole symmetric case $y(t)$ and $z(t)$ can be chosen to be purely real, in the generic case, we show $\abs{y(t)}$ and $\abs{z(t)}$, since, when $\lambda_a \lambda_b \neq 0$, both the real and imaginary parts of the order parameters oscillate about zero individually. Additionally, we can clearly see beating in the dynamics of both order parameters, which we attribute to the interactions between the two species. It appears that when one of the order parameters' oscillations reaches its peak amplitude, the other is at its minimum.

\subsection{Hybrid cases}

In the hybrid cases, we have $\lambda_a \lambda_b = 0$ and $\lambda_a \neq \lambda_b$. In this derivation, we choose $\lambda_a \neq 0$ and $\lambda_b =0$. The other case follows similarly. Substituting into Eqs. \eqref{eq:simplified_EOMs_a} and \eqref{eq:simplified_EOMs_b} we get
\begin{subequations}
\begin{align}
0 = \kappa_a^2 \ddot{y} & + \ii \lambda_a \dot{y}  +  \delta_a y + u^a \abs{y}^2 y+  2u^{ab}y \abs{z}^2,  \\
0 = \kappa_b^2 \ddot{z} & + \delta_b z + u^b \abs{z}^2 z  +  2 u^{ab} \abs{y}^2 z,
\end{align}
\end{subequations}
and we again seek solutions of the form $y(t) = \rho(t) \e^{\ii \theta (t)}$ and $z(t) = \sigma(t) \e^{\ii\phi(t)}$. The procedure is the same as in the particle-hole symmetric case for species $b$ and the generic case for species $a$. We can thus write down the final set of coupled differential equations immediately as
\begin{subequations}
\begin{align}
0 = \kappa_a^2 \ddot{\rho} & - \frac{\lambda_a^2 \rho_0^4}{4\kappa_a^2 \rho^3} + \frac{\lambda_a^2}{4\kappa_a^2}\rho + \delta_a \rho  + u^a \rho^3 + 2u^{ab} \rho\,\sigma^2, \label{eq:numeric_form_hybrid_case_a} \\
0 =\kappa_b^2 \ddot{\sigma} & + \delta_b \sigma  + u^b\sigma^3 + 2 u^{ab} \rho^2 \sigma. \label{eq:numeric_form_hybrid_case_b}
\end{align}
\end{subequations}

\begin{figure}
\includegraphics[width = 0.47\textwidth]{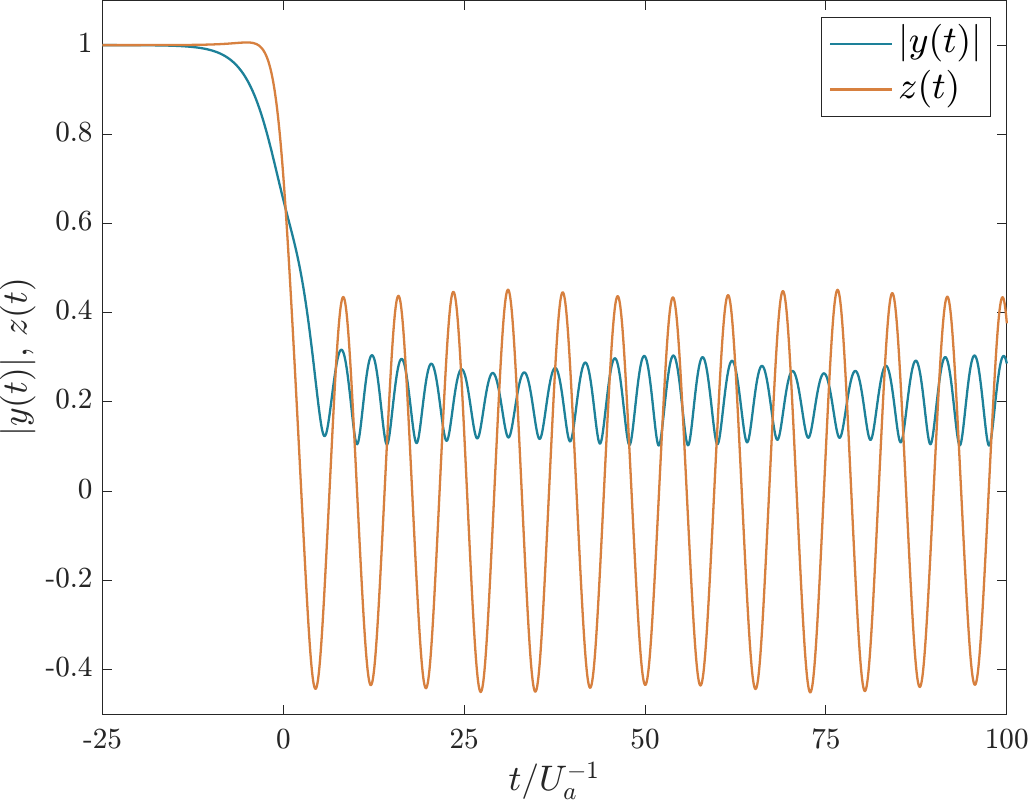}
\caption{Dynamics of $\abs{y(t)}$ and $z(t)$ normalized to unity in one of the hybrid cases. The parameters are $\beta U_a= 100$, $U_b = 0.89\, U_a$, $V = 0.72\, U_a$, $\lambda_b = 0$, $\mu = 0.297488\, U_a$, $\tau_{a,Q} = 5/U_a$, $\tau_{b,Q}=2/U_a$, and we take $\delta_{a,0} = \delta_{b,0} = 1.85\, U_a$, and $\delta_{a,1} = J_{a,0}(\mu) = 0.2332\, U_a$ and $\delta_{b,1} = J_{b,0}(\mu) = 0.2197\, U_a$.}
\label{fig:Hybrid}
\end{figure}

We solve Eqs.~\eqref{eq:numeric_form_hybrid_case_a} and \eqref{eq:numeric_form_hybrid_case_b} numerically and show $\abs{y(t)}$ and $z(t)$ in Fig. \ref{fig:Hybrid}. It can easily be seen that the dynamics have characteristics of both the particle-hole symmetric and the generic cases. Both order parameters display periodicities, however, $\abs{y(t)}$ oscillates about a non-zero value, while $z(t)$ oscillates about zero. Additionally, we can see that beats are present, and just like in the generic case, they appear to be out of phase such that the maximum amplitude of $\abs{y(t)}$ occurs at the time for which the amplitude of $z(t)$ reaches its minimum, and vice versa. When averaged over a full cycle of beats at times $t\gg \tau_{b,Q}$, $\<z(t)\> \simeq 0$, as expected in the MI phase.


\section{Discussion and Conclusions}
\label{sec:Disc}

In this paper, we derived a strong-coupling effective theory for the two-component BHM using the Schwinger-Keldysh technique, generalizing previous work for the single-component case \cite{Kennett2011}. This approach is a one-particle irreducible (1PI) formalism that allows for the description of equilibrium properties and out-of-equilibrium dynamics in both the SF and MI phases.

The derivation of this 1PI action is the main result of this paper. Using this action, we obtain the mean-field phase boundary at zero and finite temperatures. Similarly to previous authors \cite{Altman2003, Kuklov2003, Kuklov2004, Isacsson2005, Barman2015}, we find the possibility of hybrid phases with one species in a MI phase and the other in a SF phase. We show that care is required in taking the zero temperature limit and determine how the phase boundary depends on $U_a$, $U_b$, and $V$ when $U_a \neq U_b$ up to the fourth Mott-lobe, correcting some results in the literature \cite{Iskin2010, Chen2003, Bai2020}. The zero-temperature phase boundary contains abrupt jumps between Mott-lobes, which leads to the possibility of a $\mathrm{MI}(n) \rightarrow \mathrm{MI}(n+1)$ transition for non-zero values of the hopping parameters.

Other studies have shown that instead of a MI phase, the Mott-lobes with odd occupation numbers represent one of two new phases not present in the single-species case, most notably when $U_a  = U_b$. For repulsive inter-species interactions ($V>0$) the system can develop a counterflow SF phase \cite{Altman2003, Kuklov2003, Hu2009}, while attractive interactions ($V<0$) can give rise to a pair SF phase \cite{Kuklov2004, Hu2009, Chen2010, Menotti2010}. In order to investigate these additional phases within our approach, we would need to calculate two-point correlations. This requires a two-particle irreducible (2PI) theory which has been determined in the single species case for both clean \cite{Fitzpatrick2018PRA, Fitzpatrick2018NPB} and disordered situations \cite{Mokhtari2023}, and the 1PI action obtained here is essential for deriving the 2PI strong-coupling theory for the two-component BHM, which we intend to address in future work.

Another future direction is to add disorder to the two-component theory here. This would allow us to address the thermalization studied experimentally in Ref.~\cite{Rubio2019}. The strong-coupling approach here is also in principle generalizable to study the OOE dynamics of the extended BHM \cite{Iskin2009} at both the 1PI and 2PI levels.

Our second main result is the saddle-point equations of motion of the SF order parameters. We study the OOE dynamics of the SF order parameters in the low-frequency, long-wavelength limit as the hopping parameters $J_a$ and $J_b$ are varied as a function of time at fixed chemical potential such that we cross from the SF phase to the MI phase. When quenching through the particle-hole symmetric point, the resulting order parameter dynamics average to zero as expected in the MI phase. In the generic and hybrid cases, we found a beating in the dynamics that we attribute to the interactions between the two species. 
Note that in the long-time limit of the order parameter dynamics, one might expect the oscillation amplitude to decay, which is not seen in our numerical solutions. This is an artifact of the mean field approximation and should be rectified by the inclusion of terms with spatial dependence, which goes beyond the calculations presented here.
In future work we intend to extend the formalism presented here to study the out-of-equilibrium dynamics of temporal and spatial correlations as well as those of the order parameters.

\section{Acknowledgements}

F. R. B. and M. P. K. acknowledge that the research presented here was conducted on the traditional, ancestral, and unceded territories of the Coast Salish peoples, including the s\textschwa lilw\textschwa ta\textbeltl  ~(Tsleil-Waututh), k\textsuperscript{w}ik\textsuperscript{w}\textschwa \`{\textcrlambda}\textschwa m (Kwikwetlem), S\textsubbar{k}w\textsubbar{x}w\'u7mesh \'Uxwumixw (Squamish), and x\textsuperscript{w}m\textschwa\texttheta k\textsuperscript{w}\textschwa \`y\textschwa m (Musqueam) Nations.

The authors acknowledge support from NSERC, and thank Ali Mokhtari-Jazi for helpful discussions and an anonymous referee for insightful questions.

\begin{appendix}

\section{Mean-field phase boundary}
\label{app:MFPB}

One way to determine the mean-field phase boundary between the SF and MI phases is to determine when the coefficients of the quadratic terms in the action in Eq. \eqref{eq:S^1_eff} vanish \cite{Kennett2011}. To do this, we note that
\begin{equation}
\tau^1_{\alpha_1\beta_1} \hat{G}^x_{\beta_1 \beta_2}(t_1, t_2) \tau^1_{\alpha_2 \beta_2} =  \begin{pmatrix}
\mathcal{G}^{x,K}_0(t_1, t_2) & \mathcal{G}^{x,R}_0(t_1, t_2) \\
\mathcal{G}^{x,A}_0(t_1, t_2) & 0
\end{pmatrix}, 
\end{equation}
where $\mathcal{G}^{x,R}_0$, $\mathcal{G}^{x,A}_0$, and $\mathcal{G}^{x,K}_0$ are the retarded, advanced, and Keldysh propagators, respectively. The subscript 0 indicates that the propagators are associated with $\hat{H}_0$. The definitions of the propagators are
\begin{align}
\mathcal{G}^{x,K}_0(t, t') = {} & \mathcal{G}^{x,<}_0 (t',t) + \mathcal{G}^{x,>}_0 (t,t'), \nn \\
\mathcal{G}^{x,R}_0(t, t') = {} & \theta(t - t') \left[ \mathcal{G}^{x,>}_0 (t,t') - \mathcal{G}^{x,<}_0 (t',t) \right], \nn \\
\mathcal{G}^{x,A}_0(t, t') = {} & \theta(t' - t) \left[ \mathcal{G}^{x,<}_0 (t',t) - \mathcal{G}^{x,>}_0 (t,t') \right] \nn,
\end{align}
with
\begin{align}
\mathcal{G}^{x,<}_0 (t',t) = {} & -\ii \frac{\Trace\left[ \hat{x}^\dagger(t') \hat{x}(t) \hat{\rho}_0 \right]}{Z}, \nn \\
\mathcal{G}^{x,>}_0 (t,t') = {} & -\ii \frac{\Trace\left[ \hat{x}(t) \hat{x}^\dagger(t') \hat{\rho}_0 \right]}{Z}. \nn 
\end{align}
These expressions can be evaluated using the interaction representation
\begin{align}
\hat{x}^\dagger(t') = {} & \e^{\ii  \hat{H}_0 t'} \hat{x}^\dagger \e^{-\ii \hat{H}_0 t'} 
\quad \mathrm{and} \quad
\hat{x}(t) = {}  \e^{\ii \hat{H}_0 t} \hat{x} \e^{-\ii \hat{H}_0 t}. \nn
\end{align}
Note that we can treat this as a single-site problem in the atomic limit with the partition function
\begin{equation}
Z = \Trace\left( \hat{\rho}_0 \right) =  \Trace\left(\e^{-\beta \hat{H}_0}\right) = \sum_{p,q=0}^\infty \e^{-\beta E_{p,q}}, \nn
\end{equation}
so that we obtain at inverse temperature $\beta$
\begin{align}
\mathcal{G}^{a,<}_0 (t',t) = {} & -\frac{\ii}{Z} \sum_{p,q=0}^\infty p \e^{\ii (E_{p,q} -E_{p-1,q} ) (t'-t) } \e^{-\beta E_{p,q}}, \label{eq:G_lesser_a_AppendixA} \\
\mathcal{G}^{b,<}_0 (t',t) = {} & -\frac{\ii}{Z} \sum_{p,q=0}^\infty q \e^{\ii (E_{p,q} -E_{p,q-1} ) (t'-t) } \e^{-\beta E_{p,q}}, \label{eq:G_lesser_b_AppendixA}
\end{align}
where we recalled $\hat{a}\ket{p,q} = \sqrt{p} \ket{p-1,q}$ and $\hat{a}^\dagger \ket{p,q} = \sqrt{p+1} \ket{p+1,q}$,  $\hat{b}\ket{p,q} = \sqrt{q} \ket{p,q-1}$ and $\hat{b}^\dagger \ket{p,q} = \sqrt{q+1} \ket{p,q+1}$, and $\hat{H}_0 \ket{p,q} = E_{p,q} \ket{p,q}$ with
\begin{equation}
E_{p,q} = -\mu (p+q) + \frac{U_a}{2}p(p-1) + \frac{U_b}{2} q(q-1) + Vpq. \nn
\end{equation}
Similar expressions can be found for $\mathcal{G}^{a,>}_0 (t,t')$ and $\mathcal{G}^{b,>}_0 (t,t')$. Hence the retarded and Keldysh Green's functions for species $a$ take the form
\begin{widetext}
\begin{align}
\mathcal{G}^{a,R}_0 (t, t') = & - \frac{\ii}{Z} \theta (t - t')  \sum_{p,q=0}^\infty \e^{-\beta E_{p,q}}  \left[ (p+1) \e^{\ii (E_{p,q}-E_{p+1,q}) (t-t') }  - p \e^{\ii (E_{p-1,q} - E_{p,q}) (t - t') }  \right] , \label{eq:G_retarded_AppendixA} \\
\mathcal{G}^{a,K}_0 (t, t') = & - \frac{\ii}{Z} \sum_{p,q=0}^\infty \e^{-\beta E_{p,q}}  \left[ (p+1) \e^{\ii (E_{p,q} - E_{p+1,q} ) (t - t') } + p \e^{ \ii (E_{p-1,q} - E_{p,q}) (t - t') } \right], \label{eq:G_Keldysh_AppendixA}
\end{align}
\end{widetext}
respectively. 

Unlike the single species case, the Green's functions do not simplify to the same extent in the zero temperature limit. Instead of a single term in each sum being selected, all terms that have the same $n=p+q$ must be considered, as discussed in more detail below. The equivalent expressions to Eqs. \eqref{eq:G_lesser_a_AppendixA} -- \eqref{eq:G_Keldysh_AppendixA} for species $b$ can be obtained similarly.

When we Fourier transform in space and time the quadratic part of $S^I_\mathrm{int}$, Eq.~\eqref{eq:S^1_eff}, we get for species $a$
\begin{align}
-& \int_{-\infty}^\infty \frac{\dd{\omega}}{2\pi} \sum_\mathbf{k} \psi^*_{\alpha}(\omega, \mathbf{k}) \left( 2J_{a,\mathbf{k}} \right)^{-1} \tau^1_{\alpha \beta} \psi^{\phantom{*}}_{\beta}(\omega, \mathbf{k}) \nn \\
& - \int_{-\infty}^{\infty} \frac{\dd{\omega}}{2\pi} \sum_\mathbf{k} \psi^*_{\alpha_1 \beta_1}(\omega, \mathbf{k}) \tau^1_{\alpha_1 \beta_1} G^{1a,c}_{\beta_1 \beta_2}(\omega) \tau^1_{\alpha_2 \beta_2} \psi^{\phantom{*}}_{\alpha_2}(\omega, \mathbf{k}),
\end{align}
where for a $d$-dimensional cubic lattice with lattice spacing $\tilde{a}$
\begin{align*}
J_{a,\mathbf{k}}(t) = {} & \left[ J_{a,0} + j_a(t) \right] \sum_{j=1}^d \cos(k_j \tilde{a}) \\
\simeq {} & \left( d - \frac{1}{2} k^2 \tilde{a}^2 \right) \left[ J_{a,0} + j_a(t) \right],
\end{align*}
assuming that $k\tilde{a} \ll 1$.

If we set $j(t) = 0$, we can locate the phase boundary in the $\omega, k \rightarrow 0$ limit by noting that the coefficients of the $\psi^*_q \psi^{\phantom{*}}_c$ term in the action vanish when
\begin{equation}
\frac{1}{2d J_{a,0}} + \mathcal{G}^{a,R}_0(\omega = 0) = 0. \label{eq:MFPB}
\end{equation}
The retarded propagator at finite temperature is obtained from Fourier transforming Eq. \eqref{eq:G_retarded_AppendixA}:
\begin{align}
\mathcal{G}^{a,R}_0 (\omega) = \frac{1}{Z} \sum_{p,q = 0}^\infty \e^{-\beta E_{p,q}} \left[ \frac{p+1}{\mu - U_a p - Vq + \omega + \ii 0^+} \right. \nn \\
\left. {} - \frac{p}{\mu - U_a (p-1) - V q + \omega + \ii 0^+} \right]. \label{eq:G_retarted_fourier}
\end{align}
The advanced propagator may be obtained from
\begin{equation}
\mathcal{G}^{a,A}_0(\omega) = \left[ \mathcal{G}^{a,R}_0(\omega) \right]^*,
\end{equation}
and the Keldysh propagator is
\begin{align}
\mathcal{G}^{a,K}_0(\omega) = & - \frac{2\pi \ii}{Z} \sum_{p,q=0} \e^{-\beta E_{p,q}}  \left[ (p+1) \delta(\mu - U_a p - Vq + \omega) \right. \nn\\
& \left. \hspace*{1.5cm} + p \delta\left( \mu - U_a (p-1) - Vq + \omega \right) \right].
\end{align}
Note that for $q=0$, we recover the single-species result \cite{Kennett2011}. The phase boundary for species $b$ can be derived in the exact same way.

As discussed in Sec.~\ref{sec:MFPB}, we need to be careful in evaluating the Mott lobes in the zero temperature limit. We can determine the phase boundary of each Mott lobe individually by rearranging Eq.~\eqref{eq:MFPB},
\begin{equation}
2dJ_a = \frac{-Z}{\sum_{p,q=0}^\infty \e^{-\beta E_{p,q}} \left[ \frac{p+1}{\mu-U_a p - Vq} - \frac{p}{\mu - U_a (p-1) -Vq} \right]},
\end{equation}
and considering each occupation number separately. We showed the analytic expressions for Mott lobes for $n=1$ and $n=2$ in Sec.~\ref{sec:MFPB}. Here, we also show $n=0$, $n=3$, and $n=4$. Expressions for the Mott lobes of species $b$ can be obtained similarly.

\subsection{Zeroth Mott-lobe}

For the zeroth Mott-lobe, $n=0$, i.e. $p=q=0$, and Eq. \eqref{eq:MFPB} reduces to
\begin{equation}
2dJ_a = -\mu,
\end{equation}
which is simply the phase boundary to the vacuum state.

\subsection{Third Mott-lobe}

The derivation for $n=3$ follows similarly to the $n=2$ case shown in Sec.~\ref{sec:Second_Lobe}. We simply state the result for the phase boundary:
\begin{widetext}
\begin{equation}
2dJ_a = \left\lbrace \begin{array}{lll}
\vphantom{\Bigg\lbrace} \dfrac{-(1 + \delta_{U_a,V} + \delta_{U_b+2V,3U_a} + \delta_{U_a,U_b} )}{ M(3,0) + \delta_{U_a,V} M(2,1)  + \delta_{U_b+2V,3U_a} M(1,2) + \delta_{U_a,U_b} M(0,3) }, & \mathrm{if}~ U_a \leq U_b, V \\
\vphantom{\Bigg\lbrace} \dfrac{-(1 + \delta_{U_a,V} + \delta_{U_a+2V,3U_b} + \delta_{U_a,U_b} )}{ \delta_{U_a,V} M(3,0) + M(2,1)  + \delta_{U_a,U_b} M(1,2) + \delta_{U_a+2V,3U_b} M(0,3) }, & \mathrm{if}~ V \leq U_a \leq U_b \\
\vphantom{\Bigg\lbrace} \dfrac{-(1 + \delta_{U_b,V} + \delta_{U_b+2V,3U_a} + \delta_{U_a,U_b} )}{ \delta_{U_b+2V,3U_a} M(3,0) + \delta_{U_a,U_b} M(2,1)  +  M(1,2) + \delta_{U_b,V}   M(0,3) }, & \mathrm{if}~ V \leq U_b \leq U_a \\
\vphantom{\Bigg\lbrace} \dfrac{-(1 + \delta_{U_b,V} + \delta_{U_a+2V,3U_b} + \delta_{U_a,U_b} )}{ \delta_{U_a,U_b} M(3,0) + \delta_{U_a+2V,3U_b} M(2,1)  +  \delta_{U_b,V} M(1,2) +   M(0,3) }, & \mathrm{if}~  U_b \leq U_a, V \\
\end{array} \right. , \label{eq:MottLobe3}
\end{equation}
where $M(p,q) = \left[ \frac{p+1}{\mu - U_a p - Vq}  - \frac{p}{\mu - U_a (p-1) - Vq}\right]$. The first and last lines of Eq.~\eqref{eq:MottLobe3} show that when $V$ is bigger than $U_a$ or $U_b$, only $M(3,0)$ or $M(0,3)$, respectively, remain, which indicates phase separation.

\subsection{Fourth Mott-lobe}

The expression for the fourth Mott lobe also follows similarly:
\begin{equation}
2dJ_a = \left\lbrace
\begin{array}{ll}
\vphantom{\Bigg\lbrace}  \dfrac{-( 1 + \delta_{U_a,V} + \delta_{U_b+4V,5U_a} + \delta_{U_b+V,2U_a} + \delta_{U_a,U_b} )}{ M(4,0) + \delta_{U_a,V} M(3,1) + \delta_{U_b+4V,5U_a} M(2,2) + \delta_{U_b+V,2U_a} M(1,3) + \delta_{U_a,U_b} M(0,4) }, & \mathrm{if}~  U_a \leq U_b, V   \\
\vphantom{\Bigg\lbrace} \dfrac{-( 1 + \delta_{U_a,V} + \delta_{U_a+V,2U_b} + \delta_{U_b+V,2U_a} + \delta_{U_a,U_b} )}{ \delta_{U_a,V} M(4,0) +  M(3,1) + \delta_{U_b+V,2U_a} M(2,2) + \delta_{U_a,U_b}  M(1,3) +  \delta_{U_a+V,2U_b} M(0,4) }, & \mathrm{if}~ \left\lbrace\begin{array}{l}
V \leq U_a \leq U_b \\ 2U_a \leq U_b + V
\end{array} \right.  \\
\vphantom{\Bigg\lbrace} \dfrac{-( 1 + \delta_{U_b+4V,5U_a} + \delta_{U_b+V,2U_a} + \delta_{U_a+V,2U_b} + \delta_{U_a+4V,5U_b} )}{ \delta_{U_b+4V,5U_a}M(4,0) + \delta_{U_b+V,2U_a} M(3,1) +  M(2,2) + \delta_{U_a+V,2U_b} M(1,3) + \delta_{U_a+4V,5U_b} M(0,4) }, & \mathrm{if}~ \left\lbrace\begin{array}{l}
V \leq U_a, U_b \\ U_a + V \leq 2U_b \\ U_b + V \leq 2U_a
\end{array} \right.  \\
\vphantom{\Bigg\lbrace} \dfrac{-( 1 + \delta_{U_b+V, 2U_a} + \delta_{U_a,U_b} + \delta_{U_a+V,2U_b} + \delta_{U_b,V} )}{  \delta_{U_b+V, 2U_a} M(4,0) + \delta_{U_a,U_b} M(3,1) + \delta_{U_a+V,2U_b} M(2,2) +  M(1,3) + \delta_{U_b,V} M(0,4) }, & \mathrm{if}~ \left\lbrace\begin{array}{l}
V \leq U_b \leq U_a \\ 2U_b \leq U_a + V
\end{array} \right.  \\
\vphantom{\Bigg\lbrace} \dfrac{-( 1 + \delta_{U_a,U_b} + \delta_{U_a+V,2U_b} + \delta_{U_a+4V,5U_b} + \delta_{U_b,V} )}{ \delta_{U_a,U_b} M(4,0) + \delta_{U_a+V,2U_b} M(3,1) + \delta_{U_a+4V,5U_b} M(2,2) + \delta_{U_b,V} M(1,3) +  M(0,4) }, & \mathrm{if}~  U_b \leq U_a, V  
\end{array}
\right. . \label{eq:Fourth_Mott_Lobe}
\end{equation}


\section{Quartic couplings in the action of the effective theory}
\label{app:couplings}

The quartic couplings in Eq.~\eqref{eq:S2_eff} are given by
\begin{align}
u^x_{\alpha^{\phantom{1}}_1 \alpha^{\phantom{1}}_2 \alpha^{\phantom{1}}_3 \alpha^{\phantom{1}}_4}(t^{\phantom{1}}_1, t^{\phantom{1}}_2, t^{\phantom{1}}_3, t^{\phantom{1}}_4) = & - \frac{1}{4} \int_{-\infty}^\infty \dd{t^{\phantom{1}}_5} \dd{t'_5} \dd{t^{\phantom{1}}_6} \dd{t'_6} G^{2x}_{\alpha^{\phantom{1}}_5 \alpha^{\phantom{1}}_6 \alpha_6' \alpha_5'}(t^{\phantom{1}}_5, t^{\phantom{1}}_6, t_6', t_5') \nn \\
& \times  \left\lbrace \left[ G^x_{\alpha^{\phantom{1}}_{5}\alpha^{\phantom{1}}_{1}}(t^{\phantom{1}}_{5}, t^{\phantom{1}}_{1})  \right]^{-1}
\left[ G^x_{\alpha^{\phantom{1}}_{6}\alpha^{\phantom{1}}_{2}}(t^{\phantom{1}}_{6}, t^{\phantom{1}}_{2})  \right]^{-1}
\left[ G^x_{\alpha^{\phantom{1}}_{3}\alpha'_{5}}(t^{\phantom{1}}_{3}, t'_{5})  \right]^{-1}
\left[ G^x_{\alpha^{\phantom{1}}_{4}\alpha'_{6}}(t^{\phantom{1}}_{4}, t'_{6})  \right]^{-1}\right. \nn \\
& + \left. \left[ \left( \alpha^{\phantom{1}}_5, t^{\phantom{1}}_5 \right) \leftrightarrow \left(  \alpha^{\phantom{1}}_6, t^{\phantom{1}}_6 \right) \right] + \left[ \left(  \alpha^{\phantom{1}}_3, t^{\phantom{1}}_3 \right) \leftrightarrow \left(  \alpha^{\phantom{1}}_4, t^{\phantom{1}}_4 \right) \right] + \left[ \left(  \alpha^{\phantom{1}}_5, t^{\phantom{1}}_5 \right) \leftrightarrow \left(  \alpha^{\phantom{1}}_6, t^{\phantom{1}}_6 \right)  ; \left(  \alpha^{\phantom{1}}_3, t^{\phantom{1}}_3 \right) \leftrightarrow \left(  \alpha^{\phantom{1}}_4, t^{\phantom{1}}_4 \right)\right] \vphantom{\left[ G^a_{\alpha^{\phantom{1}}_{4}\alpha'_{6}}(t^{\phantom{1}}_{4}, t'_{6})  \right]^{-1}}  \right\rbrace, \nn \\\label{eq:u^a}
\end{align}
with $x=a$ or $b$, and
\begin{align}
u^{ab}_{\alpha^{\phantom{1}}_1 \alpha^{\phantom{1}}_2 \alpha^{\phantom{1}}_3 \alpha^{\phantom{1}}_4}(t^{\phantom{1}}_1, t^{\phantom{1}}_2, t^{\phantom{1}}_3, t^{\phantom{1}}_4) 
= & - \frac{1}{4} \int_{-\infty}^\infty \dd{t^{\phantom{1}}_5} \dd{t_5'} \dd{t^{\phantom{1}}_6} \dd{t_6'} G^{1a+1b, c}_{\alpha^{\phantom{1}}_5 \alpha^{\phantom{1}}_6 \alpha_6' \alpha_5'} (t^{\phantom{1}}_5, t^{\phantom{1}}_6, t_6', t_5') \nn \\
& \times\left[ G^a_{\alpha^{\phantom{1}}_5 \alpha^{\phantom{1}}_1}(t^{\phantom{1}}_5, t^{\phantom{1}}_1)  \right]^{-1} \left[ G^a_{\alpha^{\phantom{1}}_2 \alpha_5'}(t^{\phantom{1}}_2, t_5')  \right]^{-1} \left[ G^b_{\alpha^{\phantom{1}}_6 \alpha^{\phantom{1}}_3}(t^{\phantom{1}}_6, t^{\phantom{1}}_3)  \right]^{-1}  \left[ G^b_{\alpha^{\phantom{1}}_4 \alpha_6'}(t^{\phantom{1}}_4, t_6')  \right]^{-1}. \label{eq:u^ab}
\end{align}
\end{widetext}
Note that $u^a$, $u^b$, and $u^{ab}$ are non-local in time.

\section{Parameters in the equations of motion}
\label{app:EOMs}

There are three parameters per boson species that enter the equations of motion:
\begin{align}
\nu_x = \left. \left[ \mathcal{G}^{x,R}_0 \right]^{-1}\right\vert_{\omega=0}&; \quad \lambda_x = \left. - \pdv{\omega} \left[  \mathcal{G}_0^{x,R} \right]^{-1} \right\vert_{\omega=0} ; \nn \\
\kappa_x^2 = & \left. \frac{1}{2} \pdv[2]{\omega} \left[  \mathcal{G}_0^{x,R} \right]^{-1} \right\vert_{\omega = 0}.
\end{align}
For species $a$ we can evaluate them from Eq. \eqref{eq:G_retarted_fourier} to give (expressions for species $b$ can be obtained similarly)
\begin{equation}
\nu_a = \frac{Z}{\sum_{p,q=0}^\infty \e^{-\beta E_{p,q}} \left[ \frac{p+1}{\mu - U_a p - Vq}  - \frac{p}{\mu - U_a (p-1) - Vq}  \right]}, \label{eq:nu_a_AppendixB}
\end{equation}
\begin{align}
\lambda_a = {} & - \frac{\nu_a^2 }{Z} \sum\limits_{p,q=0}^\infty \e^{-\beta E_{p,q}} \nn \\
& \times \left\lbrace \frac{p+1}{[\mu - U_a p - Vq]^2}  - \frac{p}{[\mu - U_a (p-1) - Vq]^2} \right\rbrace, \label{eq:lambda_a_AppendixB}
\end{align}
and
\begin{align}
\kappa_a^2 = {} & \frac{\lambda_a^2}{\nu_a} - \frac{\nu_a^2}{Z} \sum_{p,q=0}^\infty \e^{-\beta E_{p,q}} \nn \\
& \times \left\lbrace \frac{ p+1 }{[\mu - U_a p - Vq]^3}  - \frac{p}{[\mu - U_a (p-1) - Vq]^3}  \right\rbrace. \label{eq:kappa_a_AppendixB}
\end{align}
Note that for $q=0$, Eq. \eqref{eq:lambda_a_AppendixB} corrects Eq. (C2) in Ref. \cite{Kennett2011} by a factor of (-1). The three different interaction parameters obtained by taking the low-frequency limit are
\begin{widetext}
\begin{align}
u^a = {} & - \frac{\nu_a^4}{2Z} \sum_{p,q=0}^\infty \e^{-\beta E_{p,q}}\left\lbrace \frac{4(p+1)(p+2)}{(E_{p,q} - E_{p+1,q})^2 (E_{p,q} - E_{p+2,q})} 
+\frac{4p(p-1)}{(E_{p,q} - E_{p-1,q})^2 (E_{p,q} - E_{p-2,q}) }  \right.  \nn\\
& \left. - \frac{4(p+1)^2}{(E_{p,q} - E_{p+1,q})^3} 
- \frac{4p^2}{(E_{p,q} - E_{p-1,q})^3} 
- \frac{4p(p+1)}{(E_{p,q} - E_{p-1,q})^2 (E_{p,q} - E_{p+1,q}) } 
- \frac{4p(p+1)}{(E_{p,q} - E_{p-1,q})(E_{p,q} - E_{p+1,q})^2} \right\rbrace, \label{eq:u^a_low_frequency}
\end{align}
\begin{align}
u^b = {} & - \frac{\nu_b^4}{2Z} \sum_{p,q=0}^\infty \e^{-\beta E_{p,q}} \left\lbrace \frac{4(q+1)(q+2)}{ (E_{p,q} - E_{p,q+1})^2 (E_{p,q} - E_{p,q+2}) } +\frac{4q(q-1)}{(E_{p,q} - E_{p,q-1})^2 (E_{p,q} - E_{p,q-2}) }  \right. \nn\\
& \left. - \frac{4(q+1)^2}{(E_{p,q} - E_{p,q+1})^3} - \frac{4q^2}{(E_{p,q} - E_{p,q-1})^3} - \frac{4q(q+1)}{(E_{p,q} - E_{p,q-1})^2 (E_{p,q} - E_{p,q+1})} - \frac{4q(q+1)}{(E_{p,q} - E_{p,q-1}) (E_{p,q} - E_{p,q+1})^2} \right\rbrace, \label{eq:u^b_low_frequency}
\end{align}
and
\begin{align}
u^{ab}  = {} & -\frac{\nu_a^2 \nu_b^2}{8Z} \sum_{p,q=0}^\infty \e^{ -\beta E_{p,q}}  \nn \\
& \times  \left\lbrace ~ \frac{(p+1)(q+1)}{E_{p,q}-E_{p+1,q+1}}\left[ \frac{1}{E_{p,q}-E_{p+1,q}} +  \frac{1}{E_{p,q}-E_{p,q+1}}  \right]^2  +  \frac{(p+1)q}{E_{p,q}-E_{p+1,q-1}}\left[ \frac{1}{E_{p,q}-E_{p+1,q}} +  \frac{1}{E_{p,q}-E_{p,q-1}}  \right]^2 \right. \nn  \\
& \quad+  \frac{p(q+1)}{E_{p,q}-E_{p-1,q+1}}\left[ \frac{1}{E_{p,q}-E_{p-1,q}} +  \frac{1}{E_{p,q}-E_{p,q+1}}  \right]^2 +  \frac{pq}{E_{p,q}-E_{p-1,q-1}}\left[ \frac{1}{E_{p,q}-E_{p-1,q}} +  \frac{1}{E_{p,q}-E_{p,q-1}}  \right]^2 \nn \\
& \quad - \frac{(p+1)(q+1)}{\left(E_{p,q}-E_{p+1,q}\right) \left(E_{p,q}-E_{p,q+1}\right)}\left[ \frac{1}{E_{p,q}-E_{p+1,q}} +  \frac{1}{E_{p,q}-E_{p,q+1}}  \right] \nn \\
& \quad - \frac{(p+1)q}{\left(E_{p,q}-E_{p+1,q}\right) \left(E_{p,q}-E_{p,q-1}\right)}\left[ \frac{1}{E_{p,q}-E_{p+1,q}} +  \frac{1}{E_{p,q}-E_{p,q-1}}  \right]  \nn \\
& \quad - \frac{p(q+1)}{\left(E_{p,q}-E_{p-1,q}\right) \left(E_{p,q}-E_{p,q+1}\right)}\left[ \frac{1}{E_{p,q}-E_{p-1,q}} +  \frac{1}{E_{p,q}-E_{p,q+1}}  \right] \nn  \\
& \quad \left. -  \frac{pq}{\left(E_{p,q}-E_{p-1,q}\right) \left(E_{p,q}-E_{p,q-1}\right)}\left[ \frac{1}{E_{p,q}-E_{p-1,q}} +  \frac{1}{E_{p,q}-E_{p,q-1}}  \right] \right\rbrace. \label{eq:u^ab_low_frequency}
\end{align}
\end{widetext}
The intra-species interaction terms in Eqs. \eqref{eq:u^a_low_frequency} and \eqref{eq:u^b_low_frequency}  reduce to the single species ones found in Ref.~\cite{Kennett2011} for $q=0$ and $p=0$, respectively.

\section{Special cases for $u^{ab}$}

In Eq. \eqref{eq:u^ab_low_frequency}, we find the differences 
\begin{equation}
E_{p,q} - E_{p+1,q-1} = - U_a p + U_b (q-1) + V(p-q+1),
\end{equation} 
and
\begin{equation}
E_{p,q}- E_{p-1,q+1} = U_a (p-1) - U_b q - V(p-q+1),
\end{equation} 
in the denominators of the second and third terms inside the braces. These differences can vanish independently of the chemical potential $\mu$ for specific $p$ and $q$ given specific values of the various interaction strengths. Specifically, there are six cases in which these differences can vanish:
\begin{enumerate}
\item $U_a, U_b \neq V$ and $p+q = 1$,
\item $U_a = U_b \neq V$ and $\abs{p-q} = 1$,
\item $U_a = U_b = V$, any/all $p$ and $q$,
\item $U_a = V \neq U_b$, any $p$, and $q=0$ and $q=1$, 
\item $U_a \neq U_b = V$, any $q$, and $p=0$ and $p=1$, 
\item $c_1 U_a + c_2 U_b + c_3 V = 0$ with $c_1, c_2, c_3 \in \mathbb{Z}$.
\end{enumerate}
Note that unless $U_a$, $U_b$, and/or $V$ are irrational numbers, we can never avoid case 6. However, for the $n<5$ Mott-lobes we consider in this work, we can ignore this case by choosing values of $U_a$, $U_b$, and $V$ such that $\abs{c_1}, \abs{c_2}, \abs{c_3} >5$, so that the values of $p$ and $q$ that would introduce any divergences are not part of the relevant phase boundary expressions.

The other five cases, however, result in apparently divergent terms that need to be dealt with in order for us to get sensible results. In fact, these divergences were artificially introduced when taking the low frequency limit. To avoid this, we must separate out the relevant terms from the sum before taking the low frequency limit. Below we show how to take care of these terms in $u^{ab}$.

\begin{widetext}
\subsection{Case 1: $U_a, U_b \neq V$ and $p+q=1$} \label{Case1}

The two terms for $p+q=1$ are
\begin{align}
& \left. \e^{ -\beta E_{p,q}} \left\lbrace ~  \frac{(p+1)q}{E_{p,q}-E_{p+1,q-1}}\left[ \frac{1}{E_{p,q}-E_{p+1,q}} +  \frac{1}{E_{p,q}-E_{p,q-1}}  \right]^2 +  \frac{p(q+1)}{E_{p,q}-E_{p-1,q+1}}\left[ \frac{1}{E_{p,q}-E_{p-1,q}} +  \frac{1}{E_{p,q}-E_{p,q+1}}  \right]^2 \right\rbrace \right\vert_{p+q=1} \nn \\
& \rightarrow {}  2 \e^{- \beta E_{1,0}} \left[ \frac{1}{(E_{0,1} - E_{1,1})} + \frac{1}{(E_{0,1} - E_{0,0})}\right] \left[ \frac{1}{(E_{0,1} - E_{1,1})^2} + \frac{1}{(E_{0,1} - E_{0,0})^2}\right]  \nn \\
& = {} 2 \e^{\beta \mu} \left[ \frac{1}{(\mu - V)} - \frac{1}{\mu}\right] \left[ \frac{1}{(\mu - V)^2} + \frac{1}{\mu^2}\right]. \label{eq:case1}
\end{align}

\subsection{Case 2: $U_a = U_b \neq V$ and $\abs{p-q} = 1$} \label{Case2}

\begin{align}
& \sum_{p,q=0}^\infty \e^{ -\beta E_{p,q}} \left\lbrace ~  \frac{(p+1)q}{E_{p,q}-E_{p+1,q-1}}\left[ \frac{1}{E_{p,q}-E_{p+1,q}} +  \frac{1}{E_{p,q}-E_{p,q-1}}  \right]^2 \right. \nn \\
& \left. \left. \hspace{2cm}  + \frac{p(q+1)}{E_{p,q}-E_{p-1,q+1}}\left[ \frac{1}{E_{p,q}-E_{p-1,q}} +  \frac{1}{E_{p,q}-E_{p,q+1}}  \right]^2 \right\rbrace \right\vert_{\abs{p-q}=1} \nn \\
&\rightarrow \sum_{p=0}^\infty \e^{-\beta E_{p,q}} \left\lbrace (p+1)q \left[ \frac{1}{(E_{p,q} - E_{p+1,q})} + \frac{1}{(E_{p,q} - E_{p,q-1})} \right] \left[ \frac{1}{(E_{p,q} - E_{p+1,q})^2} + \frac{1}{(E_{p,q} - E_{p,q-1})^2} \right] \vphantom{\left[ \frac{1}{E_{p,q} - E_{p,q+1}} + \frac{1}{E_{p,q} - E_{p+1,q}}  \right]^2}  \right.  \nn \\
& \left. \left. \hspace{2cm} + \frac{p(q+1)}{2(V-U)} \left[ \frac{1}{E_{p,q} - E_{p+1,q}} + \frac{1}{E_{p,q} - E_{p,q+1}}  \right]^2 \right\rbrace \right\vert_{q=p+1} \nn \\
& \quad + \sum_{q=0}^\infty \e^{-\beta E_{p,q}} \left\lbrace p(q+1) \left[ \frac{1}{E_{p,q} - E_{p,q+1}} + \frac{1}{E_{p,q} - E_{p-1,q}} \right] \left[ \frac{1}{(E_{p,q} - E_{p,q+1})^2} + \frac{1}{(E_{p,q} - E_{p-1,q})^2} \right]   \vphantom{\left[ \frac{1}{E_{p,q} - E_{p+1,q}} + \frac{1}{E_{p,q} - E_{p,q+1}} \right]^2}     \right. \nn \\
& \left. \left. \hspace{2cm} + \frac{(p+1)q}{2(V-U)} \left[ \frac{1}{E_{p,q} - E_{p+1,q}} + \frac{1}{E_{p,q} - E_{p,q+1}} \right]^2 \right\rbrace \right\vert_{p=q+1} \nn \\
& =  \sum_{p=0}^\infty \e^{-\beta E_{p,p+1}}  \left\lbrace 2(p+1)^2 \left[ \frac{1}{(E_{p,p+1} - E_{p+1,p+1})} + \frac{1}{(E_{p+1,p} - E_{p,p})}  \right] \left[ \frac{1}{(E_{p,p+1} - E_{p+1,p+1})^2} + \frac{1}{(E_{p+1,p} - E_{p,p})^2}  \right] \vphantom{\left[ \frac{1}{(E_{p,p+1} - E_{p,p+2})} + \frac{1}{(E_{p,+1} - E_{p+1, p+1})} \right]^2} \right.  \nn \\
& \left. \hspace{2cm} + \frac{p(p+2)}{V-U} \left[ \frac{1}{(E_{p,p+1} - E_{p,p+2})} + \frac{1}{(E_{p,p+1} - E_{p+1, p+1})} \right]^2 \right\rbrace ,
\end{align}
where $U_a = U_b =U$. Note that the denominator in the second term enforces $V\neq U$ which was required from the start. Additionally, for $p=0$, we recover Eq.~\eqref{eq:case1}.

\subsection{Case 3: $U_a = U_b = V$ for all $p$ and $q$} \label{Case3}

\begin{align}
& \sum_{p,q=0}^\infty \e^{ -\beta E_{p,q}}  \left\lbrace \frac{(p+1)q}{E_{p,q}-E_{p+1,q-1}}\left[ \frac{1}{E_{p,q}-E_{p+1,q}} +  \frac{1}{E_{p,q}-E_{p,q-1}}  \right]^2 +  \frac{p(q+1)}{E_{p,q}-E_{p-1,q+1}}\left[ \frac{1}{E_{p,q}-E_{p-1,q}} +  \frac{1}{E_{p,q}-E_{p,q+1}}  \right]^2 \right\rbrace  \nn \\
& \rightarrow \sum_{p,q=0}^\infty \e^{-\beta E_{p,q}} \left[ (p+1)q +p(q+1) \right] \left[ \frac{1}{(E_{p,q} - E_{p+1,q})} + \frac{1}{ (E_{p,q} -E_{p,q-1}) }  \right] \left[ \frac{1}{(E_{p,q} - E_{p+1,q})^2} + \frac{1}{ (E_{p,q} -E_{p,q-1})^2 }  \right] 
\end{align}

\subsection{Case 4: $U_a = V \neq U_b$, any $p$, and $q=0$ and $q=1$} \label{Case4}


Note that $E_{p+1,0} = E_{p,1}$ for $U_a = V$, so we can combine the $q=0$ and $q=1$ terms into a single expression:
\begin{align}
& \sum_{p=0}^\infty \e^{ -\beta E_{p,q}} \left. \left\lbrace \frac{(p+1)q}{E_{p,q}-E_{p+1,q-1}}\left[ \frac{1}{E_{p,q}-E_{p+1,q}} +  \frac{1}{E_{p,q}-E_{p,q-1}}  \right]^2  +  \frac{p(q+1)}{E_{p,q}-E_{p-1,q+1}}\left[ \frac{1}{E_{p,q}-E_{p-1,q}} +  \frac{1}{E_{p,q}-E_{p,q+1}}  \right]^2 \right\rbrace \right|_{\begin{matrix} q = 0 \\ q = 1 
\end{matrix}} \nn \\
& \rightarrow \sum_{p=0}^\infty \e^{-\beta E_{p+1,0}}  2(p+1)  \left[ \frac{1}{E_{p+1,0} - E_{p,0}} + \frac{1}{E_{p+1,0} - E_{p+1,1}}  \right] \left[ \frac{1}{(E_{p+1,0} - E_{p,0})^2} + \frac{1}{(E_{p+1,0} - E_{p+1,1})^2}  \right] \vphantom{\left( \frac{1}{E_{p+1,q} - E_{p+1,q+1}} + \frac{1}{E_{p+1,q} - E_{p,q}} \right)^2}.
\end{align}

\subsection{Case 5: $U_b = V \neq U_a$, any $q$, and $p=0$ and $p=1$} \label{Case5}

This is essentially the same as case 4 up to a relabelling in variables/dummy indices:
\begin{align}
& \sum_{q=0}^\infty \e^{ -\beta E_{p,q}} \left. \left\lbrace \frac{(p+1)q}{E_{p,q}-E_{p+1,q-1}}\left[ \frac{1}{E_{p,q}-E_{p+1,q}} +  \frac{1}{E_{p,q}-E_{p,q-1}}  \right]^2   +  \frac{p(q+1)}{E_{p,q}-E_{p-1,q+1}}\left[ \frac{1}{E_{p,q}-E_{p-1,q}} +  \frac{1}{E_{p,q}-E_{p,q+1}}  \right]^2 \right\rbrace \right|_{\begin{matrix} p = 0 \\ p = 1 
\end{matrix}} \nn \\
& \rightarrow \sum_{q=0}^\infty \e^{-\beta E_{0,q+1}}  2(q+1) \left[ \frac{1}{E_{0,q+1} - E_{0,q}} + \frac{1}{E_{0,q+1} - E_{1,q+1}}  \right] \left[ \frac{1}{(E_{0,q+1} - E_{0,q})^2} + \frac{1}{(E_{0,q+1} - E_{1,q+1})^2}  \right] \vphantom{\left( \frac{1}{E_{p+1,q} - E_{p+1,q+1}} + \frac{1}{E_{p+1,q} - E_{p,q}} \right)^2}.
\end{align}

\end{widetext}

\begin{widetext}
\section{Evaluation of the four-point function}
\label{app:four-point}

To evaluate the four time correlation functions, there are 24 basic correlations we need
\begin{align}
B^{a b b^\dagger a^\dagger}(t_1, t_2,t_3,t_4) = {} & \frac{1}{Z} \Trace\left[ \e^{-\beta \hat{H}} \hat{a}(t_1) \hat{b}(t_2) \hat{b}^\dagger(t_3) \hat{a}^\dagger(t_4) \right] \nn \\ 
= {} & \frac{1}{Z}		
\sum_{p,q = 0}^\infty (p+1)(q+1) \e^{\ii E_{p,q} (t_1-t_4 +\ii\beta) 
+ \ii E_{p+1,q}(t_2 + t_4 - t_1 - t_3) 
+ \ii E_{p+1,q+1} (t_3-t_2)  }, \\																						
B^{a^\dagger b^\dagger b a}(t_4, t_3,t_2,t_1) = {} & \frac{1}{Z} \sum_{p,q = 0}^\infty   pq             
\e^{\ii E_{p,q} (t_4 - t_1 + \ii\beta) 
+ \ii E_{p - 1, q}(t_3 + t_1 - t_4 - t_2) 
+ \ii E_{p - 1, q - 1} (t_2 - t_3)  }, \\																					
B^{a b a^\dagger b^\dagger}(t_1, t_2,t_4,t_3) = {} & \frac{1}{Z} 
\sum_{p,q = 0}^\infty (p+1)(q+1) \e^{\ii E_{p,q} (t_1-t_3 +\ii\beta) 
+ \ii E_{p+1,q}(t_2-t_1) +\ii E_{p,q+1}(t_3-t_4) 
+ \ii E_{p+1,q+1} (t_4-t_2)  }, \\ 																						
B^{b^\dagger a^\dagger b a }(t_3, t_4,t_2,t_1) = {} & \frac{1}{Z} \sum_{p,q = 0}^\infty   pq             
\e^{\ii E_{p,q} (t_3 - t_1 + \ii\beta) 
+ \ii E_{p, q-1}(t_4 - t_3) 
+ \ii E_{p-1, q}(t_1 - t_2) 
+ \ii E_{p - 1, q - 1} (t_2 - t_4)  }, 	\\																				
B^{a b^\dagger b a^\dagger}(t_1, t_3,t_2,t_4) = {} & \frac{1}{Z} \sum_{p,q = 0}^\infty   (p+1)q             
\e^{\ii E_{p,q} (t_1 - t_4 + \ii\beta) 
+ \ii E_{p + 1, q}(t_3 +t_4 - t_1 - t_2) 
+ \ii E_{p + 1, q - 1} (t_2 - t_3)  }, \\																				
B^{a^\dagger b b^\dagger a}(t_4, t_2,t_3,t_1) = {} & \frac{1}{Z} \sum_{p,q = 0}^\infty   p(q+1)             
\e^{\ii E_{p,q} (t_4 - t_1 + \ii\beta) 
+ \ii E_{p - 1, q}(t_2 +t_1 - t_4 - t_3) 
+ \ii E_{p - 1, q + 1} (t_3 - t_2)  }, \\																				
B^{a b^\dagger a^\dagger b}(t_1, t_3,t_4,t_2) = {} & \frac{1}{Z} \sum_{p,q = 0}^\infty   (p+1)q             
\e^{\ii E_{p,q} (t_1 - t_2 + \ii\beta) 
+ \ii E_{p + 1, q}(t_3 - t_1) 
+ \ii E_{p, q - 1}(t_2 - t_4) 
+ \ii E_{p + 1, q - 1} (t_4 - t_3)  }, \\																				
B^{b a^\dagger b^\dagger a}(t_2, t_4,t_3,t_1) = {} & \frac{1}{Z} \sum_{p,q = 0}^\infty   p(q+1)             
\e^{\ii E_{p,q} (t_2 - t_1 + \ii\beta) 
+ \ii E_{p, q+1}(t_4 - t_2) 
+ \ii E_{p-1, q}(t_1 - t_3) 
+ \ii E_{p - 1, q + 1} (t_3 - t_4)  }, \\ 																				
B^{aa^\dagger bb^\dagger}(t_1, t_4,t_2,t_3) = {} & \frac{1}{Z} 
\sum_{p,q = 0}^\infty (p+1)(q+1) 
\e^{\ii E_{p,q} (t_1+t_2-t_4-t_3 +\ii\beta) 
+\ii E_{p+1,q} (t_4-t_1) 
+\ii E_{p,q+1} (t_3-t_2) }, \\ 																							
B^{b^\dagger b a^\dagger a }(t_3, t_2,t_4,t_1) = {} & \frac{1}{Z} \sum_{p,q = 0}^\infty   pq           
\e^{\ii E_{p,q} (t_3 +t_4 - t_2 - t_1 + \ii\beta) 
+ \ii E_{p, q-1}(t_2 - t_3) 
+ \ii E_{p-1, q}(t_1 - t_4)  }, \\																						
B^{a a^\dagger b^\dagger b}(t_1, t_4,t_3,t_2) = {} & \frac{1}{Z} \sum_{p,q = 0}^\infty   (p+1)q             
\e^{\ii E_{p,q} (t_1 + t_3 - t_4 - t_2 +\ii\beta) 
+ \ii E_{p + 1,q}(t_4 - t_1) 
+ \ii E_{p, q - 1}(t_2 - t_3) }, \\																						
B^{b b^\dagger a^\dagger a}(t_2, t_3,t_4,t_1) = {} & \frac{1}{Z} \sum_{p,q = 0}^\infty   p(q+1)             
\e^{\ii E_{p,q} (t_2 + t_4 - t_3 - t_1 +\ii\beta) 
+ \ii E_{p,q+1}(t_3 - t_2) 
+ \ii E_{p-1, q}(t_1 - t_4) }, 																							
\end{align}
\begin{align}
B^{b a b^\dagger a^\dagger}(t_2, t_1,t_3,t_4) = {} & \frac{1}{Z} 
\sum_{p,q = 0}^\infty (p+1)(q+1) 
\e^{\ii E_{p,q} (t_2-t_4 +\ii\beta) 
+ \ii E_{p,q+1}(t_1-t_2) 
+\ii E_{p+1,q}(t_4-t_3) 
+ \ii E_{p+1,q+1} (t_3-t_1)  }, \\																						
B^{a^\dagger b^\dagger a b }(t_4, t_3,t_1,t_2) = {} & \frac{1}{Z} \sum_{p,q = 0}^\infty   pq             
\e^{\ii E_{p,q} (t_4 - t_2 + \ii\beta) 
+ \ii E_{p - 1, q}(t_3 - t_4) 
+ \ii E_{p, q - 1}(t_2 - t_1) 
+ \ii E_{p - 1, q - 1} (t_1 - t_3)  }, \\																					
B^{b a a^\dagger b^\dagger}(t_2, t_1,t_4,t_3) = {} & \frac{1}{Z}
\sum_{p,q = 0}^\infty (p+1)(q+1) \e^{\ii E_{p,q} (t_2-t_3 +\ii\beta) 
+ \ii E_{p,q+1}(t_1 + t_3 - t_2 - t_4) 
+ \ii E_{p+1,q+1} (t_4-t_1)  }, \\																						
B^{b^\dagger a^\dagger a b}(t_3, t_4,t_1,t_2) = {} & \frac{1}{Z} \sum_{p,q = 0}^\infty   pq             
\e^{\ii E_{p,q} (t_3 - t_2 + \ii\beta) 
+ \ii E_{p, q-1}(t_4 + t_2 - t_3 - t_1) 
+ \ii E_{p - 1, q - 1} (t_1 - t_4)  }, \\																					
B^{bb^\dagger aa^\dagger}(t_2, t_3,t_1,t_4) = {} &  \frac{1}{Z}
\sum_{p,q = 0}^\infty (p+1)(q+1) 
\e^{\ii E_{p,q} (t_2+t_1-t_3-t_4 +\ii\beta) 
+\ii E_{p,q+1} (t_3-t_2) 
+\ii E_{p+1,q} (t_4-t_1) }, \\																							
B^{a^\dagger a b^\dagger b }(t_4, t_1,t_3,t_2) = {} & \frac{1}{Z} \sum_{p,q = 0}^\infty   pq           
\e^{\ii E_{p,q} (t_4 +t_3 - t_1 - t_2 + \ii\beta) 
+ \ii E_{p - 1, q}(t_1 - t_4) 
+ \ii E_{p, q - 1}(t_2 - t_3)  }, \\																						
B^{b a^\dagger a b^\dagger}(t_2, t_4,t_1,t_3) = {} & \frac{1}{Z} \sum_{p,q = 0}^\infty   p(q+1)             
\e^{\ii E_{p,q} (t_2 - t_3 + \ii\beta) 
+ \ii E_{p, q+1}(t_4 +t_3 - t_2 - t_1) 
+ \ii E_{p - 1, q + 1} (t_1 - t_4)  }, 	\\																			
B^{b^\dagger a a^\dagger b}(t_3, t_1,t_4,t_2) = {} & \frac{1}{Z} \sum_{p,q = 0}^\infty   (p+1)q             
\e^{\ii E_{p,q} (t_3 - t_2 + \ii\beta) 
+ \ii E_{p, q-1}(t_1 +t_2 - t_3 - t_4) 
+ \ii E_{p + 1, q - 1} (t_4 - t_1)  }, \\																				
B^{b^\dagger a b a^\dagger }(t_3, t_1,t_2,t_4) = {} & \frac{1}{Z} \sum_{p,q = 0}^\infty   (p+1)q             
\e^{\ii E_{p,q} (t_3 - t_4 + \ii\beta) 
+ \ii E_{p, q-1}(t_1 - t_3) 
+ \ii E_{p+1, q}(t_4 - t_2) 
+ \ii E_{p + 1, q - 1} (t_2 - t_1)  }, 	\\																				
B^{a^\dagger b a b^\dagger }(t_4, t_2,t_1,t_3) = {} & \frac{1}{Z} \sum_{p,q = 0}^\infty   p(q+1)             
\e^{\ii E_{p,q} (t_4 - t_3 + \ii\beta) 
+ \ii E_{p - 1, q}(t_2 - t_4) 
+ \ii E_{p, q + 1}(t_3 - t_1) 
+ \ii E_{p - 1, q + 1} (t_1 - t_2)  }, \\																				
B^{b^\dagger b a a^\dagger }(t_3, t_2,t_1,t_4) = {} & \frac{1}{Z} \sum_{p,q = 0}^\infty   (p+1)q             
\e^{\ii E_{p,q} (t_3 +t_1 - t_2 - t_4 + \ii\beta) 
+ \ii E_{p, q-1}(t_2 - t_3) 
+ \ii E_{p+1, q}(t_4 - t_1) }, \\																						
B^{a^\dagger a b b^\dagger }(t_4, t_1,t_2,t_3) = {} &\frac{1}{Z} \sum_{p,q = 0}^\infty   p(q+1)             
\e^{\ii E_{p,q} (t_4 +t_2 - t_1 - t_3 + \ii\beta) 
+ \ii E_{p - 1, q}(t_1 - t_4) 
+ \ii E_{p, q + 1}(t_3 - t_2) }. 																						
\end{align}

In addition, we need the two-point correlations:
\begin{align}
C^{aa^\dagger}(t_1, t_4) = {} & \frac{1}{Z} \Trace \left[ \e^{-\beta \hat{H} } \hat{a}(t_1) \hat{a}^\dagger(t_4)  \right] \nn \\
= {} & \frac{1}{Z} \sum_{p,q = 0}^\infty (p+1) \e^{iE_{p,q} (t_1-t_4 + \ii\beta) +\ii E_{p+1,q} (t_4-t_1)} =  \ii \mathcal{G}_0^{a,>}(t_1, t_4), \\
C^{a^\dagger a}(t_4, t_1) = {} &  \frac{1}{Z} \sum_{p,q = 0}^\infty p \e^{iE_{p,q} (t_4-t_1 + \ii\beta) +\ii E_{p-1,q} (t_1-t_4)}  =  \ii \mathcal{G}_0^{a,<}(t_4, t_1), \\
C^{bb^\dagger }(t_2, t_3) = {} & \frac{1}{Z} \sum_{p,q = 0}^\infty (q+1) \e^{iE_{p,q} (t_2-t_3 + \ii\beta) +\ii E_{p,q+1} (t_3-t_2)} =  \ii \mathcal{G}_0^{b,>}(t_2, t_3) , \\
C^{b^\dagger b}(t_3, t_2)= {} & \frac{1}{Z} \sum_{p,q = 0}^\infty q \e^{iE_{p,q} (t_3-t_2 + \ii\beta) +\ii E_{p,q-1} (t_2-t_3)} = \ii \mathcal{G}_0^{b,<}(t_3, t_2), \\
C^{a b} (t_1,t_4) = {} & 0 \nn \\
= {} & C^{a b^\dagger} (t_1,t_3) = C^{a^\dagger b} (t_2,t_4) = C^{a^\dagger b^\dagger} (t_2,t_3).
\end{align}

The actual expressions for the four point function are rather tiresome to derive but are given here for completeness, where we use the notation $\theta_{ij} = \theta(t_i-t_j)$:
\begin{align}
G_{qqqc}  = \frac{\ii}{2} & \left\lbrace 
	   - \theta_{43}\theta_{32}\theta_{21} \left[ B^{\1 \2 \3 \4}\tb{1}{2}{3}{4} - B^{\4 \3 \2 \1}\tb{4}{3}{2}{1}\right] \right. \nn \\  
& + \theta_{42}\theta_{21}\theta_{43} \left[ B^{\1 \2 \4 \3}\tb{1}{2}{4}{3} - B^{\3 \4 \2 \1}\tb{3}{4}{2}{1}\right] \nn \\ 
&  - \theta_{42}\theta_{23}\theta_{31} \left[ B^{\1 \3 \2 \4}\tb{1}{3}{2}{4} - B^{\4 \2 \3 \1}\tb{4}{2}{3}{1}\right]  \nn\\
& + \theta_{43}\theta_{31}\theta_{42} \left[  B^{\1 \3 \4 \2}\tb{1}{3}{4}{2} - B^{\2 \4 \3 \1}\tb{2}{4}{3}{1}\right] \nn \\
&  - \theta_{42}\theta_{23}\theta_{41} \left[ B^{\1 \4 \2 \3}\tb{1}{4}{2}{3} - B^{\3 \2 \4 \1}\tb{3}{2}{4}{1}\right] \nn \\
&  - \theta_{43}\theta_{32}\theta_{41} \left[ B^{\1 \4 \3 \2}\tb{1}{4}{3}{2} - B^{\2 \3 \4 \1}\tb{2}{3}{4}{1}\right] \nn \\
&  - \theta_{43}\theta_{31}\theta_{12} \left[ B^{\2 \1 \3 \4}\tb{2}{1}{3}{4} - B^{\4 \3 \1 \2}\tb{4}{3}{1}{2}\right] \nn \\
&  + \theta_{41}\theta_{12}\theta_{43} \left[ B^{\2 \1 \4 \3}\tb{2}{1}{4}{3} - B^{\3 \4 \1 \2}\tb{3}{4}{1}{2} \right] \nn \\
&  - \theta_{41}\theta_{13}\theta_{32} \left[ B^{\2 \3 \1 \4}\tb{2}{3}{1}{4} - B^{\4 \1 \3 \2}\tb{4}{1}{3}{2} \right] \nn \\
&  - \theta_{41}\theta_{13}\theta_{42} \left[ B^{\2 \4 \1 \3}\tb{2}{4}{1}{3} - B^{\3 \1 \4 \2}\tb{3}{1}{4}{2} \right] \nn \\
&  - \theta_{42}\theta_{21}\theta_{13} \left[ B^{\3 \1 \2 \4}\tb{3}{1}{2}{4} - B^{\4 \2 \1 \3}\tb{4}{2}{1}{3}\right]  \nn \\
& \left. - \theta_{41}\theta_{12}\theta_{23} \left[ B^{\3 \2 \1 \4}\tb{3}{2}{1}{4} - B^{\4 \1 \2 \3}\tb{4}{1}{2}{3} \right]  \right\rbrace, 
\end{align}

\begin{align}
G_{qqcc}  = \frac{\ii}{2} & \left\lbrace 
	      \theta_{32} \theta_{21} \left[ B^{\1 \2 \3 \4}\tb{1}{2}{3}{4} - B^{\4 \3 \2 \1}\tb{4}{3}{2}{1}\right] \right. \nn \\  
& + \theta_{42} \theta_{21} \left[ B^{\1 \2 \4 \3}\tb{1}{2}{4}{3} - B^{\3 \4 \2 \1}\tb{3}{4}{2}{1}\right] \nn \\ 
& + \theta_{31 }\left( \theta_{23} - \theta_{24} \right) \left[ B^{\1 \3 \2 \4}\tb{1}{3}{2}{4} - B^{\4 \2 \3 \1}\tb{4}{2}{3}{1}\right]  \nn\\
&  -  \theta_{31}\theta_{42} \left[  B^{\1 \3 \4 \2}\tb{1}{3}{4}{2} - B^{\2 \4 \3 \1}\tb{2}{4}{3}{1}\right] \nn \\
&  + \theta_{41}\left( \theta_{24} - \theta_{23} \right) \left[ B^{\1 \4 \2 \3}\tb{1}{4}{2}{3} - B^{\3 \2 \4 \1}\tb{3}{2}{4}{1}\right] \nn \\
&  - \theta_{41} \theta_{32} \left[ B^{\1 \4 \3 \2}\tb{1}{4}{3}{2} - B^{\2 \3 \4 \1}\tb{2}{3}{4}{1}\right] \nn \\
&  + \theta_{31}\theta_{12} \left[ B^{\2 \1 \3 \4}\tb{2}{1}{3}{4} - B^{\4 \3 \1 \2}\tb{4}{3}{1}{2}\right] \nn \\
&  + \theta_{41} \theta_{12} \left[ B^{\2 \1 \4 \3}\tb{2}{1}{4}{3} - B^{\3 \4 \1 \2}\tb{3}{4}{1}{2} \right] \nn \\
&  + \theta_{32} \left( \theta_{13} - \theta_{14} \right) \left[ B^{\2 \3 \1 \4}\tb{2}{3}{1}{4} - B^{\4 \1 \3 \2}\tb{4}{1}{3}{2} \right] \nn \\
&  + \theta_{42} \left( \theta_{14} - \theta_{13} \right) \left[ B^{\2 \4 \1 \3}\tb{2}{4}{1}{3} - B^{\3 \1 \4 \2}\tb{3}{1}{4}{2} \right] \nn \\
&  + \left( \theta_{31} \theta_{12}\theta_{24} +\theta_{42} \theta_{21}\theta_{13} \right) \left[ B^{\3 \1 \2 \4}\tb{3}{1}{2}{4} - B^{\4 \2 \1 \3}\tb{4}{2}{1}{3}\right]  \nn \\
& \left. + \left( \theta_{32} \theta_{21}\theta_{14}  + \theta_{41} \theta_{12}\theta_{23} \right) \left[ B^{\3 \2 \1 \4}\tb{3}{2}{1}{4} - B^{\4 \1 \2 \3}\tb{4}{1}{2}{3} \right]  \right\rbrace  \nn \\
- \ii & \theta_{41} \theta_{32} \left[ C^{\4 \1}\tc{4}{1} - C^{\1\4} \tc{1}{4}  \right]  \left[ C^{\2 \3}\tc{2}{3} - C^{\3\2} \tc{3}{2}  \right],
\end{align}
\begin{align}
G_{qcqq}  = \frac{\ii}{2} & \left\lbrace 
	   - \theta_{23}\theta_{34}\theta_{21} \left[ B^{\1 \2 \3 \4}\tb{1}{2}{3}{4} - B^{\4 \3 \2 \1}\tb{4}{3}{2}{1}\right] \right. \nn \\  
&  - \theta_{24}\theta_{43}\theta_{21} \left[ B^{\1 \2 \4 \3}\tb{1}{2}{4}{3} - B^{\3 \4 \2 \1}\tb{3}{4}{2}{1}\right] \nn \\ 
& + \theta_{23}\theta_{31}\theta_{24} \left[ B^{\1 \3 \2 \4}\tb{1}{3}{2}{4} - B^{\4 \2 \3 \1}\tb{4}{2}{3}{1}\right]  \nn\\
&  - \theta_{24}\theta_{43}\theta_{31} \left[  B^{\1 \3 \4 \2}\tb{1}{3}{4}{2} - B^{\2 \4 \3 \1}\tb{2}{4}{3}{1}\right] \nn \\
& + \theta_{24}\theta_{41}\theta_{23} \left[ B^{\1 \4 \2 \3}\tb{1}{4}{2}{3} - B^{\3 \2 \4 \1}\tb{3}{2}{4}{1}\right] \nn \\
&  - \theta_{23}\theta_{34}\theta_{41} \left[ B^{\1 \4 \3 \2}\tb{1}{4}{3}{2} - B^{\2 \3 \4 \1}\tb{2}{3}{4}{1}\right] \nn \\
&  + \theta_{21}\theta_{13}\theta_{34} \left[ B^{\2 \1 \3 \4}\tb{2}{1}{3}{4} - B^{\4 \3 \1 \2}\tb{4}{3}{1}{2}\right] \nn \\
&  + \theta_{21}\theta_{14}\theta_{43} \left[ B^{\2 \1 \4 \3}\tb{2}{1}{4}{3} - B^{\3 \4 \1 \2}\tb{3}{4}{1}{2} \right] \nn \\
&  + \theta_{23}\theta_{31}\theta_{14} \left[ B^{\2 \3 \1 \4}\tb{2}{3}{1}{4} - B^{\4 \1 \3 \2}\tb{4}{1}{3}{2} \right] \nn \\
&  + \theta_{24}\theta_{41}\theta_{13} \left[ B^{\2 \4 \1 \3}\tb{2}{4}{1}{3} - B^{\3 \1 \4 \2}\tb{3}{1}{4}{2} \right] \nn \\
&  + \theta_{21}\theta_{13}\theta_{24} \left[ B^{\3 \1 \2 \4}\tb{3}{1}{2}{4} - B^{\4 \2 \1 \3}\tb{4}{2}{1}{3}\right]  \nn \\
& \left. - \theta_{21}\theta_{14}\theta_{23} \left[ B^{\3 \2 \1 \4}\tb{3}{2}{1}{4} - B^{\4 \1 \2 \3}\tb{4}{1}{2}{3} \right]  \right\rbrace, 
\end{align}
\begin{align}
G_{qcqc}  = \frac{\ii}{2} & \left\lbrace 
	      \theta_{21} \left( \theta_{32} - \theta_{34} \right) \left[ B^{\1 \2 \3 \4}\tb{1}{2}{3}{4} - B^{\4 \3 \2 \1}\tb{4}{3}{2}{1}\right] \right. \nn \\  
&  - \theta_{21} \theta_{43} \left[ B^{\1 \2 \4 \3}\tb{1}{2}{4}{3} - B^{\3 \4 \2 \1}\tb{3}{4}{2}{1}\right] \nn \\ 
& + \theta_{23 } \theta_{31} \left[ B^{\1 \3 \2 \4}\tb{1}{3}{2}{4} - B^{\4 \2 \3 \1}\tb{4}{2}{3}{1}\right]  \nn\\
& + \theta_{43}\theta_{31} \left[  B^{\1 \3 \4 \2}\tb{1}{3}{4}{2} - B^{\2 \4 \3 \1}\tb{2}{4}{3}{1}\right] \nn \\
&  - \theta_{41} \theta_{23}  \left[ B^{\1 \4 \2 \3}\tb{1}{4}{2}{3} - B^{\3 \2 \4 \1}\tb{3}{2}{4}{1}\right] \nn \\
& + \theta_{41} \left( \theta_{34} - \theta_{32} \right) \left[ B^{\1 \4 \3 \2}\tb{1}{4}{3}{2} - B^{\2 \3 \4 \1}\tb{2}{3}{4}{1}\right] \nn \\
& + \left( \theta_{21}\theta_{13}\theta_{34} + \theta_{43}\theta_{31}\theta_{12}  \right) \left[ B^{\2 \1 \3 \4}\tb{2}{1}{3}{4} - B^{\4 \3 \1 \2}\tb{4}{3}{1}{2}\right] \nn \\
& + \theta_{43} \left( \theta_{14} - \theta_{12} \right) \left[ B^{\2 \1 \4 \3}\tb{2}{1}{4}{3} - B^{\3 \4 \1 \2}\tb{3}{4}{1}{2} \right] \nn \\
& + \left( \theta_{23}\theta_{31}\theta_{14} - \theta_{41}\theta_{13}\theta_{32} \right) \left[ B^{\2 \3 \1 \4}\tb{2}{3}{1}{4} - B^{\4 \1 \3 \2}\tb{4}{1}{3}{2} \right] \nn \\
& + \theta_{41}\theta_{13}  \left[ B^{\2 \4 \1 \3}\tb{2}{4}{1}{3} - B^{\3 \1 \4 \2}\tb{3}{1}{4}{2} \right] \nn \\
& + \theta_{21}\theta_{13}  \left[ B^{\3 \1 \2 \4}\tb{3}{1}{2}{4} - B^{\4 \2 \1 \3}\tb{4}{2}{1}{3}\right]  \nn \\
& \left. + \theta_{23} \left( \theta_{12} - \theta_{14} \right) \left[ B^{\3 \2 \1 \4}\tb{3}{2}{1}{4} - B^{\4 \1 \2 \3}\tb{4}{1}{2}{3} \right]  \right\rbrace  \nn \\
- \ii & \theta_{41} \theta_{32} \left[ C^{\4 \1}\tc{4}{1} - C^{\1\4} \tc{1}{4}  \right]  \left[ C^{\2 \3}\tc{2}{3} - C^{\3\2} \tc{3}{2}  \right],
\end{align}
\begin{align}
G_{qccq}  = \frac{\ii}{2} & \left\lbrace 
	  -  \theta_{21}  \theta_{34}  \left[ B^{\1 \2 \3 \4}\tb{1}{2}{3}{4} - B^{\4 \3 \2 \1}\tb{4}{3}{2}{1}\right] \right. \nn \\  
& + \theta_{21} \left( \theta_{42} - \theta_{43} \right) \left[ B^{\1 \2 \4 \3}\tb{1}{2}{4}{3} - B^{\3 \4 \2 \1}\tb{3}{4}{2}{1}\right] \nn \\ 
& -  \theta_{31} \theta_{24} \left[ B^{\1 \3 \2 \4}\tb{1}{3}{2}{4} - B^{\4 \2 \3 \1}\tb{4}{2}{3}{1}\right]  \nn\\
& + \theta_{31} \left( \theta_{43} - \theta_{42} \right) \left[  B^{\1 \3 \4 \2}\tb{1}{3}{4}{2} - B^{\2 \4 \3 \1}\tb{2}{4}{3}{1}\right] \nn \\
& +  \theta_{24} \theta_{41}  \left[ B^{\1 \4 \2 \3}\tb{1}{4}{2}{3} - B^{\3 \2 \4 \1}\tb{3}{2}{4}{1}\right] \nn \\
& +  \theta_{34} \theta_{41}  \left[ B^{\1 \4 \3 \2}\tb{1}{4}{3}{2} - B^{\2 \3 \4 \1}\tb{2}{3}{4}{1}\right] \nn \\
& + \theta_{34} \left( \theta_{13} - \theta_{12}  \right) \left[ B^{\2 \1 \3 \4}\tb{2}{1}{3}{4} - B^{\4 \3 \1 \2}\tb{4}{3}{1}{2}\right] \nn \\
& + \left( \theta_{21}  \theta_{14}\theta_{43} - \theta_{34} \theta_{41} \theta_{12} \right) \left[ B^{\2 \1 \4 \3}\tb{2}{1}{4}{3} - B^{\3 \4 \1 \2}\tb{3}{4}{1}{2} \right] \nn \\
& +  \theta_{31}\theta_{14} \left[ B^{\2 \3 \1 \4}\tb{2}{3}{1}{4} - B^{\4 \1 \3 \2}\tb{4}{1}{3}{2} \right] \nn \\
& + \left( \theta_{24} \theta_{41}\theta_{13} +\theta_{31}\theta_{14}\theta_{42} \right) \left[ B^{\2 \4 \1 \3}\tb{2}{4}{1}{3} - B^{\3 \1 \4 \2}\tb{3}{1}{4}{2} \right] \nn \\
& + \theta_{24} \left( \theta_{12} - \theta_{13} \right) \left[ B^{\3 \1 \2 \4}\tb{3}{1}{2}{4} - B^{\4 \2 \1 \3}\tb{4}{2}{1}{3}\right]  \nn \\
& \left. + \theta_{21} \theta_{14} \left[ B^{\3 \2 \1 \4}\tb{3}{2}{1}{4} - B^{\4 \1 \2 \3}\tb{4}{1}{2}{3} \right]  \right\rbrace  ,
\end{align}
\begin{align}
G_{qccc}  = \frac{\ii}{2} & \left\lbrace 
	   -   \theta_{21} \left( \theta_{32} - \theta_{23}\theta_{34} \right) \left[ B^{\1 \2 \3 \4}\tb{1}{2}{3}{4} - B^{\4 \3 \2 \1}\tb{4}{3}{2}{1}\right] \right. \nn \\  
& - \theta_{21} \left( \theta_{42} - \theta_{24}\theta_{43} \right)\left[ B^{\1 \2 \4 \3}\tb{1}{2}{4}{3} - B^{\3 \4 \2 \1}\tb{3}{4}{2}{1}\right] \nn \\ 
& - \theta_{31} \left( \theta_{23} - \theta_{32}\theta_{24} \right) \left[ B^{\1 \3 \2 \4}\tb{1}{3}{2}{4} - B^{\4 \2 \3 \1}\tb{4}{2}{3}{1}\right]  \nn\\
& - \theta_{31} \left( \theta_{43} - \theta_{34}\theta_{42} \right)\left[  B^{\1 \3 \4 \2}\tb{1}{3}{4}{2} - B^{\2 \4 \3 \1}\tb{2}{4}{3}{1}\right] \nn \\
& - \theta_{41} \left( \theta_{24} - \theta_{42}\theta_{24} \right)  \left[ B^{\1 \4 \2 \3}\tb{1}{4}{2}{3} - B^{\3 \2 \4 \1}\tb{3}{2}{4}{1}\right] \nn \\
& - \theta_{41} \left( \theta_{34} - \theta_{43}\theta_{32} \right) \left[ B^{\1 \4 \3 \2}\tb{1}{4}{3}{2} - B^{\2 \3 \4 \1}\tb{2}{3}{4}{1}\right] \nn \\
& + \left( \theta_{21}\theta_{13}\theta_{34} - \theta_{31}\theta_{12}  \right) \left[ B^{\2 \1 \3 \4}\tb{2}{1}{3}{4} - B^{\4 \3 \1 \2}\tb{4}{3}{1}{2}\right] \nn \\
&  + \left( \theta_{21}\theta_{14}\theta_{43} - \theta_{41}\theta_{12}  \right) \left[ B^{\2 \1 \4 \3}\tb{2}{1}{4}{3} - B^{\3 \4 \1 \2}\tb{3}{4}{1}{2} \right] \nn \\
&  + \left( \theta_{31}\theta_{14} - \theta_{41}\theta_{13}\theta_{32}  \right) \left[ B^{\2 \3 \1 \4}\tb{2}{3}{1}{4} - B^{\4 \1 \3 \2}\tb{4}{1}{3}{2} \right] \nn \\
&  + \left( \theta_{41}\theta_{13} - \theta_{31}\theta_{14}\theta_{42}  \right) \left[ B^{\2 \4 \1 \3}\tb{2}{4}{1}{3} - B^{\3 \1 \4 \2}\tb{3}{1}{4}{2} \right] \nn \\
&  + \left( \theta_{31}\theta_{12}\theta_{24} - \theta_{21}\theta_{13}  \right) \left[ B^{\3 \1 \2 \4}\tb{3}{1}{2}{4} - B^{\4 \2 \1 \3}\tb{4}{2}{1}{3}\right]  \nn \\
& \left. + \left( \theta_{21}\theta_{14} - \theta_{41}\theta_{12}\theta_{23}  \right) \left[ B^{\3 \2 \1 \4}\tb{3}{2}{1}{4} - B^{\4 \1 \2 \3}\tb{4}{1}{2}{3} \right]  \right\rbrace  \nn \\
- \ii & \theta_{41} \left[ C^{\4 \1}\tc{4}{1} - C^{\1\4} \tc{1}{4}  \right]  \left[ C^{\2 \3}\tc{2}{3} - C^{\3\2} \tc{3}{2}  \right],
\end{align}
\begin{align}
G_{ccqq}  = \frac{\ii}{2} & \left\lbrace 
	     \theta_{23}\theta_{34} \left[ B^{\1 \2 \3 \4}\tb{1}{2}{3}{4} - B^{\4 \3 \2 \1}\tb{4}{3}{2}{1}\right] \right. \nn \\  
& + \theta_{24}\theta_{43} \left[ B^{\1 \2 \4 \3}\tb{1}{2}{4}{3} - B^{\3 \4 \2 \1}\tb{3}{4}{2}{1}\right] \nn \\ 
& + \theta_{24} \left( \theta_{32} - \theta_{31} \right) \left[ B^{\1 \3 \2 \4}\tb{1}{3}{2}{4} - B^{\4 \2 \3 \1}\tb{4}{2}{3}{1}\right]  \nn\\
& + \left( \theta_{13} \theta_{34} \theta_{42} + \theta_{24}\theta_{43}\theta_{31} \right)\left[  B^{\1 \3 \4 \2}\tb{1}{3}{4}{2} - B^{\2 \4 \3 \1}\tb{2}{4}{3}{1}\right] \nn \\
& + \theta_{23} \left( \theta_{42} - \theta_{41} \right)  \left[ B^{\1 \4 \2 \3}\tb{1}{4}{2}{3} - B^{\3 \2 \4 \1}\tb{3}{2}{4}{1}\right] \nn \\
& + \left( \theta_{14}\theta_{43}\theta_{32} + \theta_{23}\theta_{34}\theta_{41} \right) \left[ B^{\1 \4 \3 \2}\tb{1}{4}{3}{2} - B^{\2 \3 \4 \1}\tb{2}{3}{4}{1}\right] \nn \\
& + \theta_{13}\theta_{34} \left[ B^{\2 \1 \3 \4}\tb{2}{1}{3}{4} - B^{\4 \3 \1 \2}\tb{4}{3}{1}{2}\right] \nn \\
& + \theta_{14}\theta_{43} \left[ B^{\2 \1 \4 \3}\tb{2}{1}{4}{3} - B^{\3 \4 \1 \2}\tb{3}{4}{1}{2} \right] \nn \\
& + \theta_{14} \left( \theta_{31} - \theta_{32}  \right) \left[ B^{\2 \3 \1 \4}\tb{2}{3}{1}{4} - B^{\4 \1 \3 \2}\tb{4}{1}{3}{2} \right] \nn \\
& + \theta_{13} \left( \theta_{41} - \theta_{42}  \right) \left[ B^{\2 \4 \1 \3}\tb{2}{4}{1}{3} - B^{\3 \1 \4 \2}\tb{3}{1}{4}{2} \right] \nn \\
&  - \theta_{13}\theta_{24} \left[ B^{\3 \1 \2 \4}\tb{3}{1}{2}{4} - B^{\4 \2 \1 \3}\tb{4}{2}{1}{3}\right]  \nn \\
& \left. - \left( \theta_{23}\theta_{14} - \theta_{41}\theta_{12}\theta_{23}  \right) \left[ B^{\3 \2 \1 \4}\tb{3}{2}{1}{4} - B^{\4 \1 \2 \3}\tb{4}{1}{2}{3} \right]  \right\rbrace  \nn \\
- \ii & \theta_{14}\theta_{23} \left[ C^{\1 \4}\tc{1}{4} - C^{\4\1} \tc{4}{1}  \right]  \left[ C^{\2 \3}\tc{2}{3} - C^{\3\2} \tc{3}{2}  \right],
\end{align}
\begin{align}
G_{cccq}  = \frac{\ii}{2} & \left\lbrace 
	   \theta_{34} \left( \theta_{32}\theta_{21} + \theta_{23} \right) \left[ B^{\1 \2 \3 \4}\tb{1}{2}{3}{4} - B^{\4 \3 \2 \1}\tb{4}{3}{2}{1}\right] \right. \nn \\  
& + \left(\theta_{24}\theta_{43} - \theta_{34} \theta_{42} \theta_{21} \right)\left[ B^{\1 \2 \4 \3}\tb{1}{2}{4}{3} - B^{\3 \4 \2 \1}\tb{3}{4}{2}{1}\right] \nn \\ 
& + \theta_{24} \left( \theta_{23}\theta_{31} - \theta_{32} \right) \left[ B^{\1 \3 \2 \4}\tb{1}{3}{2}{4} - B^{\4 \2 \3 \1}\tb{4}{2}{3}{1}\right]  \nn\\
& + \left( \theta_{34} \theta_{42} - \theta_{24}\theta_{43}\theta_{31} \right)\left[  B^{\1 \3 \4 \2}\tb{1}{3}{4}{2} - B^{\2 \4 \3 \1}\tb{2}{4}{3}{1}\right] \nn \\
& +  \left( \theta_{14} \theta_{42}\theta_{23} - \theta_{24}\theta_{41} \right)  \left[ B^{\1 \4 \2 \3}\tb{1}{4}{2}{3} - B^{\3 \2 \4 \1}\tb{3}{2}{4}{1}\right] \nn \\
& + \left( \theta_{14}\theta_{43}\theta_{32} - \theta_{34}\theta_{41} \right) \left[ B^{\1 \4 \3 \2}\tb{1}{4}{3}{2} - B^{\2 \3 \4 \1}\tb{2}{3}{4}{1}\right] \nn \\
& + \theta_{34} \left( \theta_{31}\theta_{12} + \theta_{13} \right) \left[ B^{\2 \1 \3 \4}\tb{2}{1}{3}{4} - B^{\4 \3 \1 \2}\tb{4}{3}{1}{2}\right] \nn \\
& + \left( \theta_{14}\theta_{43} - \theta_{34}\theta_{41}\theta_{12} \right) \left[ B^{\2 \1 \4 \3}\tb{2}{1}{4}{3} - B^{\3 \4 \1 \2}\tb{3}{4}{1}{2} \right] \nn \\
& + \theta_{14} \left( \theta_{13}\theta_{32} + \theta_{31}  \right) \left[ B^{\2 \3 \1 \4}\tb{2}{3}{1}{4} - B^{\4 \1 \3 \2}\tb{4}{1}{3}{2} \right] \nn \\
& +  \left( \theta_{24} \theta_{41}\theta_{13}  - \theta_{14}\theta_{42}  \right) \left[ B^{\2 \4 \1 \3}\tb{2}{4}{1}{3} - B^{\3 \1 \4 \2}\tb{3}{1}{4}{2} \right] \nn \\
&  + \theta_{24} \left( \theta_{21} \theta_{13} + \theta_{12} \right) \left[ B^{\3 \1 \2 \4}\tb{3}{1}{2}{4} - B^{\4 \2 \1 \3}\tb{4}{2}{1}{3}\right]  \nn \\
& \left. + \theta_{14} \left( \theta_{12}\theta_{23} + \theta_{21}  \right) \left[ B^{\3 \2 \1 \4}\tb{3}{2}{1}{4} - B^{\4 \1 \2 \3}\tb{4}{1}{2}{3} \right]  \right\rbrace  \nn \\
- \ii & \theta_{14} \left[ C^{\1 \4}\tc{1}{4} - C^{\4\1} \tc{4}{1}  \right]  \left[ C^{\2 \3}\tc{2}{3} + C^{\3\2} \tc{3}{2}  \right],
\end{align}
\begin{align}
G_{cccc}  = \frac{\ii}{2} & \left\lbrace 
	   \left( \theta_{23}\theta_{34} + \theta_{32}\theta_{21} \right) \left[ B^{\1 \2 \3 \4}\tb{1}{2}{3}{4} - B^{\4 \3 \2 \1}\tb{4}{3}{2}{1}\right] \right. \nn \\  
&+ \left( \theta_{24}\theta_{43} + \theta_{42}\theta_{21} \right) \left[ B^{\1 \2 \4 \3}\tb{1}{2}{4}{3} - B^{\3 \4 \2 \1}\tb{3}{4}{2}{1}\right] \nn \\ 
& + \left( \theta_{32}\theta_{24} + \theta_{23}\theta_{31} \right) \left[ B^{\1 \3 \2 \4}\tb{1}{3}{2}{4} - B^{\4 \2 \3 \1}\tb{4}{2}{3}{1}\right]  \nn\\
& + \left( \theta_{34}\theta_{42} + \theta_{43}\theta_{31} \right) \left[  B^{\1 \3 \4 \2}\tb{1}{3}{4}{2} - B^{\2 \4 \3 \1}\tb{2}{4}{3}{1}\right] \nn \\
& + \left( \theta_{42}\theta_{23} + \theta_{24}\theta_{41} \right) \left[ B^{\1 \4 \2 \3}\tb{1}{4}{2}{3} - B^{\3 \2 \4 \1}\tb{3}{2}{4}{1}\right] \nn \\
& + \left( \theta_{43}\theta_{32} + \theta_{34}\theta_{41} \right) \left[ B^{\1 \4 \3 \2}\tb{1}{4}{3}{2} - B^{\2 \3 \4 \1}\tb{2}{3}{4}{1}\right] \nn \\
& + \left( \theta_{13}\theta_{34} + \theta_{31}\theta_{12} \right) \left[ B^{\2 \1 \3 \4}\tb{2}{1}{3}{4} - B^{\4 \3 \1 \2}\tb{4}{3}{1}{2}\right] \nn \\
& + \left( \theta_{14}\theta_{43} + \theta_{41}\theta_{12} \right) \left[ B^{\2 \1 \4 \3}\tb{2}{1}{4}{3} - B^{\3 \4 \1 \2}\tb{3}{4}{1}{2} \right] \nn \\
& + \left( \theta_{31}\theta_{14} + \theta_{13}\theta_{32} \right) \left[ B^{\2 \3 \1 \4}\tb{2}{3}{1}{4} - B^{\4 \1 \3 \2}\tb{4}{1}{3}{2} \right] \nn \\
& + \left( \theta_{41}\theta_{13} + \theta_{14}\theta_{42} \right) \left[ B^{\2 \4 \1 \3}\tb{2}{4}{1}{3} - B^{\3 \1 \4 \2}\tb{3}{1}{4}{2} \right] \nn \\
& + \left( \theta_{12}\theta_{24} + \theta_{21}\theta_{13} \right) \left[ B^{\3 \1 \2 \4}\tb{3}{1}{2}{4} - B^{\4 \2 \1 \3}\tb{4}{2}{1}{3}\right]  \nn \\
& \left. + \left( \theta_{21}\theta_{14} + \theta_{12}\theta_{23} \right) \left[ B^{\3 \2 \1 \4}\tb{3}{2}{1}{4} - B^{\4 \1 \2 \3}\tb{4}{1}{2}{3} \right]  \right\rbrace  \nn \\
- \ii &  \left[ C^{\1 \4}\tc{1}{4} + C^{\4\1} \tc{4}{1}  \right]  \left[ C^{\2 \3}\tc{2}{3} + C^{\3\2} \tc{3}{2}  \right].
\end{align}

\end{widetext}

\end{appendix}

\bibliography{Paper_references}

\end{document}